\documentclass[lettersize,journal]{IEEEtran}
\usepackage{amsmath,amsfonts}
\usepackage{algorithmic}
\usepackage{algorithm}
\usepackage{array}
\usepackage[caption=false,font=normalsize,labelfont=sf,textfont=sf]{subfig}
\usepackage{textcomp}
\usepackage{stfloats}
\usepackage{url}
\usepackage{verbatim}
\usepackage{graphicx}
\usepackage{cite}
\usepackage{pifont}
\usepackage{amsthm}
\usepackage{cite}
\usepackage{amsmath,amssymb,amsfonts}
\usepackage{amsfonts,amssymb}
\usepackage{algorithmic}
\usepackage{tikz}
 \usepackage{multirow}
\usepackage{textcomp}
\usepackage{comment}
\usepackage{amsmath,xparse}
\usepackage{xcolor}
\usepackage{graphicx}
\usepackage{epstopdf}
\usepackage{subfig}
\usepackage{amsmath}
\usepackage{color}
\usepackage{algorithm}
\usepackage{comment}
\usepackage{setspace}

\usepackage{amsmath}
 \usepackage{enumitem}

\newtheorem{definition}{\textbf{Definition}}

\newtheorem{lemma}{\textbf{Lemma}}
\newtheorem{observation}{\textbf{Observation}}
\newcommand*{\circled}[1]{\lower.7ex\hbox{\tikz\draw (0pt, 0pt) circle (.5em) node {\makebox[1em][c]{\small #1}};}}

\def\BibTeX{{\rm B\kern-.05em{\sc i\kern-.025em b}\kern-.08em
		T\kern-.1667em\lower.7ex\hbox{E}\kern-.125emX}}
\begin{document}

\title{Blockchain Mining with Multiple Selfish Miners}

\author{\IEEEauthorblockN{Qianlan Bai, Yuedong Xu, Nianyi Liu, Xin Wang}
\thanks{ Q. Bai and X. Wang are with the School of Computer Science, Fudan University, Shanghai 200438, China.

 Y. Xu and N. Liu are with the School of Information Science and Technology, Fudan University, Shanghai 200438, China.}
\thanks{This work was supported in part by the Natural Science Foundation of China under Grant Grant 62072117, in part by the Shanghai Natural Science Foundation under the Grant 22ZR1407000, in part by Shanghai-Hong Kong Collaborative Project under Grant 18510760900 and and in part by Zhuhai Research Institute of Fudan University.}
\thanks{Yuedong Xu is the corresponding author.}}
\maketitle
	
\begin{abstract}

This paper studies a fundamental problem regarding the security of blockchain PoW consensus on how the existence of multiple misbehaving miners influences the profitability of selfish mining. Each selfish miner maintains a private chain and makes it public opportunistically for acquiring more rewards incommensurate to his Hash power. We first establish a general Markov chain model to characterize the state transition of public and private chains for Basic Selfish Mining (\emph{BSM}), and derive the stationary \emph{profitable threshold} of Hash power in closed form. It reduces from 25\% for a single attacker to below 21.48\% for two symmetric attackers theoretically, and further reduces to around 10\% with eight symmetric attackers experimentally. We next explore the profitable threshold when one of the attackers performs strategic mining based on Partially Observable Markov Decision Process (POMDP) that only half of the attributes pertinent to a mining state are observable to him. An online algorithm is presented to compute the nearly optimal policy efficiently despite the large state space and high dimensional belief space. 
The profitable threshold is much lower for the strategic attacker.
Last, we formulate a simple model of absolute mining revenue that yields an interesting observation: selfish mining is never profitable at the first difficulty adjustment period, but relies on the reimbursement of stationary selfish mining gains in future periods. The delay till being profitable of an attacker increases with the decrease of his Hash power, making blockchain miners more cautious about performing selfish mining.  
		
\end{abstract}
	
	
\section{Introduction}
Bitcoin has gained tremendous concerns as the first fully decentralized cryptocurrency since its advent in 2008. All historical transactions are recorded in a global and public data structure known as \emph{blockchain}.
The security of blockchain relies on consensus algorithms to come to an agreement among a large body of pseudonymous participants called \emph{miners} \cite{ref:whitepaper}.
Solving a Hash puzzle is deemed as a way to generate Proof-of-Work (PoW) for reaching global consensus.
Each miner competes in this ``game'', and is rewarded with cryptocurrencies if he 
finds a valid block first. The PoW consensus has been widely deployed in public blockchains, serving as the cornerstone of current major cryptocurrencies. 
	
The security of PoW is challenged by the trend of centralization of Hash power. Mining a Bitcoin block is random and it needs {more than 10 years} on average with a latest-generation ASIC chip \cite{ref:10 years}. Therefore, miners operate cooperatively to form pools to acquire a stable income rate. As a side effect, a small number of mining pools occupy a vast majority of global Hash power, placing blockchain systems at risk of being overthrown. The conventional wisdom believes that PoW is secure as long as no mining pool controls 51\% of total Hash power. However, a miner can choose to mine selfishly instead of conforming to the standard Bitcoin protocol. 
	
Selfish mining refers to a class of block publishing policies in which a miner does not release his newly found block immediately, but forks a private chain of which others are unaware. At a future epoch, he will release his private blocks strategically to obsolete the current public chain for the purpose of obtaining a higher share of valid blocks in the new public chain than his ratio of Hash power. The minimum ratio of Hash power that brings extra rewards is conventionally called \emph{profitable threshold}. Eyal and Sirer introduced the first selfish mining scheme (namely basic selfish mining, \emph{BSM}) and pointed out that the profitable threshold of \emph{BSM} is 25\% of total Hash power \cite{ref:majority}. Nayak et al. \cite{ref:stubborn} proposed the stubborn mining that improves the revenue of the selfish miner by 13.94\% compared to \emph{BSM}. 
Sapirshtein et al. modeled the optimal selfish mining as a Markov Decision Process (MDP) that reduces the selfish miner's profitable threshold to 23.21\% \cite{ref:optimal selfish mining}. Tao et al. \cite{ref:hidden} introduced the semi-selfish mining attack based on hidden MDP with the control of fork rate. Grunspan and Perez-Marco proved rigorously using martingale theory that selfish mining is an attack on the difficulty adjustment algorithm of blockchain consensus \cite{ref:on profitability of selfish mining}. Recently, a lot of efforts have been devoted to the compound attacks of selfish mining with block withholding attacks \cite{ref:FAW} \cite{ref:selfholding}, bribery attacks \cite{ref:PAW}, eclipse attacks and double spending attacks \cite{ref:on the security}.

Until very recently, the competition of multiple non-colluding selfish miners came into view. 
Liu et al. presented the \emph{publish-$n$} scheme for two selfish mining attackers and simulated their revenues as well as profitable thresholds \cite{ref:ruanna}. Bai et al. modeled the mining race among two \emph{BSM} miners and one honest miner as a Markov process \cite{ref:a deep dive}. Zhang et al. \cite{ref:n attackers1} simulated the profitable threshold in the presence of multiple selfish miners. Charlie et al. \cite{ref:squirRL} proposed SquirRL, a framework for using deep reinforcement learning (DRL) to analyze selfish mining and block withholding attacks in blockchain systems with two attackers.
Previous experimental studies, though insightful, lack theoretical understandings on the principle of competitive selfish mining. When compared to the scenario of a solitary attacker, the presence of multiple selfish mining attackers results in two noteworthy effects. First, the fierce competition among attackers will make the system state transition more complex. Second, non-colluding attackers conduct strategic mining independently, making all miners incapable of acquiring comprehensive information about the network. Consequently, it is usually believed that modeling the profitable threshold with multiple selfish miners is hard, and the optimal mining may be intractable due to the above-mentioned challenges.

In this paper, we \textbf{theoretically} investigate the profitability of selfish mining with multiple attackers by asking a sequence of key progressive questions: \emph{1) Will selfish mining become more easily profitable with multiple attackers than with a solo attacker? 2) How can we design a nearly optimal mining policy for an attacker despite the complex interactions among miners and the incomplete observations of the system state? 3) How long should a {BSM} attacker wait from the beginning of selfish mining until being profitable eventually? } Figuring out these questions will provide essential understandings of the security of blockchain PoW consensus. 

We formulate a Markov chain model to compute the  \emph{relative revenue} of \emph{BSM} attackers for the first question, which is the percentage of his \emph{valid} blocks in the public chain. Our model is very general in the sense that it can capture cases with more than two attackers or allow an attacker to hide multiple blocks privately. In particular, the latter case may cause the complicated chain-reaction release in which the publishing action of one attacker triggers that of the other attacker. We demonstrate how the existence of multiple attackers affects the system security from a theoretic perspective.
	
Answering the second question is very challenging if a strategic attacker with a nearly optimal mining policy (Alice), a \emph{BSM} attacker (Bob) and an honest miner (Henry) coexist in the system. Firstly, the interactions among three chains are more complicated. 
The state of the mining race is captured by a 10-tuple pertinent to the composition of all  chains and the forking status, as opposed to a 3-tuple in \emph{MDP}-based optimal policy for a single attacker.
 The number of actions is larger and the state-action pairs can be $10^2$ to $10^3$ times larger. 
Second, Alice as the strategic miner cannot observe the complicated state information. 
In fact, she is merely aware of 5 attributes at each state related to her private chain and the public chain. 
We formulate a family of parameterized partially observable Markov decision process (POMDP) models to characterize Alice's strategic mining policy 
with a continuous belief of the current state. 
To tackle the large state space and the high-dimensional belief space, we adopt AEMS2
to compute the nearly optimal mining policy. A binary search method similar to \cite{ref:optimal selfish mining} is used to find the maximum revenue among the family of POMDP models. 

As for the third question, we build a simple model to compute attackers' \emph{absolute revenue}, which is the average number of valid blocks received by each miner per unit of time. 
Since selfish mining is an attack on the difficulty adjustment algorithm (DAA), it is not profitable instantaneously even if the attacker's Hash power is above the profitable threshold. 
This model enables us to compute the number of DAA periods that lead to profitable selfish mining eventually. Meanwhile, we prove that the absolute revenue and relative revenue are approximately equivalent on a sufficiently long time scale. 
	
		Our major observations are summarized as below.

\begin{itemize}
\item \textbf{BSM.} The Markov chain is established to model the revenues of each miner when there are multiple basic selfish mining attackers in the system. We prove the profitable threshold of Hash power is below 21.48\% with two symmetric \emph{BSM} attackers from theory and experiment, as opposed to 25\% with a single \emph{BSM} attacker and 23.21\% with a single optimal attacker. More blocks allowed to hold privately or more attackers will reduce this threshold. 

\item \textbf{POMDP.} We propose \emph{POMDP}-based mining policy which brings more revenues to the strategic miner than \emph{BSM} and honest mining.
When the \emph{BSM} attacker (Bob) has 34\% Hash power, the other attacker's (Alice's) profitable threshold decreases from 29.44\% to about 2\% if she chooses \emph{POMDP}-based policy rather than \emph{BSM}. The designed online algorithm can rapidly and effectively compute the near optimal action under the current observable information.

\item \textbf{Profitable Delay.} We make a transient analysis of basic selfish mining and explore the benefit time. A \emph{BSM} miner receives less absolute revenue than honest mining in the first difficulty adjustment period even if his Hash power is above the profitable threshold, and his gain is achieved in future periods. \emph{BSM} is profitable after 51 rounds of difficulty adjustment (i.e. 714 days in Bitcoin) if the Hash power of two symmetric selfish miners is 22\%. This delay decreases to 5 rounds (i.e. 70 days in Bitcoin) as their Hash power accrues to 33\%, which is still quite long. 
\end{itemize}

The remainder of this paper is organized as follows. Section II describes the background of selfish mining. Section III models the relative revenue of basic selfish mining with different attackers. Section IV proposed the {\emph{POMDP}}-based mining policy and designed the efficient algorithm. The profitable time of basic selfish mining is modeled in Section V. Section VI validates the revenue model of \emph{BSM} and the performance of the {\emph{POMDP}}-based policy. Section VII introduces the related work, and Section VIII concludes our work.

\section{Selfish-Mine Strategy}
\label{sec:systemmodel}
	
In this section, we present the block-release procedure of blockchain mining in the presence of two adversarial miners. We further introduce the new features on tie-breaking and chain-reaction release. 
	
\subsection{System Description}
	
Consider a blockchain system with two misbehaving attackers \footnote{A malicious mining pool can be treated as a miner because its Hash power is the sum of the Hash power of its members, and the pool manager is responsible for broadcasting any blocks mined by its members to the network.} \emph{Alice} and \emph{Bob}, as well as an honest miner,  $\emph{Henry}$ \footnote{Multiple honest miners can be boiled down to a single miner for the sake of their linear additivity of Hash powers  and the consistency of their mining strategies.}. They compete to mine a valid block for the purpose of acquiring bitcoin-like tokens. The proof-of-work (PoW) consensus is adopted and the mining of blocks is stateless: the probability of discovering a block by a miner is proportional to his current Hash power, but inversely proportional to the current aggregate Hash power of the entire blockchain network. The blockchain system dynamically adjusts the difficulty of cryptographic puzzles such that new blocks are generated at a fixed average rate (e.g., one block per 10 minutes on average in Bitcoin). The miners maintain a globally-agreed ordered set of transactions via the adoption and the mining on the longest chain. 
The reward of each valid block is normalized as one cryptographic coin. 
For simplicity, we make the following assumptions consistent with the literature \cite{ref:majority} \cite{ref:optimal selfish mining}:
\begin{itemize}
	\item The total Hash power of the blockchain system is normalized as a unit. Then, the Hash power of a miner is represented as a fraction of the total;
	\item The block discovery time is exponentially distributed.
\end{itemize}

These assumptions are obtained based on the PoW mining mechanism \cite{ref:ass1} \cite{ref:ass2}. They allow us to succinctly express the probability of generating a block per time unit and the probability that each miner generates the next block. It is worth noting that the memoryless of the exponential distribution guarantees that the probability of generating a block in the current time unit is not affected by how much time has already passed.

The honest miner Henry who finds a valid block will release it immediately. Alice (resp. Bob) may release her blocks strategically by forcing Henry into wasting his computation. When Alice and Bob are both selfish miners, the interaction between two private chains becomes more complicated because none of them knows the other's state. In what follows, we capture all the different states that each miner may encounter. 
	
Denote by ${\alpha}_{1}$, ${\alpha}_{2}$ and ${\alpha}_{h}$ the Hash powers of Alice, Bob and Henry respectively, i.e., \begin{math}\alpha_1+\alpha_2+\alpha_h=1\end{math}. Denote by $\gamma_1$ (resp. $\gamma_2$) the proportion that 
 all except Alice's (resp. Bob's) Hash power mines after Alice's (resp. Bob's) released chain in the tie-breaking between Alice (resp. Bob) and Henry. Denote by $\theta_1$ and $\theta_2$ the probabilities that honest miners choose to mine after Alice's and Bob's chains in the three-party tie-breaking, respectively. When the blockchain system creates a new block, it is mined by pool $i$ with the probability $\alpha_i$ , $\forall i\in\{1,2,h\}$, owing to the memorylessness of Hash computations.

\subsection{Basic Selfish Mining Mode}
	
Alice maintains a private chain, and so does Bob, while Henry operates on the public chain. Alice and Bob are not aware of each other's role. We suppose that all the miners work on the same public chain at the beginning where the starting point is expressed as ``0''. The length of the private chain is kept as private information by attackers, and the length of the public chain is observed by all of them. We consider the selfish mining method proposed by \cite{ref:majority}, and our analytical approach can be generalized to a variety of other methods. 
	
The \emph{mining procedure} consists of two cases as follows.
\begin{itemize}
	\item \emph{(Public-chain mining case)} Henry always mines after the public chain. Alice or Bob also mines on the public chain if it is longer than his private chain. 
		
	\item \emph{(Private-chain mining case)} Alice (resp. Bob) continues to mine on her (resp. his) private chain if she (resp. he) discovers a new block and the private chain is now longer than the public chain. 
		
\end{itemize}
	
The \emph{release procedure} is more complicated than the mining procedure. Henry broadcasts his mined block as soon as it is discovered, while Alice and Bob will decide whether to release their mined blocks depending on the length of the public chain. 
	
\begin{itemize}
		
	\item \emph{(Forfeit case)} Alice (resp. Bob) abandons her (resp. his) private chain and conforms to mining after the public chain if the latter is longer. Henry also abandons his public chain if Alice or Bob publishes a longer chain. 
		
	\item \emph{(Risk-avoiding release case)} Alice (resp. Bob) releases her (resp. his) privately mined blocks to the public if the new block is mined by the others and the leading advantage of her private chain is no more than two blocks. 
		
	\item \emph{(Chain reaction case)} When Alice (resp. Bob) releases her (resp. his) blocks to public chain, the release of Bob's (resp. Alice's) private blocks is triggered immediately. 
		
\end{itemize}
	
The chain reaction case is the combination of the forfeit and the risk-avoiding cases, whereas the existence of chain reaction complicates evolution of the public chain. Suppose that Alice publishes her private blocks to obsolete the current public chain. After the construction of the new public chain, Bob may release his private chain to forfeit it immediately. 
	
\vspace{-0.5cm}	
\subsection{Release procedure and tie-breaking Logics}
	
The consensus on the public chain requires that it is the longest. 
We illustrate the evolution of private and public chains using examples, where $A_k$, $B_k$, and $H_k$ denote that the $k^{th}$ block belongs to Alice, Bob and Henry respectively. 
	
\begin{figure}[!ht]
		\centering
		\includegraphics[width=3in]{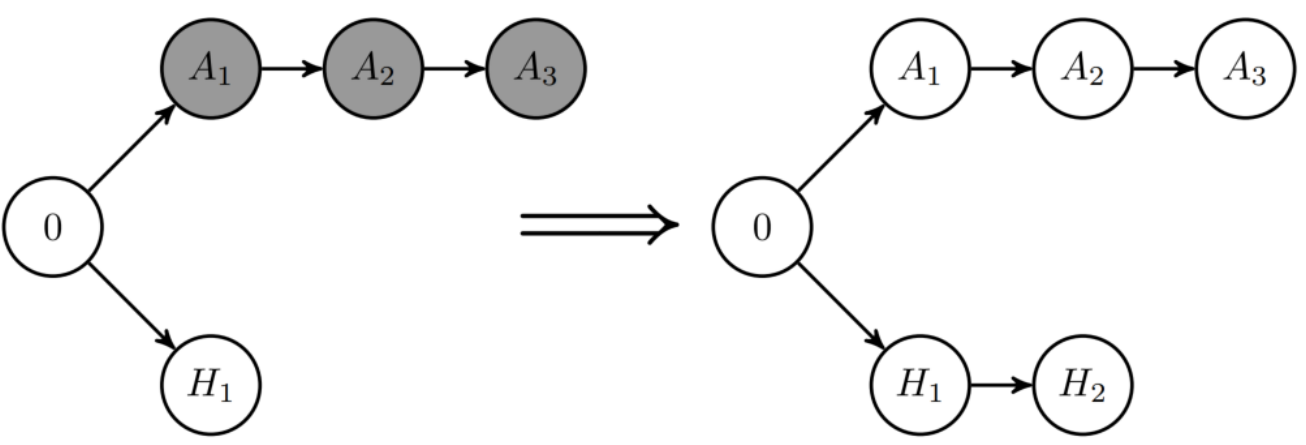}
		\vspace{-0.3cm}
		\caption{Alice's risk-avoiding release and Henry's abandonment. The blocks of private chains are in grey and those of public chains are in white.}
		\label{fig:forfeit}
		\vspace{-0.2cm}
	\end{figure}

\noindent\textbf{Risk-avoiding release case.} 
We show the risk-avoiding release of Alice's private chain in Fig.\ref{fig:forfeit} where the blocks of private chains are in grey and those of public chains are in white.  Alice is only one block ahead of Henry after the latter mines a new block for the public chain. 
Because Alice fears losing the competition, she publishes her private blocks, obsoleting Henry's public chain, 
so that both Alice and Henry mine on the new longest chain afterward. 

\begin{figure}[!ht]
		\vspace{-0.5cm}
		\centering
		\includegraphics[width=0.3\textwidth]{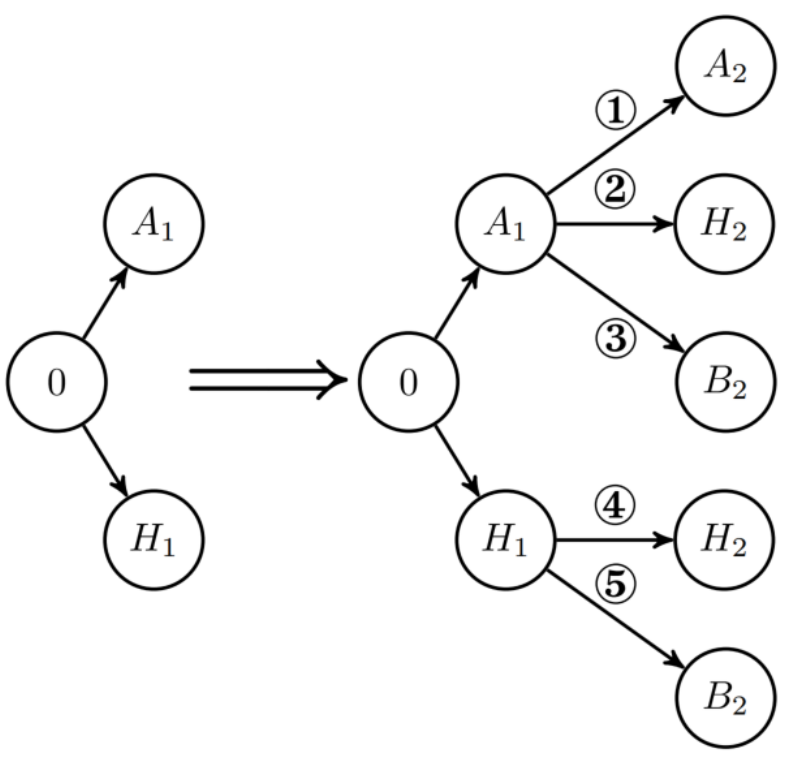}
		\vspace{-0.4cm}
		\caption{Tie-breaking case of two public chains.}
		\label{fig:tie_two}
		\vspace{-0.3cm}
	\end{figure}

\noindent\textbf{Tie-breaking resolving.} In Alice's case, if she only hides one block,
Henry may catch up with her. When it happens, Alice publishes her private blocks to compete with Henry. Thus, two public chains of the same length exist in Fig. \ref{fig:tie_two}. Since only one public chain prevails, a tie-breaking rule needs to be taken into account. The case is that the public chains of Alice and Henry have the same length, and Bob's private chain is 0. Hence, we only need to resolve the tie between Alice and Henry. All the miners are possible to mine after block $A_1$, while Bob and Henry may mine after $H_1$. There are five possibilities of extending the longest public chain, and the shorter one will be obsoleted. If Alice and Bob hide one block respectively, there will be three public chains with the same length. Alice and Bob will mine after their own chain, and Henry will choose one chain randomly. More details about tie-breaking with three public chains are described in Appendix A. 
	
	\begin{figure}[!ht]
		\centering
		\includegraphics[width=3in]{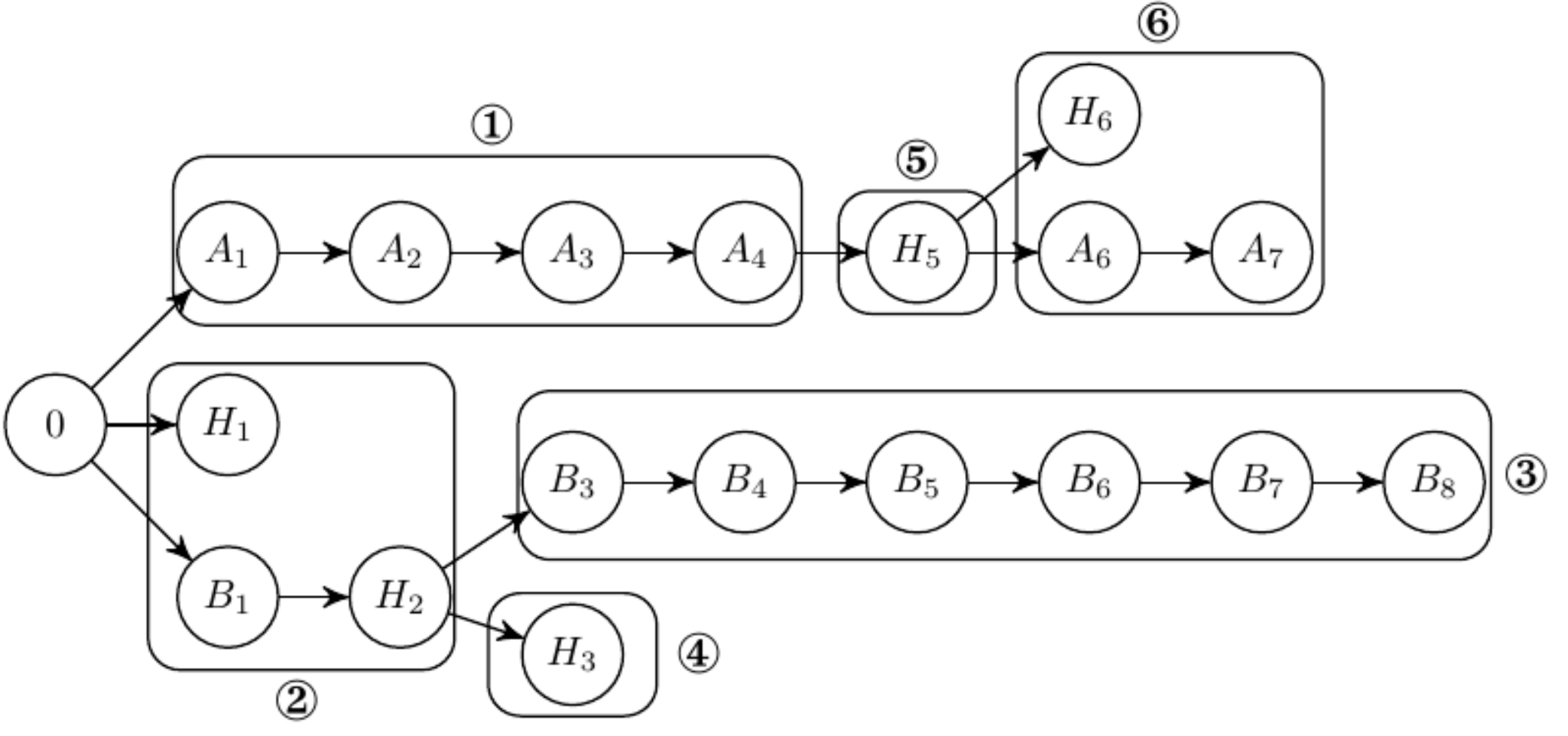}
		\vspace{-0.3cm}
		\caption{Chain reaction case.}
		\label{fig:chain_reaction}
		\vspace{-0.6cm}
	\end{figure}
	
\noindent\textbf{Chain reaction release.} 
We next introduce the chain reaction release.
Note that the chain reaction release consists of a sequence of risk-avoiding releases and tie-breaking resolvings. Fig. \ref{fig:chain_reaction} illustrates an example of how the chain reaction phenomenon is triggered.
First, Alice's private chain contains four blocks, while the lengths of Bob's private chain and Henry's public chain are 0. This process is marked as \ding{172} in Fig. \ref{fig:chain_reaction}. After a tie-breaking resolving at \ding{173}, the longer public chain contains two blocks $B_1$ and $H_2$, and the shorter one is orphaned. Bob constructs a new private chain starting from $B_3$ to $B_8$ at \ding{174}, while Henry continues to mine one block after $H_2$ at \ding{175}. From Alice's perspective, her private chain is merely one block ahead of the public chain. She releases her private blocks in order to avoid the risk of losing the race with Henry. The new public longest chain now starts from block $A_4$. Next, \ding{176} and \ding{177} constitute a new round of tie-breaking resolving between Alice and Henry, extending the public chain to block $A_7$. However, the release of $A_7$ triggers Bob to release all of his private blocks starting from $B_3$ to $B_8$. When retrospecting the whole mining process,
we observe that the winning branch switches back and forth, making the analysis of selfish mining extremely complicated. To be noted, the chain reaction occurs only when the length of the private chain is greater than three.

\section{Stationary Analysis of Basic Selfish Mining}
\label{sec:stationary}

In this section, we present Markov chain models to characterize the block-publishing dynamics with multiple selfish miners. The expected revenues are derived in explicit form.
\vspace{-0.5cm}
\subsection{Definition} 

The theme of our study is placed on the profitability of selfish mining so that the profitable measures should be clarified first. For notational simplicity, we only consider three miners: Alice, Bob and Henry. 
We denote the \textbf{mining round} as the process in which all miners start mining on the same public chain, go through a series of block-hiding and block-releasing actions, and finally reach a consensus on the public chain again.

\begin{definition}({Relative Revenue})
{Let $R_a$, $R_b$ and $R_h$ be the expected numbers of valid blocks mined by Alice, Bob and Henry in a mining round, respectively. The relative revenue of a miner, $\hat{R}_i$, is expressed as
\begin{eqnarray}
\hat{R}_i = \frac{R_i}{R_a + R_b + R_h}, \;\;\;\;\;\; i\in \{a, b, h\}. \nonumber
\end{eqnarray}
}\end{definition}
It should be emphasized that the \emph{valid} blocks are confirmed blocks in the longest chain. Selfish mining attack will increase the number of orphan blocks that leads to a drop in the valid block generation rate. The Bitcoin mining protocol will adjust the mining difficulty, so that the mining rate of the public chain is kept at one block on average every 10 minutes. Therefore, the expected revenue needs to be normalized. We leave a more rigorous analysis of the relative revenue in Section \ref{sec:transient}.

The profitability of selfish mining does not refer to the surplus that the block reward subtracts the cost of cryptographic computation. In fact, it is a contrastive measure to the honest mining that needs an objective index. 
\begin{definition}({Profitability})
{The selfish or strategic mining performed by Alice (resp. Bob) is deemed profitable if the relative revenue is higher than the normalized Hash power, i.e. $\hat{R}_a > \alpha_1$ (resp. $\hat{R}_b > \alpha_2$).
}
\end{definition}

\vspace{-0.5cm}
\subsection{Stationary Analysis for Two Attackers} 
In order to analyze the profitability of selfish mining, we need to capture the states of the system which satisfy the Markov property, including the block generation and block release.
We hereby formulate a discrete-time Markov chain model similar to \cite{ref:majority} to characterize the dynamics of the public and private chains.
We begin with the assumption that each selfish miner will release his two private blocks immediately after he has mined the second one (i.e. $N=2$). The underlying reasons are two-fold. Firstly, the simpler block-release process avoids the complicated representation of Markov states, thus allowing more tractable mathematical modeling. Secondly, the bursty release of many valid blocks in a very short time usually indicates the existence of selfish mining attacks that can be easily detected.

\begin{figure}[h]
\vspace{-0.2cm}
	\centerline{\includegraphics[height=0.19\textheight,width=0.4\textwidth]{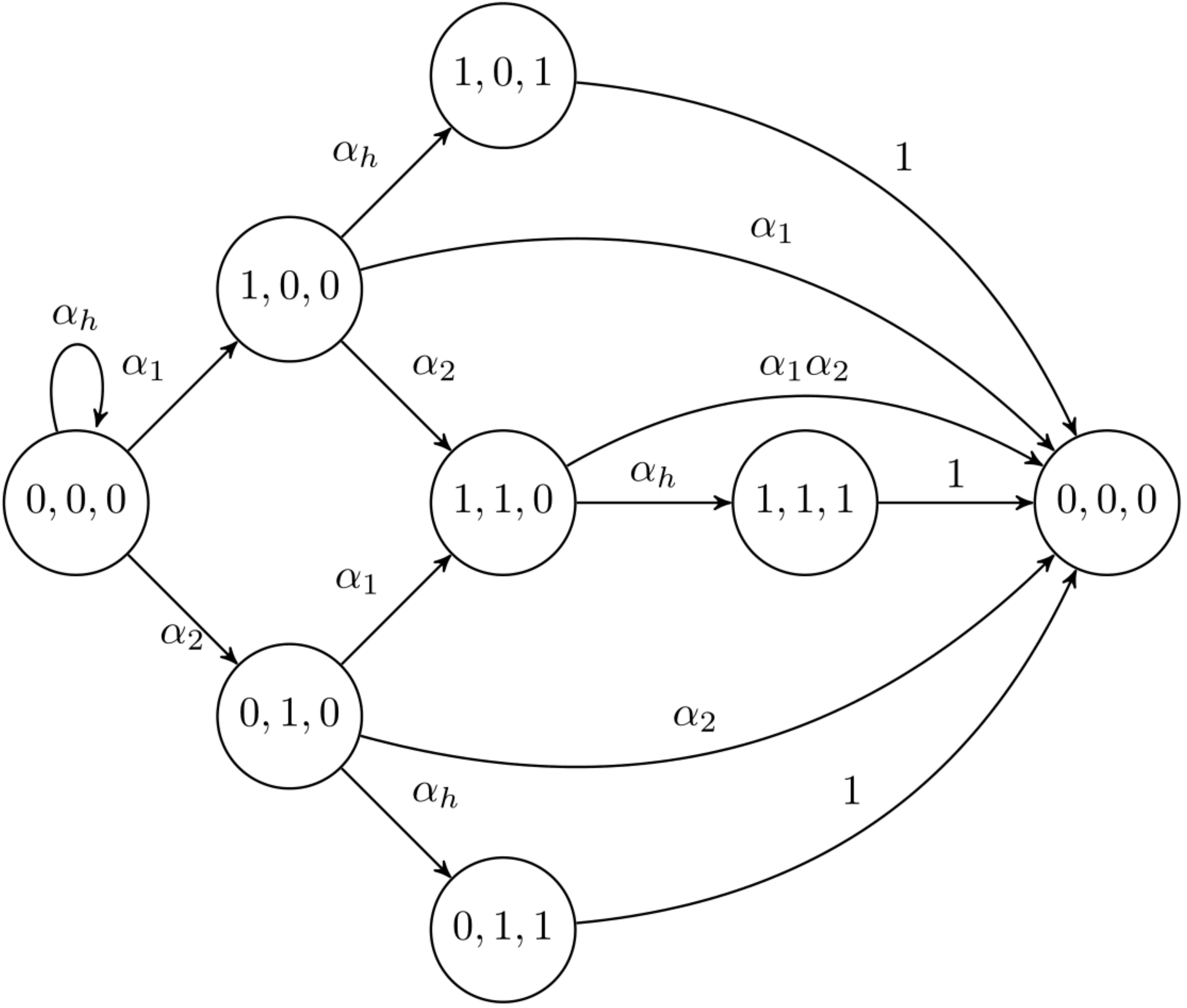}}
	\vspace{-0.2cm}
	\caption{Markov chain with less than two private blocks.}
	\label{fig:state machine2}
	\vspace{-0.3cm}
\end{figure} 

We define a \emph{state} as a three-tuple consisting of the lengths of Alice's, Bob's and Henry's chains. Fig. \ref{fig:state machine2} illustrates all the states, the state transition indicators and their transition probabilities. For instance, the transition from state $(0,0,0)$ to state $(1,0,0)$ means that Alice discovers a valid block with probability $\alpha_1$ and forks the private chain. If the maximum length of any private chain is below 2, Alice and Bob can hide their private chains and continue the selfish mining. All the transitions to state $(0,0,0)$ mean that the forked chains return to the unanimous public chain and a new round of selfish mining starts. Denote by $\mathbf{P}$ the state transition probability matrix and by $P_{ss'}$ the transition probability from state $s=(i,j,k)$ to $s'=(i',j',k')$. Let $\pi_{ijk}$ be the stationary distribution of state $(i,j,k)$. According to the \emph{detailed balance equation} \cite{ref:state} \cite{ref:markov chain}
\vspace{-0.2cm}
\begin{eqnarray}
	\pi_s = \sum_{s'} \pi_{s'} P_{s's}. 
\end{eqnarray}
We calculate the probability distribution of states and obtain the following equations:
\begin{align}
    \pi_{000}+\pi_{100}+\pi_{010}+\pi_{110}+\pi_{101}+\pi_{011}+\pi_{111}=1.
\end{align}
Then, we can compute $\pi_{000}$ as 
\begin{eqnarray}
		\pi_{000} = 
	(1{+}\alpha_1{+}\alpha_2{+}\alpha_1\alpha_h{+}2\alpha_1\alpha_2{+}\alpha_2\alpha_h {+}2\alpha_1\alpha_2\alpha_h)^{-1},
\end{eqnarray}
and $\pi_{ijk}$ at any other state $s=(i,j,k)$ in the same way. 

The transitions to state $(0,0,0)$ manifest which miner is the final winner in the current round of selfish mining. Therefore, we can compute the expected revenues of Alice, Bob and Henry that are defined as $R_a$, $R_b$ and $R_h$ respectively. Facilitated by the stationary state distributions, we calculate them as below,
\begin{align}
R_a\;&=2\alpha_1\pi_{100}+(2\alpha_1+(1-\alpha_1)\gamma_1)\pi_{101}+\alpha_1\pi_{011}\\\nonumber
&+2\alpha_1\pi_{110}+(2\alpha_1+\alpha_h\theta_1)\pi_{111};\\
  R_b\;&=2\alpha_2\pi_{010}+(2\alpha_2+(1-\alpha_2)\gamma_2)\pi_{011}+\alpha_2\pi_{101}\\\nonumber
  &+2\alpha_2\pi_{110}+(2\alpha_2+\alpha_h\theta_2)\pi_{111};  \\
    R_h\;&=(\alpha_h+(1-\alpha_1)(1-\gamma_1))\pi_{101}+\alpha_h(2-\theta_1-\theta_2)\pi_{111}\\\nonumber
    &+(\alpha_h+(1-\alpha_2)(1-\gamma_2))\pi_{011}+\alpha_h\pi_{000}.
\end{align}
Then we can obtain the expected revenue in closed form as:
	\begin{align}
	{R_a}=\;&{\pi_{000}}\cdot[2\alpha_1^2\left(1+\alpha_h \right)+\left(\alpha_2+\alpha_h \right)\alpha_1\alpha_h\gamma_1\nonumber\\
	&+\alpha_1\alpha_2\alpha_h +4\alpha_1^2\alpha_2\left(1+\alpha_h \right)+2\alpha_1\alpha_2\alpha_h^2\theta_{1}]; \\
	{R_b}=\;&{\pi_{000}}\cdot[2\alpha_2^2\left(1+\alpha_h \right)+\left(\alpha_1+\alpha_h \right)\alpha_2\alpha_h\gamma_2\nonumber\\
	&+\alpha_1\alpha_2\alpha_h +4\alpha_2^2\alpha_1\left(1+\alpha_h \right)+2\alpha_1\alpha_2\alpha_h^2\theta_{2}];\\
	{R_h}=\;&{\pi_{000}}\cdot[
	\alpha_1\alpha_h^2\left(2-\gamma_1 \right)+2\alpha_1\alpha_2\alpha_h^2\left(2{-}\theta_{1}-\theta_{2} \right)\nonumber\\
	&+\alpha_h+\alpha_2\alpha_h^2\left(2-\gamma_2 \right)+\alpha_1\alpha_2\alpha_h(2-\gamma_1-\gamma_2)].
\end{align}
The relative revenue of each miner can be given by:
\begin{align}
    \hat{R_i}=\frac{R_i}{R_a+R_b+R_h}, \forall i\in \{a,b,h\}.
\end{align}
The average number of Henry's orphaned blocks in each attack round is calculated as: 
\vspace{-0.2cm}
\begin{align}
    O_h=&\pi_{000}[(\alpha_1{+}(1{-}\alpha_1)\gamma_1)\alpha_1\alpha_h+(\alpha_2{+}(1{-}\alpha_2)\gamma_2)\alpha_2\alpha_h\nonumber\\
    &+(\alpha_1+\alpha_2+(\theta_1+\theta_2)\alpha_h)2\alpha_1\alpha_2\alpha_h].
    \label{eq:orphan two}
\end{align}
 As a special case that both selfish miners are homogeneous, i.e. $\alpha_1=\alpha_2 = \alpha < 0.5$, $\gamma_1=\gamma_2=0.5$ and $\theta_1=\theta_2=1/3$, the expected revenues can be simplified as 
\begin{eqnarray}
	&\pi_{000}=(1+4\alpha-4\alpha^3)^{-1};\nonumber
	\end{eqnarray}
	\begin{eqnarray}
	&R_a=R_b=\pi_{000}\cdot \frac{1}{6}\alpha(25\alpha+2\alpha^2+3-32\alpha^3); \label{eq:revneue formu 2_1}\\
	&R_h=\pi_{000}\cdot [(1-2\alpha)(1+3\alpha-\frac{7}{3}\alpha^2-\frac{16}{3}\alpha^3)].\label{eq:revneue formu 2_2}
\end{eqnarray}

{We can easily observe that the attackers' (resp. Henry's) expected revenues in Eq. \eqref{eq:revneue formu 2_1} (resp. Eq. \eqref{eq:revneue formu 2_2}) monotonically increase (resp. decrease) with regard to attackers' ratios of Hash power.}
\subsection{Scaling to Multiple Attackers}
\vspace{-0.1cm}
Although the profitable selfish mining demands a high Hash power, it is possible that multiple selfish miners opt in. 
The profitability of more selfish miners is obscure. 
The honest miner's share of Hash power decreases, and the competition 
among selfish miners becomes more fierce. 
Therefore, we consider a general mire scenario with $m$ $(m>2)$ \emph{BSM} miners. 
	
\begin{figure}[h]
	\vspace{-0.4cm}
	\begin{minipage}[t]{1\linewidth}
		\setlength\abovecaptionskip{-0.5pt}
		\setlength\belowcaptionskip{-1pt}
		\centering
		\includegraphics[width=1.1\textwidth]{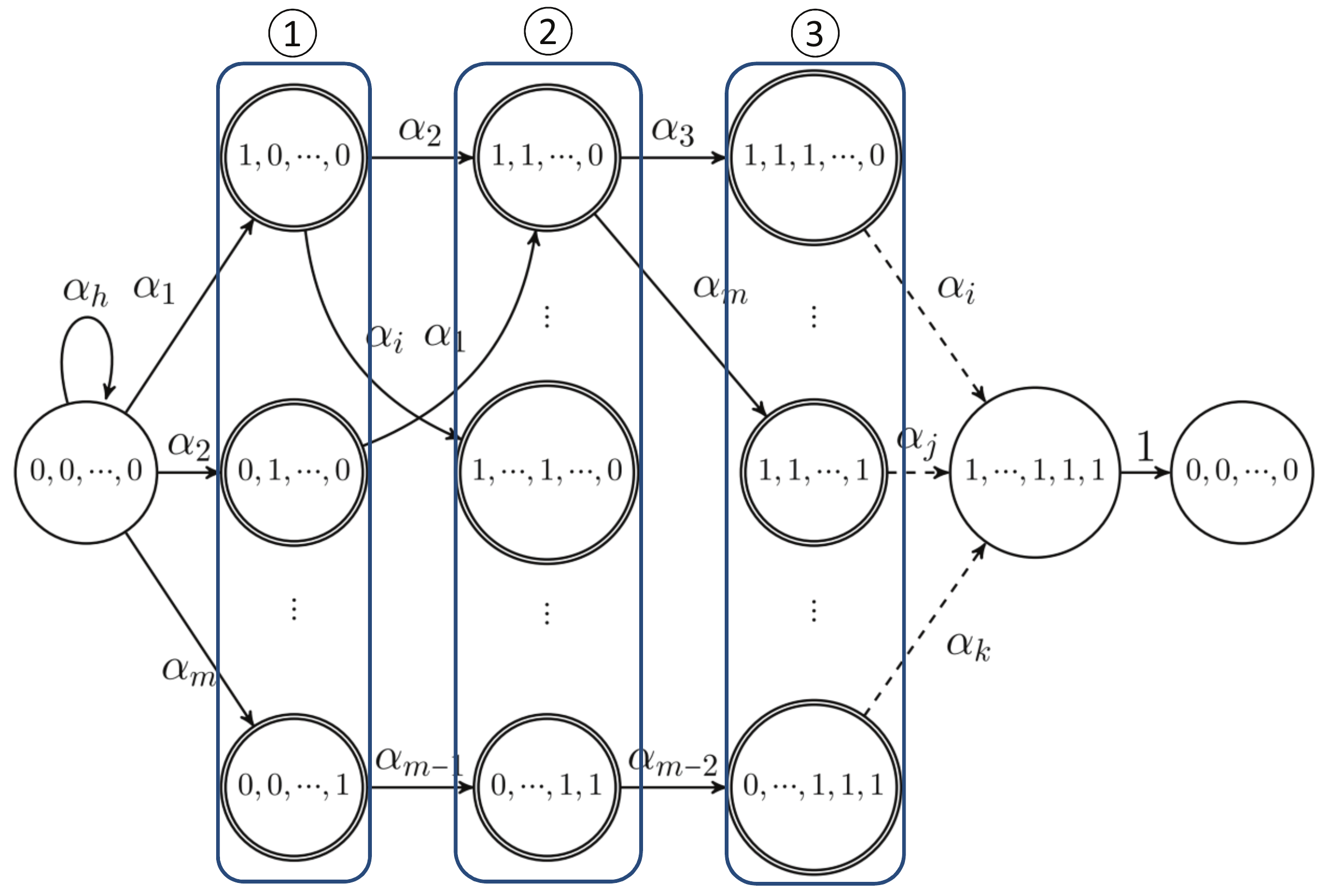}\\
		\captionsetup{justification=centering}
		\caption{Markov State Transition with $m$ attackers.}
		\label{fig:n attackers state machine}
	\end{minipage}
	\vspace{-0.3cm}
\end{figure}

We model the dynamics of the public and private chains as a Markov process likewise. Fig. \ref{fig:n attackers state machine} illustrates all the states and their transition probabilities. Recall that the assumption of the maximum private chain length also holds. Each state is expressed as a $m$-tuple, i.e. $L=\{l_1,l_2,\cdots,l_m\}$ with $l_i \in \{0, 1\}$, which consists of the lengths of each attacker's private chain. The state with $m$ zeros, denoted as $L_0$, indicates the start of the mining competition. The states with a single `1' are grouped together in which one and only one attacker has mined a valid block to build his private chain. Similarly, the states with $k$ elements of `1' indicate that the private chains of $k$ out of $m$ attackers have one valid block. Formally, we denote by $\mathcal{L}$ the set of all states with the cardinality $2^m$, and by $\mathcal{L}_k \subseteq \mathcal{L}$ the subset in which $k$ selfish miners hold their private blocks. 

The state transition probabilities are described as the following. The blockchain system can reside at the initial state with probability $\alpha_h$, i.e., the block is mined by the honest miner. Denote by $e_i$ the vector whose $i^{th}$ element is 1 and all others are zero. When attacker $i$ discovers a valid block, the blockchain system transits to a new state $(L+e_i)$ with probability $\alpha_i$. At state $L\in\mathcal{L}_k$, if the $i^{th}$ attacker who has a private block finds a valid block again, the blockchain system returns to the initial state $L_0$ (in which the state is expressed as a double circle). Otherwise, it jumps to a state in $\mathcal{L}_{k+1}$ with the probability equivalent to the relative Hash power of the selfish miner. The state transition probabilities can be expressed as:
\vspace{-0.2cm}
\begin{eqnarray}
&&\!\!\!\!\!\!\!\!\!P_{L_0L_0} = \alpha_h, \nonumber\\
&&\!\!\!\!\!\!\!\!\!P_{LL'} = \alpha_i, \;\;\forall L\in\mathcal{L}_k, \;\textrm{and}\;L' = L+e_i\in\mathcal{L}_{k+1},   \nonumber\\
&&\!\!\!\!\!\!\!\!\! P_{LL'} = \alpha_i + \alpha_h, P_{L'L_0}=1 ,\;\;\forall L\in\mathcal{L}_k, L' \notin \mathcal{L}_{k+1}. \nonumber
\end{eqnarray}
Using the detailed balance equations, we can compute the stationary distribution of each state. Specifically, the stationary probability at state $S_0$ is computed explicitly as 
\begin{eqnarray}
\pi_{L_0} = \big(1+ \sum_{k=1}^{m}\sum_{L\in\mathcal{L}_k} k! \prod_{j=1}^m( \alpha_j\cdot {1}_{l_j = 1})\big)^{-1}.
\end{eqnarray}
{ The stationary probability at state $L$ is given by:
\begin{align}
\label{eq:orphan one}
    &\pi_{L}=k!\cdot\pi_{L_0}\cdot \prod_{j=1}^m( \alpha_j\cdot {1}_{l_j = 1}),\forall L\in \mathcal{L}_k.
\end{align}
}
The revenues of all miners are computed based on the stationary state distribution and the particular transition paths to state $L_0$. If all miners have the same $\gamma$ in tie-breaking cases, their revenues can be written as:
{
\begin{align}
\label{eq:n attackers}
  \hspace{-2mm}  R_i=\sum\limits_{k=1}^{m}\sum\limits_{L\in\mathcal{L}_k}\left\{\hspace{-2mm}
             \begin{array}{lr}
             \pi_{L}\cdot [2\alpha_i+\alpha_h(2\alpha_i+\frac{1-\sum\limits_{j\in L}\alpha_j\cdot {1}_{l_j = 1}}{k+1})],  &l_i=1; \\
             \\
              \alpha_h\cdot \pi_{L}\cdot \alpha_i, & l_i=0. \\
              \end{array}
\right.\nonumber
\end{align}
\begin{align}
\hspace{-2mm}
    R_h=\big(\sum\limits_{k=1}^{m}\sum\limits_{L\in\mathcal{L}_k}\alpha_h\pi_{L}(\alpha_h+\frac{1-\sum\limits_{j\in L}\alpha_j\cdot {1}_{l_j = 1}}{k+1})\big)+\pi_{L_0}\cdot \alpha_h.
\end{align}
}
The relative revenue of each miner can be given by:
\begin{align}
    \hat{R_i}=\frac{R_i}{\sum\limits_{j=1}^{m}R_j+R_h}, \forall i\in \{1,\cdots,m,h\}.
\end{align}
As a special case that all selfish miners are homogeneous and their Hash power is $\alpha$,
the stationary probability at state $S_0$ can be simplified as
\begin{align}
  \pi_{L_0} = \big(1+ \sum_{k=1}^{m}A_m^k \alpha^k\big)^{-1},  
\end{align}
where $A_m^k$ is $k-$permutations of $m$.
The stationary probability at state $L$ can be represented as
\begin{align}
     &\pi_{L}=k!\cdot\pi_{L_0}\cdot \alpha^k,\forall L\in \mathcal{L}_k.
\end{align}
We can obtain the expected revenue of each miner as
    \begin{align}
    \frac{R_h}{\pi_{L_0}}&=\sum\limits_{k=1}^{m}A_m^k\alpha^k\alpha_h(\alpha_h+\frac{1-k\alpha}{k+1})+\alpha_h.
\end{align}
\begin{align}
    \frac{R_i}{\pi_{L_0}}
    &=\sum\limits_{k=1}^{m}A_m^k \alpha^k[\frac{k}{m}(2\alpha+\alpha_h\frac{1+\alpha}{k+1})+\alpha_h\alpha],
    \end{align}
Hence, the relative revenue can be rewritten as
\begin{align}
    &\hat{R_i}=R_i/(m\cdot R_i+R_h), \forall i\in\{1,\cdots,m\},\\
    &\hat{R_h}=R_h/(m\cdot R_i+R_h).
\end{align}

 The revenue of all miners when each private chain can hide more than one block is modeled and the detailed results are shown in Appendix B. 
\section{Optimal strategy under multiple attackers}
\label{sec:mdpandpomdp}
The basic selfish mining (\emph{BSM}) policy restricts the choices of withholding and releasing blocks. In this section, we present the 
\emph{more profitable}
selfish mining strategy (\emph{POMDP}-based mining policy) for two attackers when one of them chooses the basic selfish mining and the other chooses to be strategic. 
\subsection{{MDP}-based policy for an Upper Bound of Revenue}
The limitations of basic selfish mining are intuitive. An attacker is ``conservative'' to adopt the public chain when his private chain slightly lags behind it, and is ``less wise'' to override the public chain when it is catching up.  
 The optimal selfish mining problem with a single attacker was raised in \cite{ref:optimal selfish mining}\cite{ref:on the security} that modeled the mining race as a Markov decision process. 
 This strategic selfish mining policy lowers the profitable threshold of Hash power.

The optimal selfish mining in the presence of two attackers (Alice adopts the \emph{more profitable} mining policy and Bob adopts the \emph{BSM} mining policy)
is far more challenging than that with a single attacker. First, the state and state transition are greatly augmented. Alice has to incorporate the status of multiple chains in the state other than merely the lengths of the chains. 
Second, Alice cannot acquire the information regarding Bob's private chain.
To tackle these difficulties, we begin with the assumption that Alice has full information about the private chain of Bob. The optimal policy of Alice with full information can be solved based on an MDP model (\emph{OPT} policy), and the corresponding revenue will be used as the upper bound of revenue when the private chain of Bob is unknown. Meanwhile, the MDP model offers the principle of designing optimal policy with partially observable states.

\subsubsection{Main components} 
We formulate an MDP model for the strategic attacker as the four-tuple
$\mathcal{M} = <\mathcal{S}, \mathcal{A}, \mathcal{P}, \mathcal{R}>$ where 
$\mathcal{S}$ denotes the state space, $\mathcal{A}$ denotes the action space, $\mathcal{P}$ corresponds to the transition matrix, and $\mathcal{R}$ corresponds to the reward matrix. 

\textbf{State:} The state space $\mathcal{S}$ is defined as a 10-tuple in the form $<loc, fork, l_1, l_2, h_1, h_2, h_3, u_1, u_2, u_3>$. 
The attribute $loc\in \{1, 2, 3\}$ indicates the branch that Henry is working on. 
If $loc=1$ (resp. $loc=2$), Henry is mining on the public chain that also contains Alice's (resp. Bob's) blocks at the current mining round. If $loc=3$, the longest public chain contains only Henry's blocks. Note that Alice's and Bob's blocks are mutually exclusive on the public chain at the same mining round because one attacker will not accept the blocks of the other before this attack round ends. 
The attribute $fork$ obtains six values, dubbed as $\{ir$, $r$, $f_{12}$, $f_{13}$, $f_{23}$, $f_{123}\}$, where $r$ represents that Alice can release blocks to compete with the current public chain when $h_3>0$ while $ir$ represents that she can not. If multiple miners are competing on the public chain, $fork$ takes four values $\{f_{12}, f_{13}, f_{23}, f_{123}\}$, indicating that (Alice, Bob), (Alice, Henry), (Bob, Henry) and (Alice, Bob, Henry) are in the competition respectively. 

The notation $h_1$ indicates the distance between the starting position believed by Alice and the real starting position. Similarly, $h_2$ and $h_3$ indicate these distances of Bob and Henry. We record $h_1$, $h_2$ and $h_3$ because the blocks between the real and the believed starting positions influence which chain will prevail finally and how the revenues are calculated. The notation $l_1$ (resp. $l_2$) represents the number of unreleased blocks at Alice's (resp. Bob's) private chain. $\mu_1$ denotes the number of Henry's blocks between the real starting position and the Alice-believed starting position.
$\mu_2$ and $\mu_3$ are defined for Bob and Henry in the same way.
Similar to \cite{ref:optimal selfish mining}, we limit the lengths of private and public chains in a mining round so as to confine the size of state space, i.e. $l_1, l_2 \leq N$ and $h_3\leq h_{3,max}$.

\textbf{Action:} An action is the number of blocks that Alice publishes under a particular state. We define Alice's action space $\mathcal{A}$ as $\mathcal{A}=\{adopt, 0, 1, \cdots, l_1\}$ in which $adopt$ means that Alice gives up her private chain, $0$ means that Alice chooses to \emph{wait}, and $l_1$ is the current number of blocks held by Alice privately. The action taken by Alice has certain reasonable restrictions: if $l_1$ reaches $N$, Alice must release at least one block; if $h_3$ reaches $h_{3,max}$, Alice either chooses ``adopt'' or to release no less than $(h_3-h_1)$ blocks to end this mining round.

\textbf{State Transition:} We define the state transition function as $\mathrm{Pr}(s'|s, a\in\mathcal{A})$, the probability that the state $s$ jumps to $s'$ under action $a$. The state transition is triggered by the mining of a new block, and is determined by who discovers it. All the transitions are summarized in Appendix Table \ref{table:MDP}.

\textbf{Reward Function:} The purpose of ``optimal'' mining is to acquire a larger share of confirmed blocks on the public chain.
Recall that the relative revenue of a miner is the fraction of his blocks on the public chain for a long period. Obviously, the relative revenue cannot be measured under each state-action pair, and cannot be taken as the corresponding immediate reward. Sapirshtein et al. \cite{ref:optimal selfish mining} transform the (long-term) relative revenue into the family of (one-shot) absolute revenues parameterized by the weight $\rho \in [0, 1]$. 

This ingenious transformation operates as the following. Define a transformation function $w_\rho: \mathbb{N}^3 \to \mathbb{R}$ related to Alice's instantaneous reward:
	\begin{align}
	\label{eq:w rho}
	w_\rho^i(r_1^i,r_2^i,r_h^i)=(1-\rho)\cdot r_{1}^i - \rho \cdot (r_{2}^i+r_{h}^i),
\end{align}
where $r_1^i$, $r_2^i$ and $r_h^i$ represent the instantaneous rewards of Alice, Bob and Henry at step $i$ (analogous to time $t$ in classical MDP). We reformulate the original MDP model as $\mathcal{M}_{\rho} = <\mathcal{S}, \mathcal{A}, \mathcal{P}, w_\rho(r_1, r_2, r_h)>$. The underlying reason of such transformation is that instead of maximizing the relative revenue, we choose to maximize the expected fictitious reward $w_\rho(r_1, r_2, r_h)$. For any admissible policy $\pi$, the mean reward denoted by $v_{\rho}^{\pi}$ is characterized as: 
	\begin{align}
	v_{\rho}^{\pi}=\mathbb{E}[\varliminf \limits_{\xi\to \infty}\frac{1}{\xi}\sum_{i=1}^{\xi}w_{\rho}(r_1^{i}(\pi),r_2^{i}(\pi),r_h^{i}(\pi))],
\end{align}
where $\xi$ is the total number of state transition steps. The optimal revenue $v_{\rho}^{*}$ is given by
\begin{align}
	v_{\rho}^{*}=\max_{\pi}\;\{v_{\rho}^{\pi}\}.
\end{align}
The equivalence between two MDPs $\mathcal{M}$ and $\mathcal{M}_{\rho}$ is indirect. Sapirshtein et al. \cite{ref:optimal selfish mining} present two propositions to guarantee their equivalence for a single selfish miner.
\begin{itemize}
	\item If $v_{\rho}^*=0$ for some $\rho\in [0,1]$, then any policy $\pi^*$ obtaining this value also maximizes the relative revenue, and the relative revenue equals to $\rho$.
	\item $v_{\rho}^{*}$ is monotonically decreasing in $\rho$. 
\end{itemize}

The above propositions tell us that by searching for an appropriate $\rho$ that yields the mean reward of $\mathcal{M}_{\rho}$ to be 0, we can obtain the optimal relative revenue of Alice in $\mathcal{M}$. We generalize this idea to the MDP with multiple selfish miners. In addition, the maximum number of steps, $\xi$, is truncated to avoid excessive computations by tolerating a very gentle loss in the optimal mean reward $v_{\rho}^{\pi}$.

\subsubsection{Algorithm}
Owing to the monotonicity of $v_{\rho}^{*}$ to $\rho$, a binary search of $\rho\in[0, 1]$ is adequate. For a given $\rho$, we utilize the \emph{value iteration} method to solve the optimal policy $\pi_{\rho}^{*}$ as \cite{ref:optimal selfish mining}. Compared with \emph{policy iteration}, the advantage of \emph{value iteration} is its fast convergence rate especially in large-scale MDPs \cite{ref:QMDP} \cite{ref:value iteration}. 

\subsection{POMDP-based policy for Optimal Mining}

The \emph{OPT} framework provides important insights into the optimal mining policy in the presence of two attackers, yet its real world deployment is unrealistic. The strategic miner Alice is assumed to know precisely the system state, while in reality Bob's private chain is not observable to Alice, and the blocks on the public chain released by Bob and Henry cannot be differentiated because of their anonymity. In light of the incomplete state information, we reformulate the optimal mining as a Partially Observable Markov Decision Process (POMDP). Before diving into details, we enumerate three key challenges: 
\begin{itemize}
	\item which subset of information is non-observable to Alice;
	\item how the mining event-driven MDP model is generalized to the POMDP model;
	\item how the \emph{POMDP}-based optimal mining policy can be computed efficiently. 
\end{itemize} 

\subsubsection{Main components}
The POMDP model is expressed as a six-tuple $\mathcal{M}_{PO}:=<\mathcal{S}, \mathcal{A}, \mathcal{P}, \mathcal{R}, \mathcal{O}, \mathcal{Z}>$, where $\mathcal{O}$ is Alice's observation space, $\mathcal{Z}(\cdot)$ is the observation function, and the remaining components inherit the same meanings as their counterparts in MDP.

\textbf{Observable Information:} In the partially observable environment, Alice cannot obtain all the attributes in $\mathcal{S}$. The length of Bob's private chain $l_2$ cannot be observed absolutely. The attribute $loc$ is non-observable either because Alice is unaware of whether Henry is mining on his own chain or Bob's chain. $h_2$, $\mu_2$ and $\mu_3$ are non-observable for that the anonymity of mining covers up the owners of the blocks on the public chain. The attribute $fork$ is observable because $ir$ and $r$ are pertinent to Alice's private chain, and the values $f_{*} \in \{f_{12}, f_{13}, f_{23}, f_{123}\}$ is obtained by counting the number of focks in competition. $l_1$, $h_1$ and $\mu_1$ are known for sure; $h_3$ is actually the length of the longest public chain. In summary, the observation space is represented as $\mathcal{O}:=<fork, l_1, h_1,h_3, \mu_1> \subset \mathcal{S}$. Given the same observation, Alice is likely to be in many possible states.

\textbf{State Transition:} The state transition function is also denoted as $\textrm{Pr}(s'|s, a\in A)$. Though taking the similar form, $\mathcal{M}_{PO}$ possesses a different state transition logic from $\mathcal{M}$. In $\mathcal{M}$, an action is triggered by the discovery of a new block, and the state transition follows. In $\mathcal{M}_{PO}$, the pure \emph{event-driven} state transition will restrict Alice from participating in the fork competition. For instance, if Bob hides one block, Alice should publish her private block while there is no observable event to trigger a fork competition. On the contrary, if Bob does not have any private block while Alice believes that the length of Bob's private chain is 1, Alice may publish one or two blocks unnecessarily. Therefore, one can see that a time-slotted plus event-driven POMDP model is appropriate to handle the intricate optimal mining problem. 

Due to the memoryless Hash computation \cite{ref:poission}, the block arrival process is actually a stationary stochastic process. If we slice this stochastic process equally with a slot duration $\Delta t$, the number of mined blocks in every slot has the same distribution. By choosing a relatively small $\Delta t$, we suppose that at most one block is mined in each slot (the chance of mining two or more blocks is rarer by orders of magnitude). Denote by $p$ the probability of generating a block in one time slot. The probabilities that Alice, Bob and Henry generate it are $\alpha_1 p$, $\alpha_2 p$ and $\alpha_h p$ respectively. Alice can estimate the Hash power $\alpha_2$ and $\alpha_h$ through mining honestly for a certain period. It is worth highlighting that our POMDP model makes the optimal mining with partially observable states feasible, and is in accordance with realistic blockchain systems. All the transitions are summarized in Appendix Table \ref{table:POMDP}.

\textbf{Observation function:} Define $\mathcal{Z}:= \mathcal{S} \times \mathcal{A} \to \Delta(\mathcal{O})$ as the observation function that specifies the relationship between system states and observations. Here, $z(s, a, o)$ is the probability that observation $o$ will be reached after Alice performs action $a$ and lands in state $s$:
\begin{align}
    {z_{t+1}}(s, a, o)=\textrm{Pr}(o_{t+1}=o | s_{t+1}=s, a_{t}=a).
\end{align}
In $\mathcal{M}_{PO}$, the observation of a state is certain, i.e,
\begin{align}
&s=<loc, fork, l_1, l_2, h_1, h_2, h_3, \mu_1, \mu_2, \mu_3>,\nonumber\\
&o=<fork,l_1,h_1,h_3,\mu_1>,\nonumber\\
&z_{t+1}(s,a,o)=1.
\end{align}

\textbf{Belief:}
Our POMDP model is pertinent to  a belief $b$ which is a probability distribution over all the possible states. Intuitively, Alice makes a guess on the current state iteratively. The belief on a particular state $s$ at time $t$ is given by:
\begin{align}
b(s)=\textrm{Pr}(s_t=s|o_t,a_{t-1},o_{t-1},\cdots,a_0,b_0).
\end{align}
The updated belief state $b'(s')$ is calculated whenever the action $a$ is taken and the observation $o$ is perceived.
\begin{align}
	\label{eq:belief update}
	b'(s')&\equiv \textrm{Pr}(s'|a,o,b)=
	\frac{\textrm{Pr}(o|s',a)\sum_s \textrm{Pr}(s'|a,s)b(s)}{\textrm{Pr}(o|a,b)},
\end{align}
where $\sum_{s\in \mathcal{S}}b(s)=1$ and $\textrm{Pr}(o|a,b)$ is a normalization factor given by
\begin{align}
	\textrm{Pr}(o|a,b)=\sum_{s'}\textrm{Pr}(o|a,s')\sum_{s}\textrm{Pr}(s'|a,s)b(s).
\end{align}

\textbf{Reward function:} We use $r_1^i(s,a), r_2^i(s,a)$ and $r_h^i(s,a)$ to denote the instantaneous rewards of Alice, Bob and Henry at step $i$ when Alice takes action $a$ at state $s$. Due to the uncertainty of system state, the expected reward functions $r_1^i(b,a), r_2^i(b,a)$ and $r_h^i(b,a)$ are constructed on the belief of instantaneous rewards,  $\forall a \in \mathcal{A}$,
    \begin{align}
        &r_k^i(b,a)=\sum\limits_{s\in \mathcal{S}}b_i(s)r_k^i(s,a),k\in\{1,2,h\}.
    \end{align}
Similar to the transformation in the MDP model, we replace the relative reward by the absolute reward  $w_{\rho}^i(r_1^i(b_i,a),r_2^i(b_i,a),r_h^i(b_i,a))$ parameterized by $\rho$ using Eq. \eqref{eq:w rho}. 

We next formulate the optimal mining as a finite-horizon average reward POMDP problem as \cite{ref:optimal selfish mining} and \cite{ref:on the security}. The expected average value function is defined as
\begin{align}
    &v_{\rho}^{\pi}=\mathbb{E}[\varliminf \limits_{\xi\to \infty}\frac{1}{\xi}\sum_{i=1}^{\xi}w_{\rho}(r_1^{i}(b_i,\pi),r_2^{i}(b_i,\pi),r_h^{i}(b_i,\pi))]. \nonumber
\end{align}
The optimal policy $\pi^*$ is a set of decision rules depending on belief-state pairs: 
\begin{align}
\label{eq:optimal policy for POMDP}
    &\pi^*=\arg\max_{\pi \in \mathcal{A}}\{v_{\rho}^{\pi}\}.
\end{align}
The parameter $\rho$ that solves $v_{\rho}^*=0$ is the relative revenue of Alice under the POMDP model. In practice, the sum of revenues over $\xi$ is truncated by a sufficiently large number $\xi_0$. Given a precision threshold $\epsilon$, $\xi_0$ needs to satisfy:
\begin{align}
       |v_{\rho}^{\pi}-\mathbb{E}[\frac{1}{\xi_0}\sum_{i=1}^{\xi_0}w_{\rho}(r_1^{i}(\pi),r_2^{i}(\pi),r_h^{i}(\pi))]|\leq\epsilon.
       \vspace{-1cm}
\end{align}
\subsection{Algorithm}
A POMDP is essentially an expanded MDP defined on belief space.
However, the belief space is a high-dimensional  continuous space that needs to be segmented into a huge number of belief states. An offline POMDP algorithm will compute the optimal actions at every belief state. Considering a large-scale POMDP like ours, the offline computation is time-consuming because of generating the rewards, updating the beliefs and constructing the optimal policy at each belief. 
For efficiency considerations, we propose using the online POMDP algorithm that explores the future belief states reachable from the current belief state. The policy construction time is often substantially shorter. 
Furthermore, three properties can be used to reduce the time of searching for $\rho$.

\begin{lemma}\emph{Under the same parameter setting, the optimal result of $\mathcal{M}$ is the upper bound of $\mathcal{M}_{PO}$.}
\end{lemma}

\emph{Proof:} 
{Among the approximation algorithms of the POMDP problem, the MDP approximation consists in approximating the value function of the POMDP by the value function of its underlying MDP \cite{ref:littman 1995}. This value function is an upper bound on the value function of the POMDP \cite{ref:online planning algorithms}}.\,\,\,\,\,$\qedsymbol$

\begin{lemma}\emph{The revenue obtained under the optimal policy based on any $\rho$ is the lower bound of the actual optimal revenue for $\mathcal{M}_{PO}$.}
\end{lemma}
\emph{Proof:} The value of $\rho$ does not affect the state transition and the instantaneous reward. All the mining policies under any $\rho$ are available, even though they are not optimal. Hence, we record the number of blocks that each miner obtains at every time step, i.e, $(r_1^i,r_2^i,r_h^i)$. After enough time steps, we can compute the relative revenue under the current $\rho$ even though it is not optimal. Therefore, the revenue of the optimal policy obtained under the current $\rho$ is the lower bound of the actual optimal revenue. \,\,\,\,\,\,$\qedsymbol$
 
 \begin{lemma}
\label{lemma:optimal policy}
\emph{There exists an optimal stationary deterministic policy for  $\mathcal{M}_{PO}$ model.}
\end{lemma}
\emph{Proof:} The states, actions and beliefs are countable because the maximum lengths of private and public chains are limited. Since a POMDP problem can be regarded as a belief MDP, the existence of an optimal stationary policy for our POMDP problem is guaranteed by Theorem 7.3.6 of \cite{ref:markov book}. \,\,\,\,\,$\qedsymbol$

   \begin{algorithm}[!htp] 
  \caption{ Algorithm for solving the POMDP.}  
  \label{alg:POMDP1}  
  \textbf{Input:}  $\mathcal{M}_{PO},\mathcal{M}$, a truncation parameter $\xi_0$, a precision value $\epsilon$; The initial belief $b_c$;\\
\textbf{Output:}$\rho$;\\
  \textbf{Static:} 
   $a^*:$ optimal action;
  \begin{algorithmic}[1]
  \STATE $ low \gets 0$;
  \STATE $\rho^*=$\emph{\footnotesize{QMDP($\mathcal{M}$)}};
  \STATE $upper \gets \rho^*$;
\WHILE{$upper-low>\epsilon$}
\STATE $\rho \gets (low+upper)/2$
  \STATE \footnotesize{RESULT}$\gets \{\}$;
  \STATE  $r_1\gets0,r_2\gets0,r_h\gets0$;\\
  \STATE $v_\rho \gets 0$, $\xi\gets \xi_0$;
  \STATE $b_c \gets$ \footnotesize{initial belief}
\WHILE { \emph{\footnotesize{$\xi>0$}} }
  \IF{$b_c \in$ \footnotesize{RESULT}}
  \STATE$a^*=$\footnotesize{RESULT}$[b_c]$
  \ELSE
  \STATE $a^* \gets AEMS2(b_c,\mathcal{M}_{PO})$
  \ENDIF
  \STATE ($r_1^i,r_2^i,r_h^i$)$\gets$ Execute $a^*$ for $b_c$.
  \STATE $r_1+=r_i^i,r_2+=r_2^i,r_h+=r_h^i$;
  \STATE $v_\rho+=w_\rho(r_1^i,r_2^i,r_h^i)$;
    \STATE \footnotesize{RESULT}$[b_c]=a^*$
  \STATE Perceive a new observation $o$
  \STATE $b_c \gets b'(b_c,a^*,o)$ \qquad $\rhd$ update algorithm is Eq. \eqref{eq:belief update}
    \STATE $\xi-=1$
  \ENDWHILE
  \STATE $R'_1=r_1/(r_1+r_2+r_h)$;
\vspace{0.1cm}
\IF{$v_{\rho}>0$}
\STATE $low \gets max(\rho,R'_1)$;
\ELSE
\STATE $upper \gets \rho$; $low\gets \max(low,R'_1)$;
\ENDIF
\ENDWHILE
    \RETURN  $\rho$;
     \end{algorithmic}  
\end{algorithm}
Our online mining algorithm is described in Algorithm \ref{alg:POMDP1}.
We calculate the optimal revenue based on binary search. The upper bound can be set as the result of the underlying MDP model according to Lemma 1 (lines 2-3). Alice will execute the optimal action based on the current $\rho$ and obtain the relative revenue (lines 16 and 24). Her optimal revenue is no less than the relative revenue according to Lemma 2 (lines 25-29). We do not have to recalculate the optimal action at every step according to Lemma 3. The online algorithm will adopt the corresponding action if the same belief state has met. Otherwise, it will calculate and store the new optimal action (lines 11-15). A block of size 1MB needs 18 seconds to reach three thousand nodes in Bitcoin \cite{ref:tradeblock}. Making timely decisions  is very important. We use AEMS2 \cite{ref:AEMS} \cite{ref:online planning algorithms}, one of the fastest POMDP algorithms, to compute the optimal action (line 14).

\section{Transient Analysis of Profitability}
\label{sec:transient}

In this section, we first describe the difficulty adjustment algorithm (DAA) in Bitcoin-like systems, and model the revenue of miners in one difficulty adjustment period. We analytically show that the extra revenue of selfish mining is originated from the DAA. 
	
\subsection{Bitcoin-like Difficulty Adjustment}

The essence of Bitcoin mining is to solve a cryptographic puzzle. 
A Bitcoin miner repeatedly enumerates a \emph{\footnotesize{NONCE}} until the head Hash is below the difficulty target. The smaller a target value is, the more difficult the discovery of a valid \emph{\footnotesize{NONCE}} will be. For a fixed target difficulty, a larger Hash power means a shorter time of finding a valid \emph{\footnotesize{NONCE}}. 

To maintain a stable block generating interval, Bitcoin introduces difficulty adjustment algorithms (DAAs) to cope with the variable Hash powers in the systems. The Bitcoin DAA is executed after 2016 blocks have been mined. It is actually a feedback control system: if the actual time of mining 2016 blocks is larger than 20160 minutes (10 minutes per block), the target difficulty decreases proportionally, and increases otherwise. 
When a miner performs selfish mining, a lot of blocks are orphaned so that the actual time to mine 2016 blocks becomes longer. In the next difficulty adjustment periods, the target difficulty is lowered down to maintain the fixed block generating rate.

\subsection{Absolute Revenue}
Previously we define the revenue of a miner in each mining round and his/her relative revenue.  However, the duration of a mining round may not be fixed all the time, and the actual number of valid blocks obtained by a miner in each unit of wall-clock time is overlooked. 
In Bitcoin system, we denote 10 minutes as the unit time, and denote a DAA period as the expected time units of mining 2016 \textbf{valid} blocks. 
	
\begin{definition}
(\emph{Absolute Revenue}) \emph{The absolute revenue is the average number of blocks obtained in each unit time. }
\end{definition}

\textbf{\#1 DAA Period.} We treat the first DAA period as the beginning of selfish mining in order to analyze the transient absolute revenues of Alice and Bob. The following normalization rule is made to simplify the analysis by eliminating the randomness of block generating interval. \textit{A block is generated every time unit at the first difficulty adjustment period, i.e., $\Delta t_1 = 1$.}

With the above assumption, we can easily compute the duration of generating 2016 valid blocks.
Let $R_{vld}$ be the total number of \textbf{valid} blocks of all miners in a mining round, and let $R_{tot}$ be the total number of blocks including the \textbf{valid and orphan} blocks in the same round. This means that a mining round occupies $R_{tot}$ time units. Denote by $T_1$ the expected time units to accomplish the first DAA period. {One can calculate the time of mining rounds that the total number of valid blocks reaches 2016.} Then, $T_1$ is equivalent to the sum of time units of these mining rounds. There exists
\begin{eqnarray}
R_{vld} &=& R_a+R_b + R_h; \quad R_{tot} = 1;\nonumber \\
E[T_1] &=& \frac{2016}{R_{vld}}\cdot R_{tot}.
\end{eqnarray}
Due to the orphaned blocks, $R_{tot}$ is greater than $R_{vld}$ so that the actual time of $T_1$ is longer than 2016 time units. 

\textbf{Subsequent DAA Periods.} After the first DAA period, the blockchain finds that the time interval of generating a valid block is longer than one time unit. Consequently, the target difficulty decreases to match the current valid Hash power in the system. Given the invariable Hash power of miners, the block generating interval $\Delta t_i$ becomes smaller for $i \geq 2$. Let $T_i$ be the expected time units of the $i^{th}$ DAA period that has $T_i = 2016$ for $i\geq 2$. This is also the goal that DAA is going to achieve. It is noted we assume $R_{tot}\leq (4\cdot R_{vld})$. 

\textbf{Absolute Revenue Over Time.} 
Recall that the absolute revenue captures the expected reward of a miner in each time unit. Since our purpose is to investigate the transient profitability of selfish mining, we define $\tilde{R}_i(K)$ as the absolute revenue of the $i^{th}$ miner over $K$ DAA periods. Therefore, we obtain the following expressions for $\forall i \in \{a, b, h\}$
\begin{eqnarray}
	\label{eq:revenue each round}
	    \tilde{R}_i(K) &=&\frac{2016\cdot K\cdot  R_i}{R_{vld}} \cdot \frac{1}{\sum\nolimits_{k=1}^K E[T_k]} \nonumber\\
	    &=& \frac{K R_i}{R_{tot} + (K-1)R_{vld}}.
\end{eqnarray}

Now we are aware that the selfish mining has a smaller absolute revenue in the first DAA period no matter whether the Hash power of the attacker is above the stationary profitable threshold or not. This claim also holds in different selfish mining policies. As $K$ increases, the absolute revenue is asymptotically close to the relative revenue, which is
\begin{align}
  \tilde{R}_i(K)= \frac{K R_i}{R_{tot} + (K-1)R_{vld}} \approx \frac{K R_i}{R_{vld} + (K-1)R_{vld}}.
\end{align}

This fully justifies the use of relative revenue to represent absolute revenue in previous models. At the same time, a selfish miner can hopefully reimburse his/her loss in the first DAA period by his/her extra revenue in the future DAA periods. With our absolute revenue model, we can characterize how much time is needed to make selfish mining profitable eventually.

\begin{figure}[t]
	\begin{minipage}[t]{1\linewidth}
			\setlength\abovecaptionskip{-0.5pt}
			\setlength\belowcaptionskip{-1pt}
			\centering
			\includegraphics[width=0.8\textwidth,height=0.2\textheight]{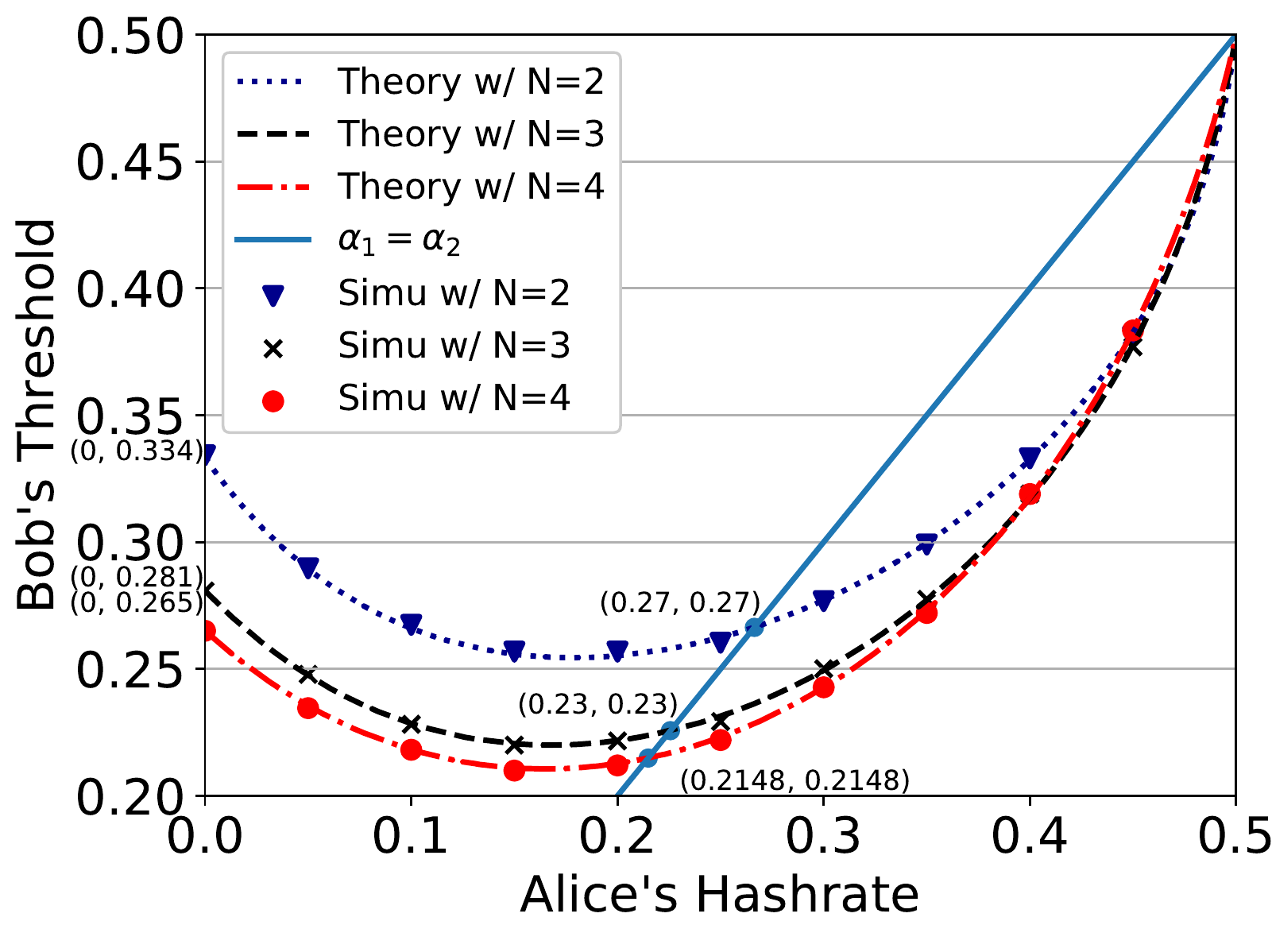}\\
			\caption{Bob's threshold under the influence of Alice’s Hashrate.}
			\label{pic:exp_basic_group2}
		\end{minipage}
		\hspace{0.1cm}
		\vspace{-0.5cm}
	\end{figure}	
\section{Evaluation}
\label{sec:evaluation}

In this section, we develop an event-driven simulator for basic selfish mining and a time-driven simulator for \emph{POMDP}-based selfish mining $\footnote{Mining strategies can be implemented in real systems possibly because miners’ decision-making relies on public information.}$. Comprehensive experiments validate the correctness of our models and reveal important properties regarding the profitability of selfish mining \cite{ref:cloudlab}. 
\begin{figure*}
\begin{minipage}[t]{0.33\linewidth}
\setlength\abovecaptionskip{3pt}
\setlength\belowcaptionskip{-1pt}
\centering
{\includegraphics[width=0.9\textwidth]{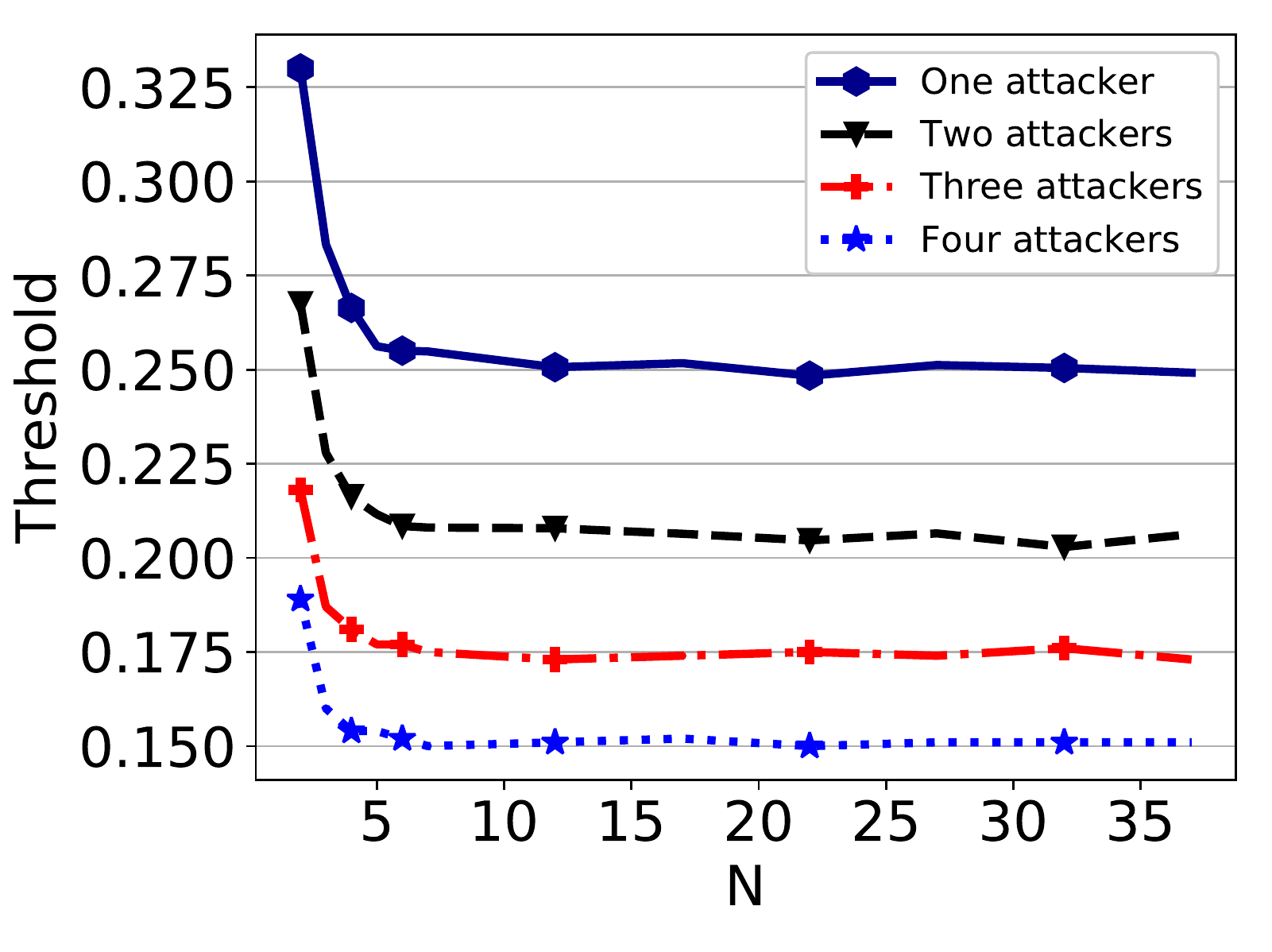}}
\caption{Profitable threshold vs the maximum number of private blocks ($N$).}
\label{fig:attackers threshold N}
\end{minipage}
\begin{minipage}[t]{0.33\linewidth}
\setlength\abovecaptionskip{3pt}
\setlength\belowcaptionskip{-1pt}
\centering
\includegraphics[width=0.9\textwidth]{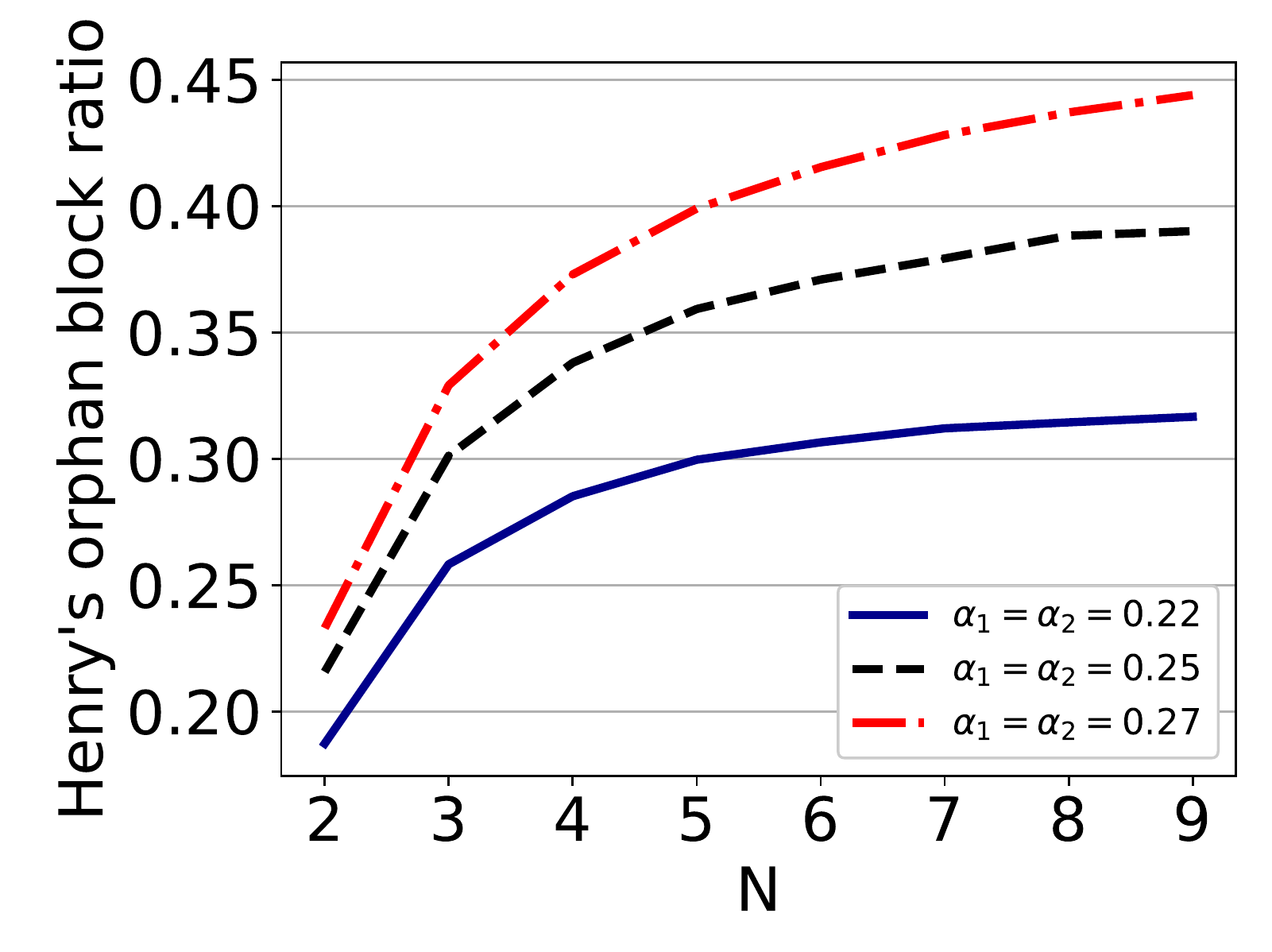}
\caption{Henry's orphan block ratio vs the maximum number of private blocks ($N$).}
\label{fig:orphan block ratio N}
\end{minipage}
\begin{minipage}[t]{0.33\linewidth}
\setlength\abovecaptionskip{3pt}
\setlength\belowcaptionskip{-1pt}
\centering
\includegraphics[width=0.9\textwidth]{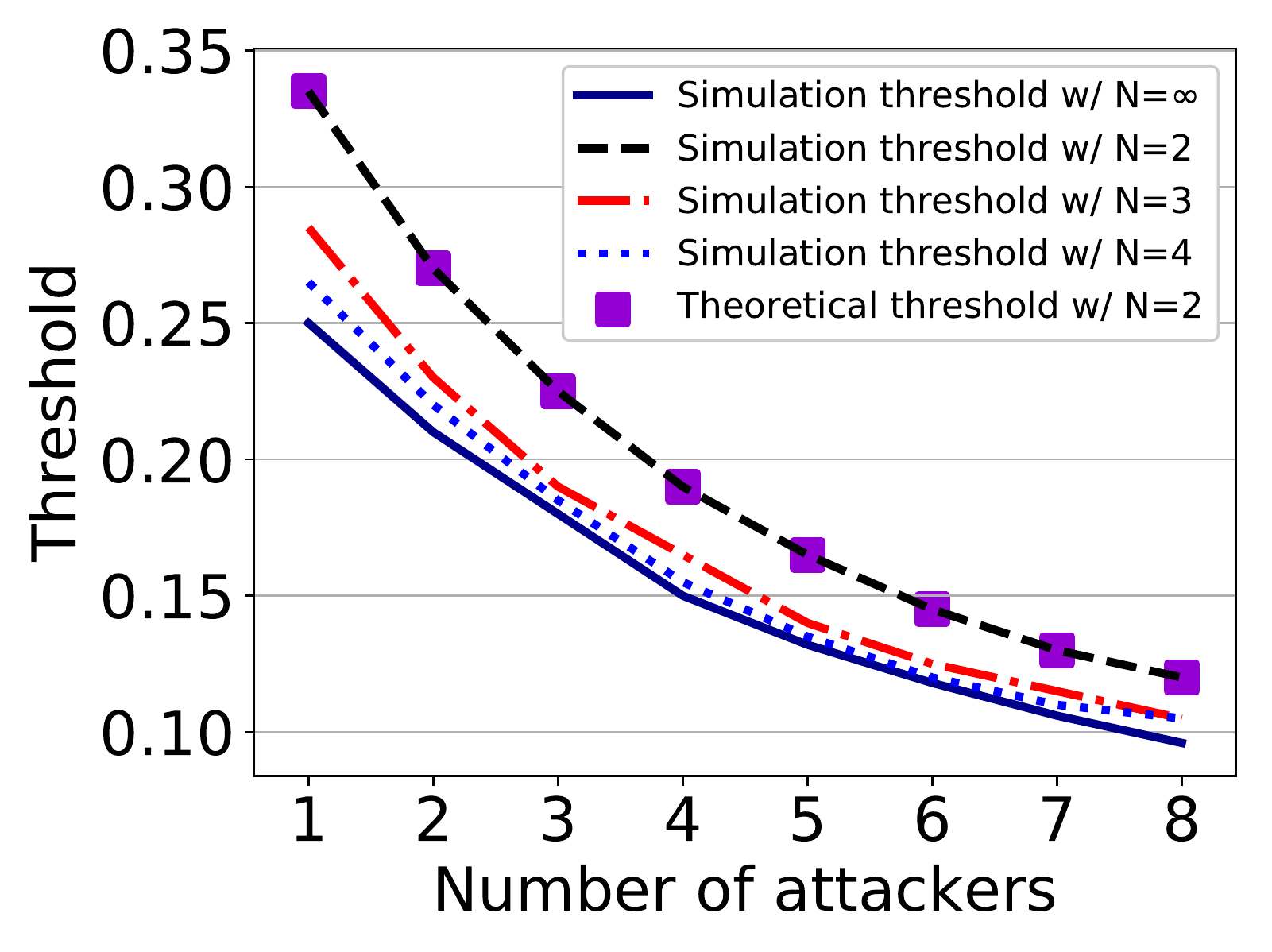}
\caption{Profitable threshold vs the number of attackers ($m$).}
\label{fig:n attackers threshold}
\end{minipage}
\vspace{-0.5cm}
\end{figure*}

\subsection{Basic Selfish Mining}
\begin{observation} {\textit {When there are multiple attackers in Bitcoin-like systems, the attackers' profitable thresholds decrease and the system security is degraded.}}
\end{observation}

We illustrate Bob's profitable threshold of selfish mining in Fig. \ref{pic:exp_basic_group2} as Alice's Hash power increases from 0 to nearly 0.5. To avoid involving too many control variables, the tie-breaking parameters are set to $\gamma_{1}{=}\gamma_{2}{=}\frac{1}{2}$ and $\theta_{1}{=}\theta_{2}{=}\frac{1}{3}$. 
One can observe from three curves with different $N$ that Bob's profitable threshold decreases at first and increases afterward. When $N=2$ and $\alpha_1 = 0.16$, Bob's profitable threshold is the lowest. Under this situation, Alice's selfish mining may yield less revenue compared with her honest mining. We further draw a $45^{\circ}$ line to indicate the profitable threshold for both Alice and Bob when their Hash power is symmetric. When $N$ is 2, 3 and 4, the profitable threshold is 26.64\%, 22.57\% and 21.48\%. In contrast, it takes the value of 33.33\%, 28.08\% and 26.50\% respectively, if there is a single attacker. An obvious conclusion is that the existence of multiple attackers makes selfish mining more easily profitable.

\begin{figure*}
 \begin{minipage}[h]{0.33\linewidth}
			\setlength\abovecaptionskip{0.5pt}
			\setlength\belowcaptionskip{-1pt}
			\centering
			\includegraphics[width=0.9\textwidth]{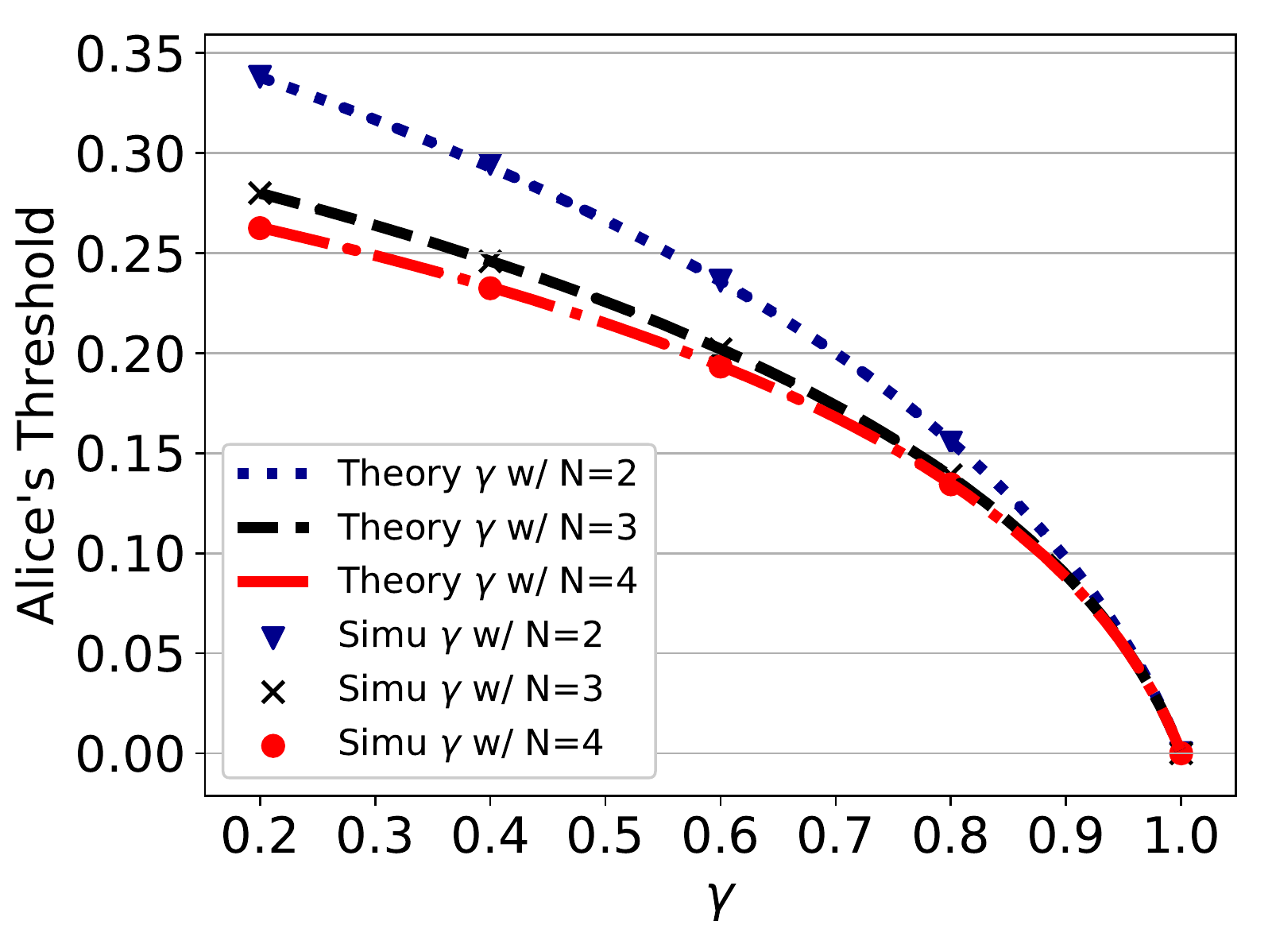}
			\caption{Profitable threshold for Alice with $\gamma$.}
			\label{fig:BSM gamma}
\end{minipage}
 \begin{minipage}[h]{0.33\linewidth}
			\setlength\abovecaptionskip{0.5pt}
			\setlength\belowcaptionskip{-1pt}
			\centering
			\includegraphics[width=0.9\textwidth]{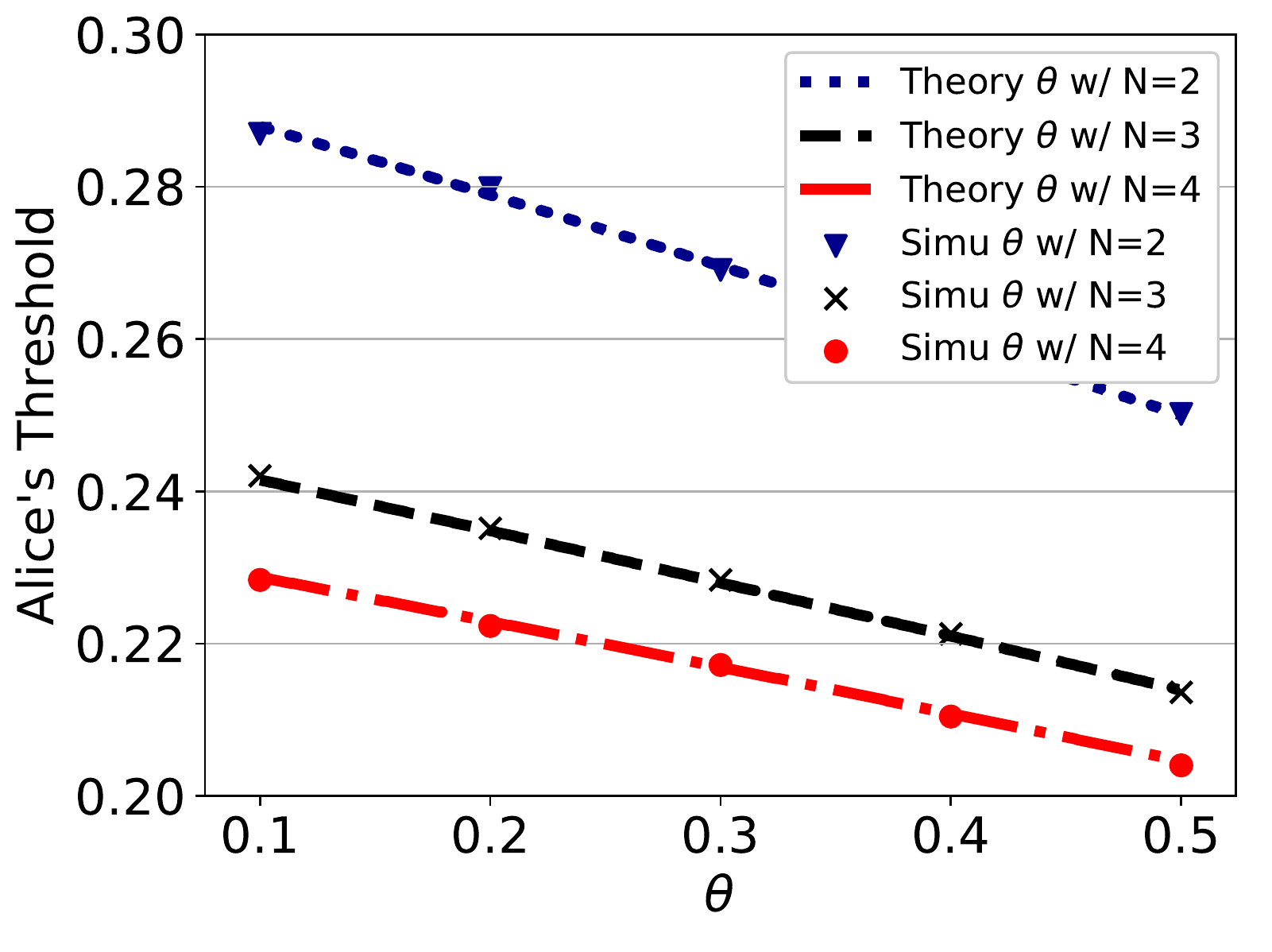}
			\caption{Profitable threshold for Alice with $\theta$.}
			\label{fig:BSM theta}
\end{minipage}
 \begin{minipage}[h]{0.33\linewidth}
			\setlength\abovecaptionskip{0.5pt}
			\setlength\belowcaptionskip{-1pt}
			\centering
			\includegraphics[width=0.9\textwidth]{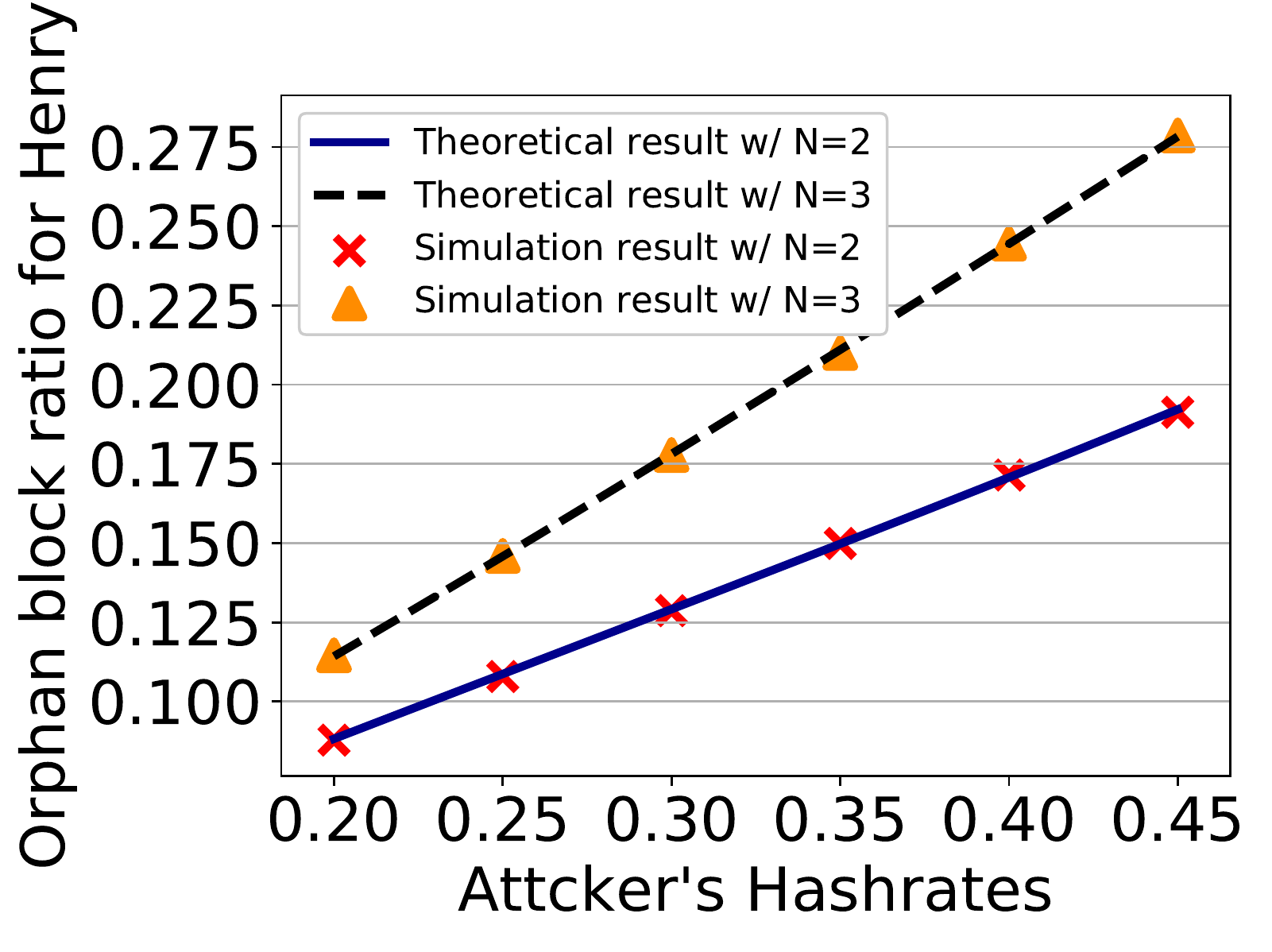}
			\caption{Estimating Bob's Hash power by observing orphan block ratio.}
			\label{fig:orphan one}
\end{minipage}
\vspace{-0.7cm}
		\end{figure*}

\begin{observation}{\textit{The profitable threshold decreases with the increase of $N$, and remains stable with the attackers of symmetric Hash power as $N\geq 4$; it also decreases when the number of attackers $m$ increases.}}
\end{observation}

We evaluate the profitable thresholds of \emph{BSM} with different $N$ and $m$. Our purposes are twofold: one is analyzing the interplay between this threshold and the environment variables, and the other is justifying the use of $N\leq 4$ in the mathematical modeling. The Hash powers of all the attackers are identical, and the competing chains are indistinguishable upon the tie-breaking rules. 

Fig. \ref{fig:attackers threshold N} shows the relationship between $N$ and the profitable threshold. The cases with 1, 2, 3 and 4 symmetric attackers are expressed in solid, dashed, dash-dotted and dotted lines. One can observe that the profitable threshold decreases remarkably for different $m$ as $N$ increases from 2 to 4. The event-driven simulations exhibit a stable profitable threshold when Alice and Bob can hold more than 5 private blocks. For instance, this threshold converges to 25\% for a sufficiently large $N$ with a single attacker, which is in line with \cite{ref:majority}. With two symmetric attackers ($m=2$), its value is 20.60\% at $N=30$, slightly different from that at $N=4$. In general, Alice's or Bob's Hash power is much smaller than Henry's Hash power. The chance that Alice's or Bob's private chain takes a large lead over the public chain in a mining round is very small. Hence, it does not make an evident influence on the profitable threshold when $N$ is already large. 
Moreover, hiding a long private chain and releasing all the blocks simultaneously will make the selfish mining attack easily detected. Fig. \ref{fig:orphan block ratio N} shows Henry's orphan block ratio as $N$ increases from 2 to 9 with $m=2$. Even though Alice's and Bob's Hash powers are merely (0.22, 0.22), they cause a very high orphan block ratio to Henry, e.g., 18.73\% with $N=2$, 28.53\% with $N=4$ and 31.66\% with $N=9$. Such a high orphan ratio can easily expose  the identity of attackers. Therefore, our modeling framework only considers $N\leq 4$ though it is extensible to $N>4$. 

We use mathematical models and event-driven simulations to quantify the impact of $m$ on the profitable threshold in Fig. \ref{fig:attackers threshold N} and \ref{fig:n attackers threshold}. One can observe that the increase in the number of attackers reduces the profitable threshold, thus endangering blockchain security. For $N=4$, the profitable thresholds with $m\in\{2, 4, 8\}$ are $\{0.2148,0.155,0.11\}$. This challenges the cognition that selfish mining is less likely to happen if the Hash power of a miner is below 25\%. The primary reason that more attackers lead to smaller profitable thresholds lies in that the Hash power of the honest miner declines relatively.  Meanwhile, our model coincides with the simulation result at $N=2$ in Fig. \ref{fig:n attackers threshold}, thus validating its correctness. In Appendix C, 
we further explore the impact of $N$ on the \emph{BSM} attackers' revenues when the attackers' Hash powers are asymmetric. 

\begin{observation}{\textit{
The profitable threshold decreases with the increase of $\gamma$ and $\theta$. When $\gamma_1=\gamma_2=1$ and $N=2$, {BSM} homogeneous attackers can obtain extra revenue with very small Hash power.
}}
\end{observation}

We also explore the impact of different information propagation delays on the profitable threshold. The information propagation delay determines the proportion of other miners' Hash power after the attacker's released chain in the tie-breaking, i.e., $\gamma_1$, $\gamma_2$, $\theta_1$ and $\theta_2$. We will study the profitable threshold of two homogeneous attackers under different information propagation delays, i.e., $\gamma_1=\gamma_2=\gamma$ and $\theta_1=\theta_2=\theta$. Fig \ref{fig:BSM gamma} shows the results when $\theta=1/3$, the profitable threshold will decrease with the increase of $\gamma$.
When $\gamma$ increases to 1, the attackers' profitable threshold tends to 0. This means that when $N$ is relatively small, the attacker can obtain extra revenue through \emph{BSM} with a very small hash power if $\gamma=1$. Fig. \ref{fig:BSM theta} shows the profitable threshold decreases with the increase of $\beta$ when $\gamma=1/2$. It can be observed the less propagation delay leads to more revenues for the attacker.

Since the advent of the seminal work \cite{ref:majority}, the Bitcoin community tries to constrain mining pools to possess less than 25\% of Hash power. However, we prove that 25\% is not \emph{enough}: Bitcoin mining is fragile in the presence of multiple selfish miners. 
\begin{figure*}
	\begin{minipage}[t]{0.32\linewidth}
			\setlength\abovecaptionskip{0.5pt}
			\setlength\belowcaptionskip{-1pt}
			\centering
			\includegraphics[width=1\textwidth]{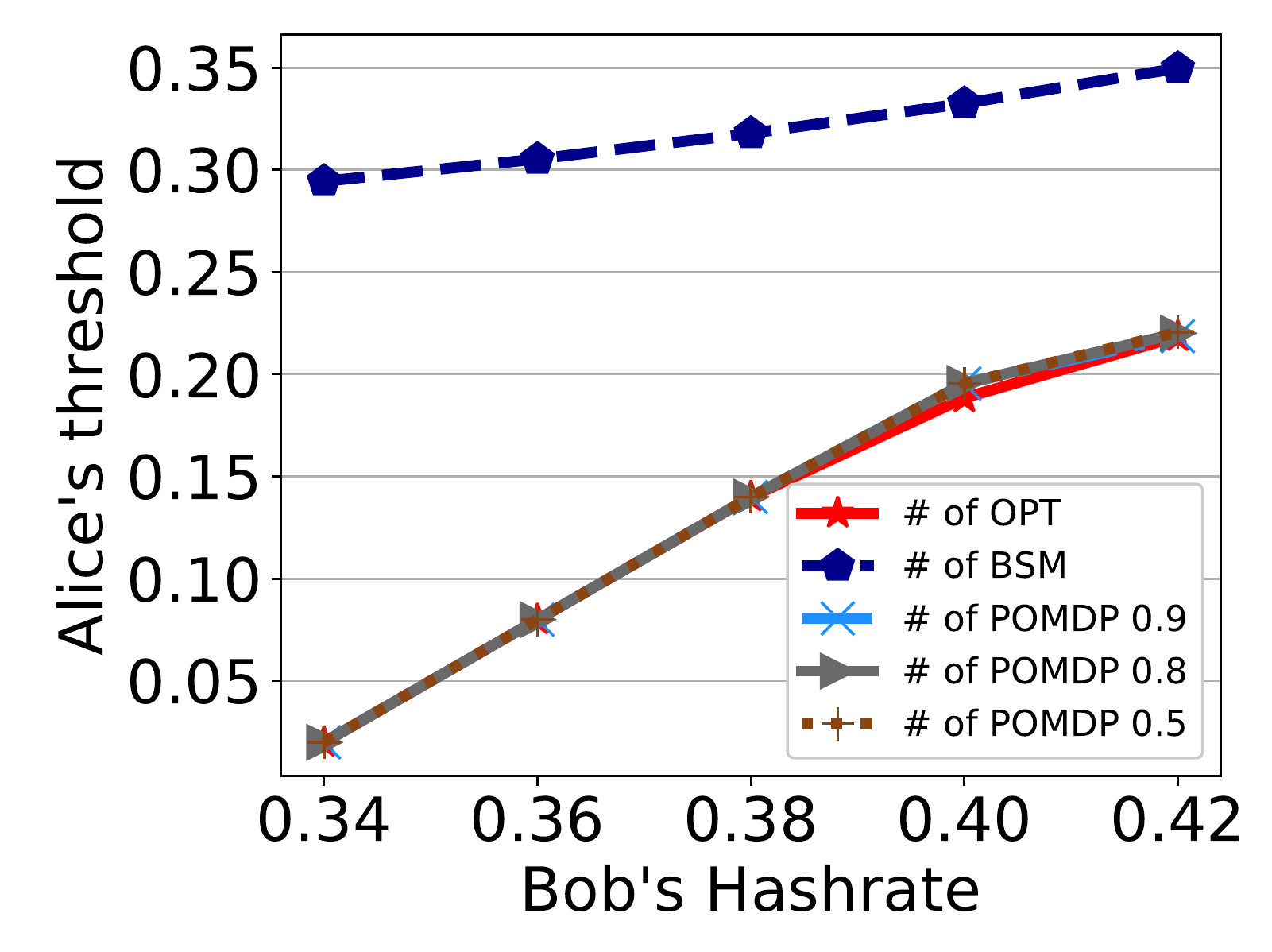}\\
			\caption{Profitable threshold for Alice when $N=2$.}
			\label{fig:OPTthreshold2}
		\end{minipage}
		\hspace{0.2cm}
		\begin{minipage}[t]{0.32\linewidth}
			\setlength\abovecaptionskip{0.5pt}
			\setlength\belowcaptionskip{-1pt}
			\centering
			\includegraphics[width=1\textwidth]{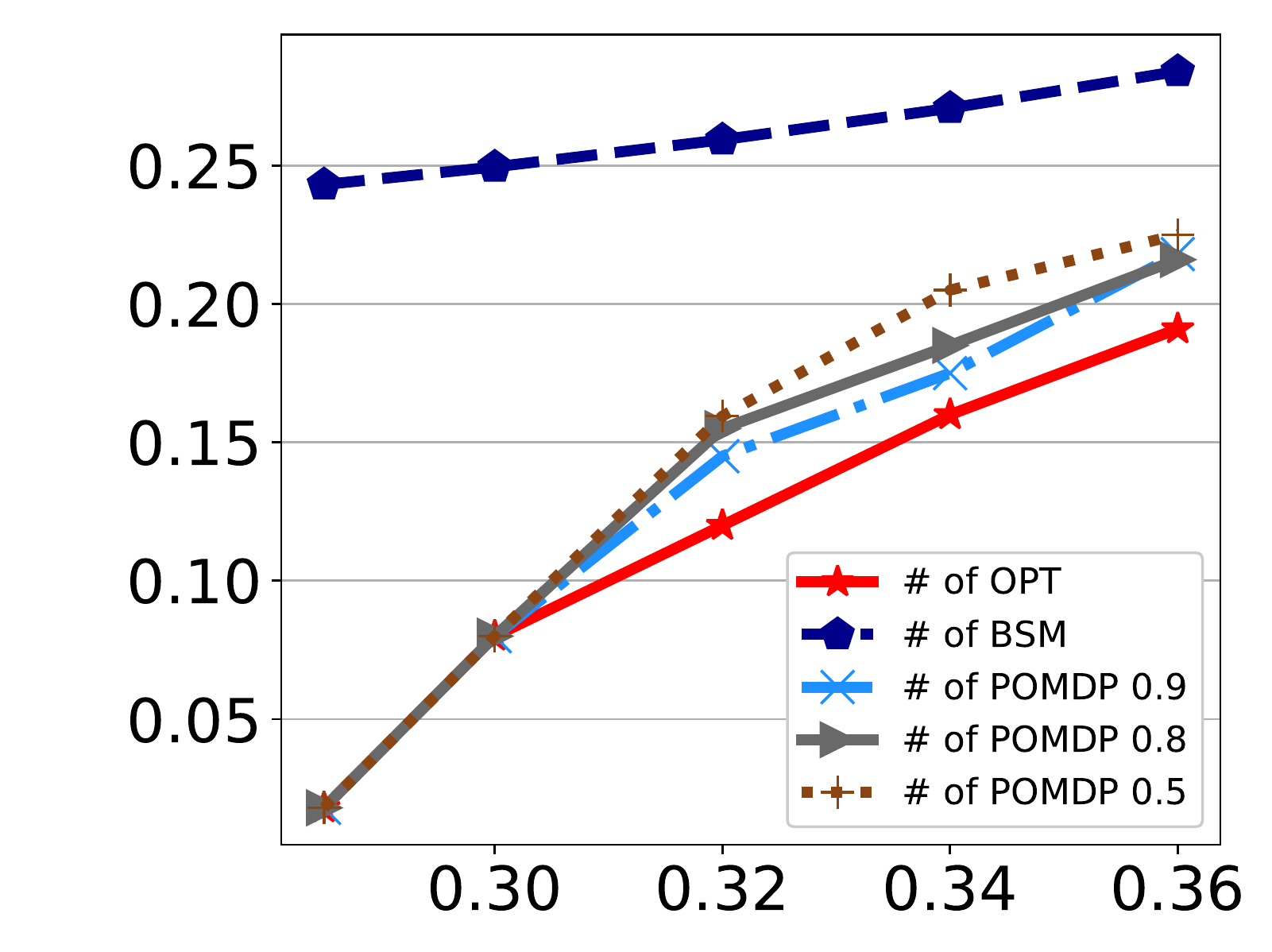}\\
			\caption{Profitable threshold for Alice when $N=3$.}
			\label{fig:OPTthreshold3}
		\end{minipage}
				\begin{minipage}[t]{0.32\linewidth}
			\setlength\abovecaptionskip{0.5pt}
			\setlength\belowcaptionskip{-1pt}
			\centering
			\includegraphics[width=1\textwidth]{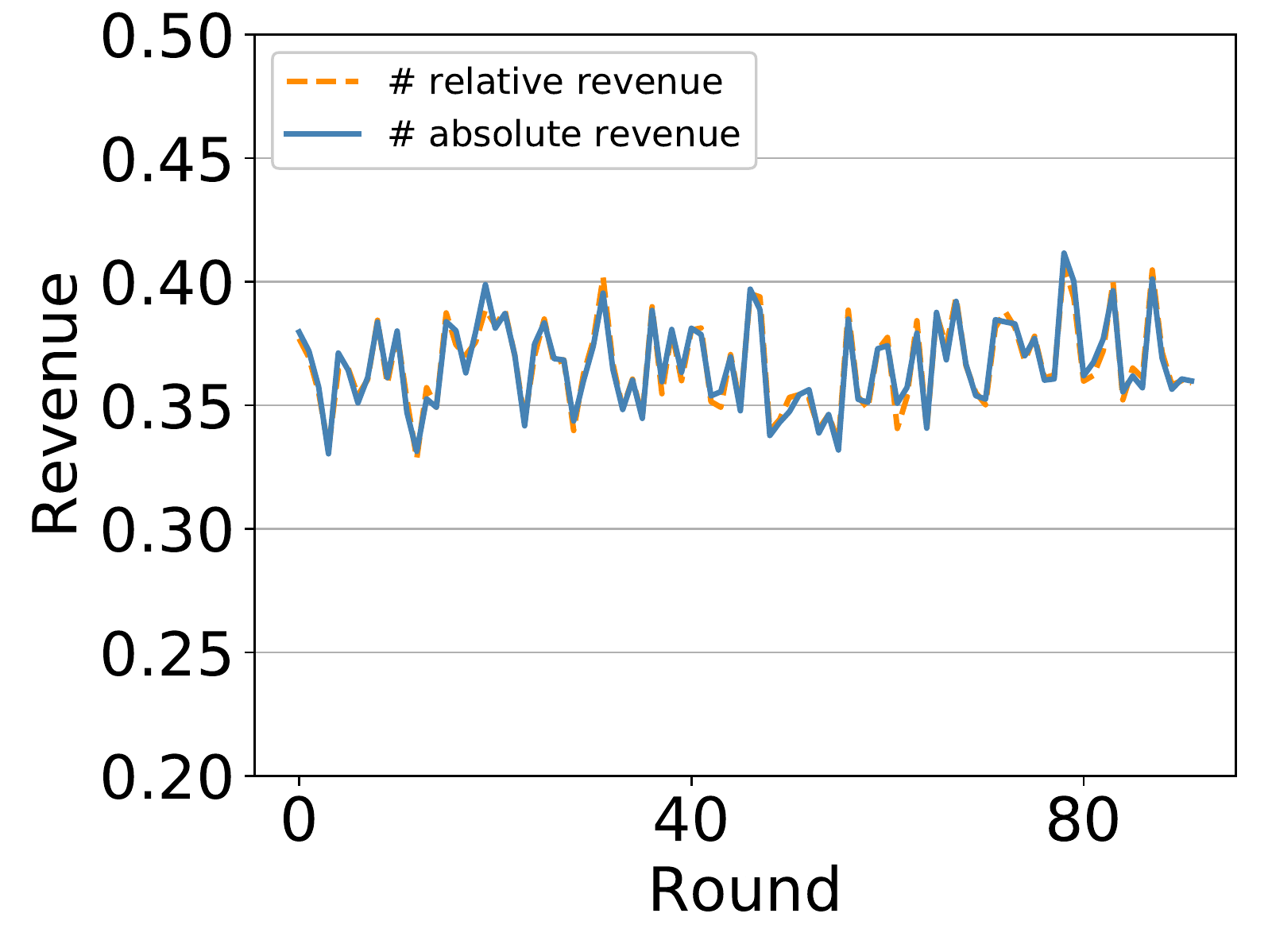}\\
			\caption{Relative revenue and absolute revenue when $\alpha_1=\alpha_2=0.33, N=4$.}
			\label{fig:relative}
		\end{minipage}
		\hspace{0.2cm}
		\vspace{-0.5cm}
		\end{figure*}
	 
					\begin{figure*}[h]
		\begin{minipage}[t]{0.33\linewidth}
			\setlength\abovecaptionskip{0.5pt}
			\setlength\belowcaptionskip{-1pt}
			\centering
			\includegraphics[width=0.8\textwidth]{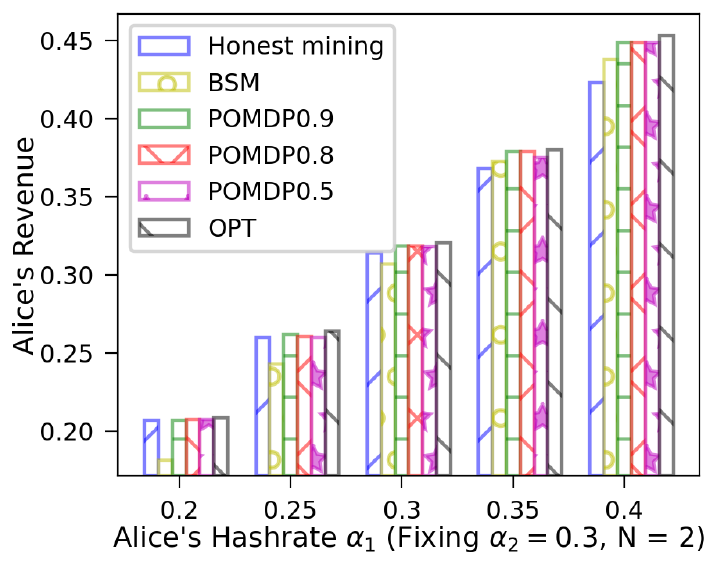}
			\caption{The revenue for Alice when $\alpha_2=0.3$, $N=2$.}
			\label{fig:POMDP21}
		\end{minipage}
				\begin{minipage}[t]{0.33\linewidth}
			\setlength\abovecaptionskip{0.5pt}
			\setlength\belowcaptionskip{-1pt}
			\centering
			\includegraphics[width=0.8\textwidth]{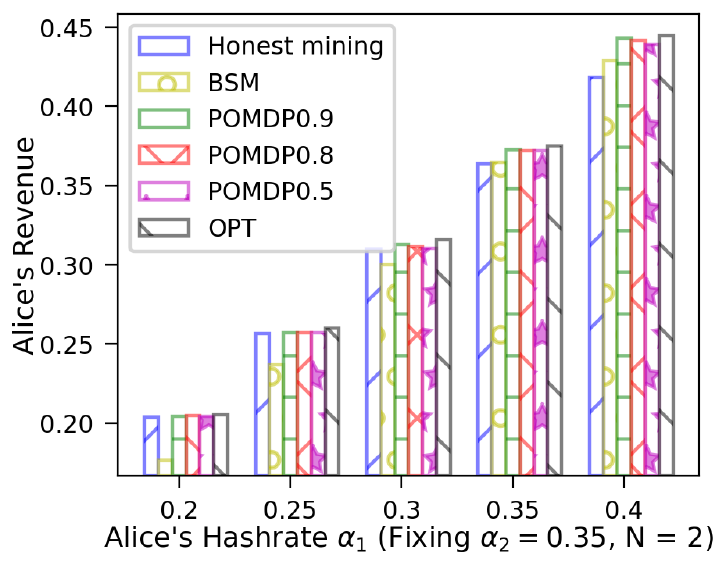}
			\caption{The revenue for Alice when $\alpha_2=0.35$, $N=2$.}
			\label{fig:POMDP2}
		\end{minipage}
				\begin{minipage}[t]{0.33\linewidth}
			\setlength\abovecaptionskip{0.5pt}
			\setlength\belowcaptionskip{-1pt}
			\centering
			\includegraphics[width=0.8\textwidth]{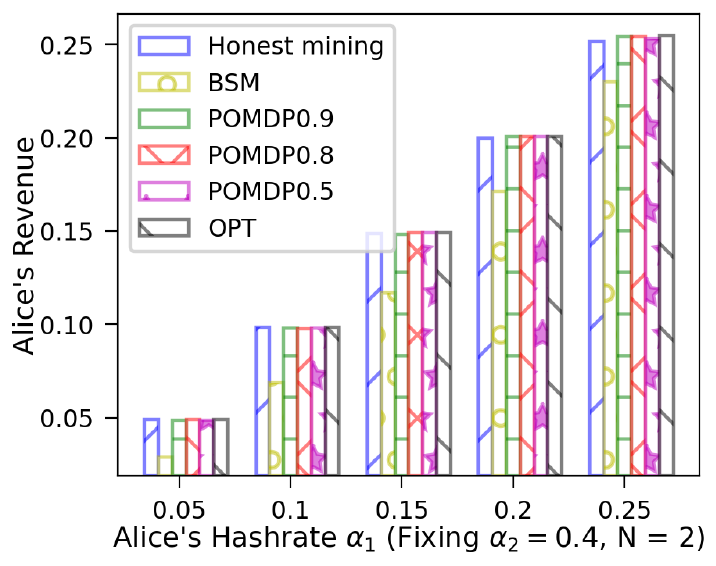}
			\caption{The revenue for Alice when $\alpha_2=0.4$, $N=2$.}
			\label{fig:POMDP23}
		\end{minipage}
		\hspace{0.2cm}
		\begin{minipage}[t]{0.33\linewidth}
			\setlength\abovecaptionskip{0.5pt}
			\setlength\belowcaptionskip{-1pt}
			\centering
			\includegraphics[width=0.8\textwidth]{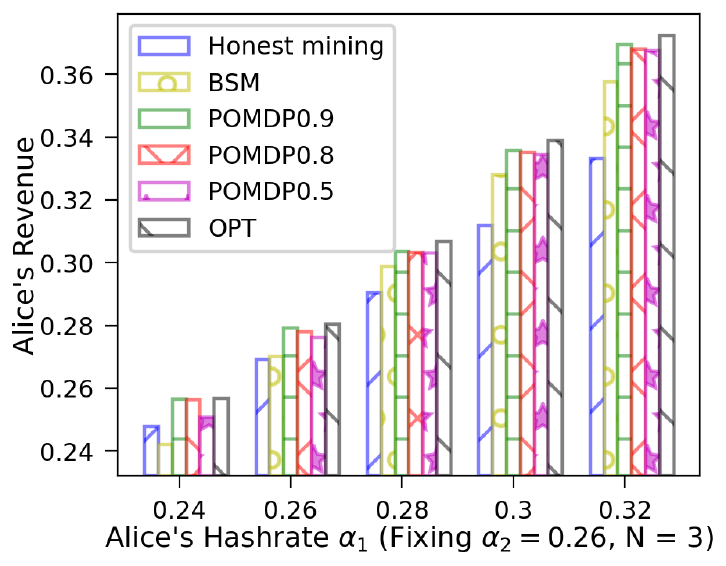}
			\caption{The revenue for Alice when $\alpha_2=0.26$, $N=3$.}
			\label{fig:POMDP31}
		\end{minipage}
				\begin{minipage}[t]{0.33\linewidth}
			\setlength\abovecaptionskip{0.5pt}
			\setlength\belowcaptionskip{-1pt}
			\centering
			\includegraphics[width=0.8\textwidth]{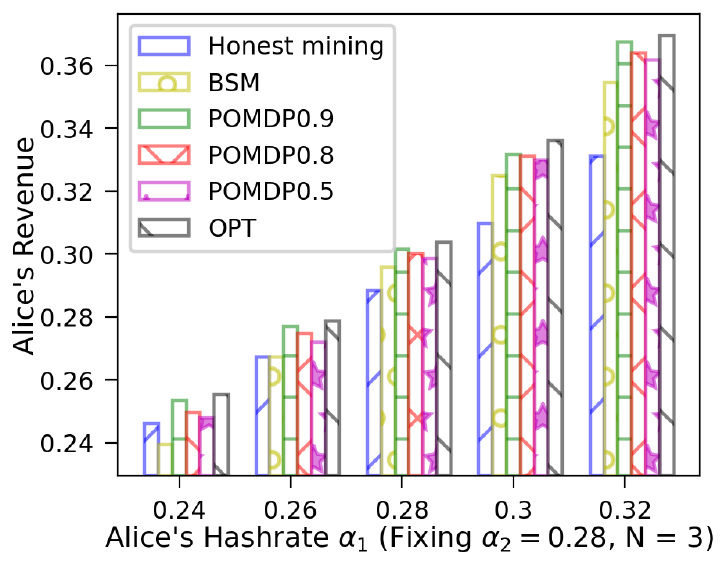}
			\caption{The revenue for Alice when $\alpha_2=0.28$, $N=3$.}
		\end{minipage}
				\begin{minipage}[t]{0.33\linewidth}
			\setlength\abovecaptionskip{0.5pt}
			\setlength\belowcaptionskip{-1pt}
			\centering
			\includegraphics[width=0.8\textwidth]{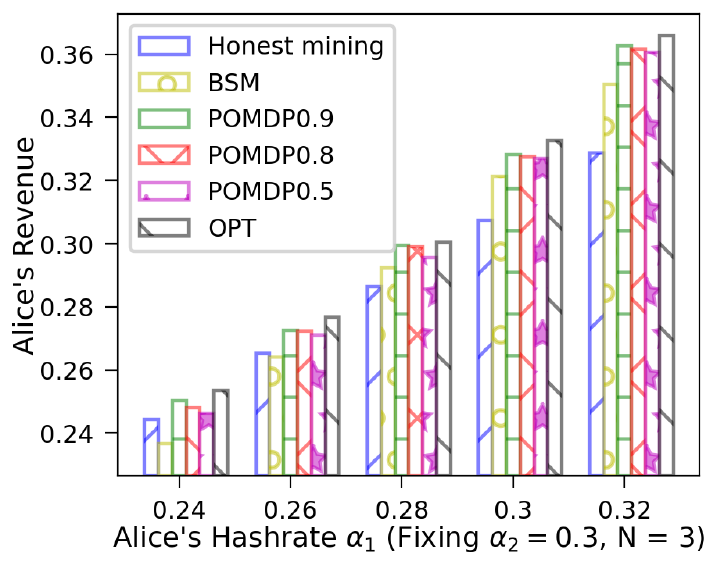}
			\caption{The revenue for Alice when $\alpha_2=0.3$, $N=3$.}
			\label{fig:POMDP33}
		\end{minipage}
		\vspace{-0.7cm}
		\hspace{0.2cm}
	\end{figure*}

\subsection{MDP and POMDP-based Mining}
The roadmap of performing optimal mining is the following. We explore its feasibility by estimating the unknown parameters. Then, we compute the optimal mining policy
and the corresponding revenue using MDP, and on this basis we compute the optimal mining policy using POMDP under partially observable states.
  
	Recall that Alice is the strategic attacker with \emph{POMDP}-based policy and Bob is the basic selfish miner (\emph{BSM}). Alice needs to compute the proportion of her Hash power to the global Hash power based on the public information \cite{ref:bitcoin.com}. Then she needs to decide whether there exists a selfish miner namely Bob, and if so, what Bob's Hash power is. Note that there has been a one-to-one mapping between Henry's orphan block ratio and Bob's Hash power. Fig. \ref{fig:orphan one} shows the orphan ratio of the honest miner as a function of Bob's Hash power with $N=2$ and $N=3$. The theoretical and experimental results match well, which indicates the feasibility of calculating Bob's Hash power through the observed orphan block ratio.
	After that, we compute the revenue upper bound of \emph{POMDP}-based policy using MDP. Analogous experiments of \emph{MDP}-based mining
can be found in Appendix D. Then, we will focus on the performances of \emph{POMDP}-based mining policy. Let the error parameter $\epsilon=0.00001$ and the execution number $\xi_0=1000000$.


	
		\begin{figure*}[h]
		\begin{minipage}[t]{0.33\linewidth}
			\setlength\abovecaptionskip{0.5pt}
			\setlength\belowcaptionskip{-1pt}
			\centering
			\includegraphics[width=0.8\textwidth]{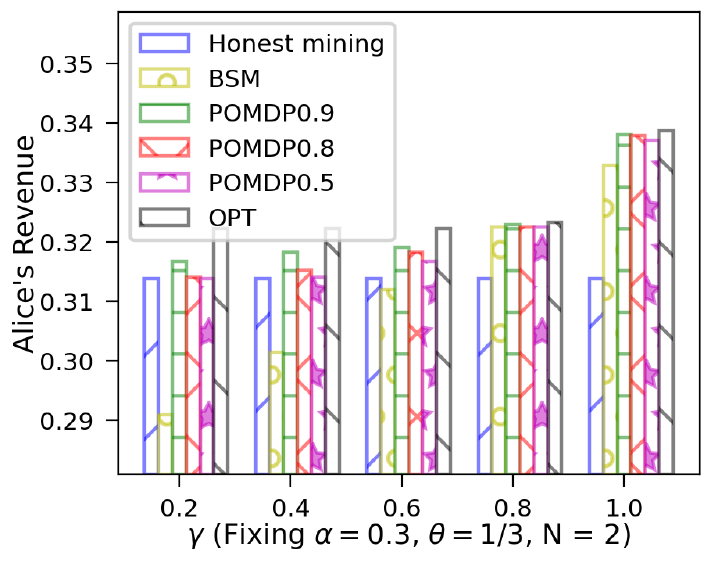}\\
			\caption{The revenue for Alice when $\alpha=0.3$, $N=2$ with different $\gamma$.}
			\label{fig:gamma21}
		\end{minipage}
				\begin{minipage}[t]{0.33\linewidth}
			\setlength\abovecaptionskip{0.5pt}
			\setlength\belowcaptionskip{-1pt}
			\centering
			\includegraphics[width=0.8\textwidth]{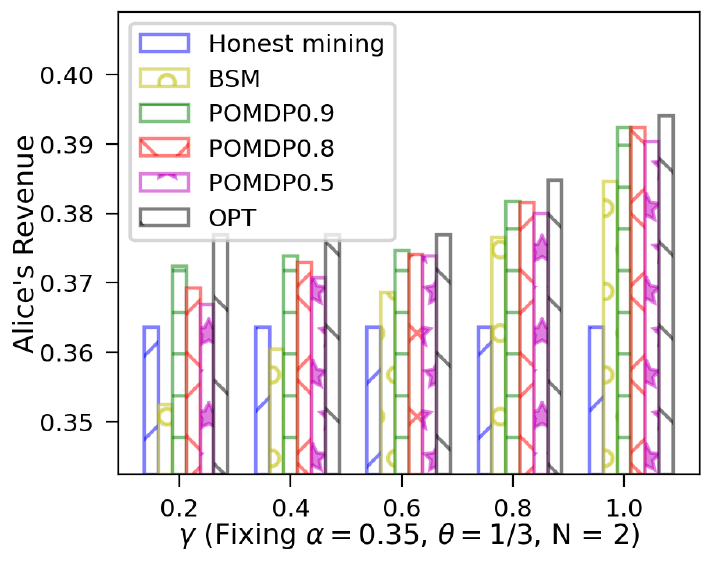}\\
			\caption{The revenue for Alice when $\alpha=0.35$, $N=2$ with different $\gamma$.}
			\label{fig:gamma22}
		\end{minipage}
				\begin{minipage}[t]{0.33\linewidth}
			\setlength\abovecaptionskip{0.5pt}
			\setlength\belowcaptionskip{-1pt}
			\centering
			\includegraphics[width=0.8\textwidth]{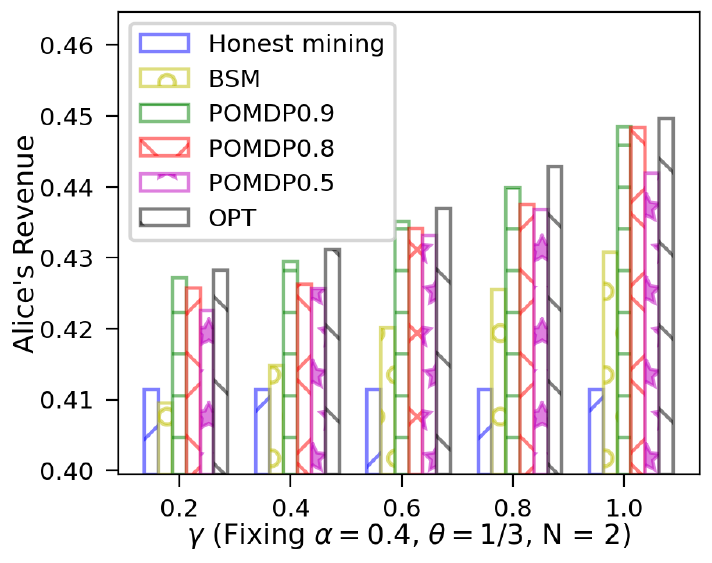}\\
			\caption{The revenue for Alice when $\alpha=0.4$, $N=2$ with different $\gamma$.}
			\label{fig:gamma23}
		\end{minipage}
		\hspace{0.2cm}
		\begin{minipage}[t]{0.33\linewidth}
			\setlength\abovecaptionskip{0.5pt}
			\setlength\belowcaptionskip{-1pt}
			\centering
			\includegraphics[width=0.8\textwidth]{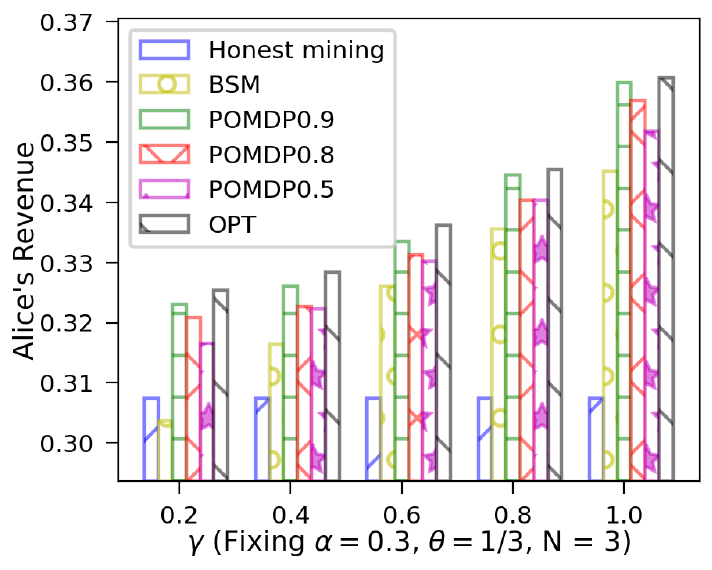}\\
			\caption{The revenue for Alice when $\alpha=0.3$, $N=3$ with different $\gamma$.}
			\label{fig:gamma31}
		\end{minipage}
				\begin{minipage}[t]{0.33\linewidth}
			\setlength\abovecaptionskip{0.5pt}
			\setlength\belowcaptionskip{-1pt}
			\centering
			\includegraphics[width=0.8\textwidth]{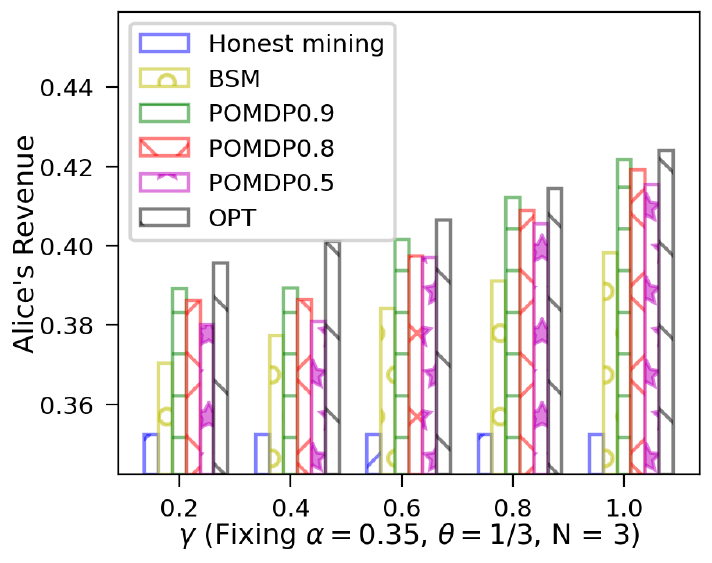}\\
			\caption{The revenue for Alice when $\alpha=0.35$, $N=3$ with different $\gamma$.}
			\label{fig:gamma32}
		\end{minipage}
				\begin{minipage}[t]{0.33\linewidth}
			\setlength\abovecaptionskip{0.5pt}
			\setlength\belowcaptionskip{-1pt}
			\centering
			\includegraphics[width=0.8\textwidth]{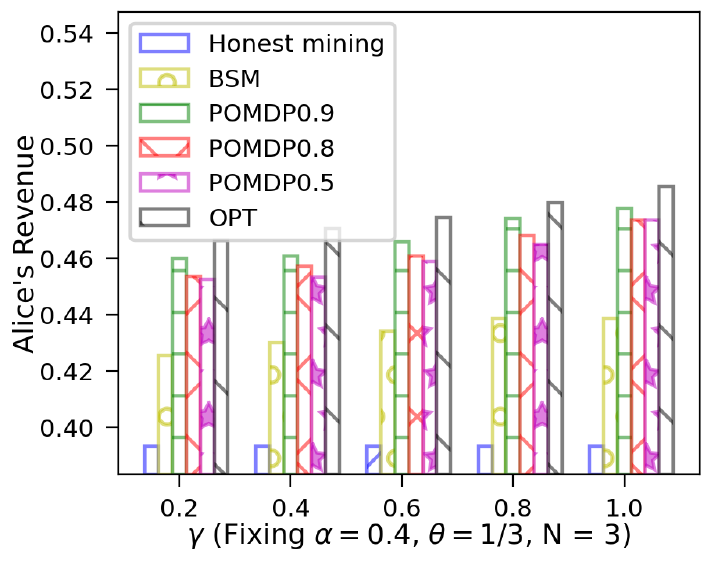}\\
			\caption{The revenue for Alice when $\alpha=0.4$, $N=3$ with different $\gamma$.}
			\label{fig:gamma33}
		\end{minipage}
		\hspace{0.2cm}
		\begin{minipage}[t]{0.33\linewidth}
			\setlength\abovecaptionskip{0.5pt}
			\setlength\belowcaptionskip{-1pt}
			\centering
			\includegraphics[width=0.8\textwidth]{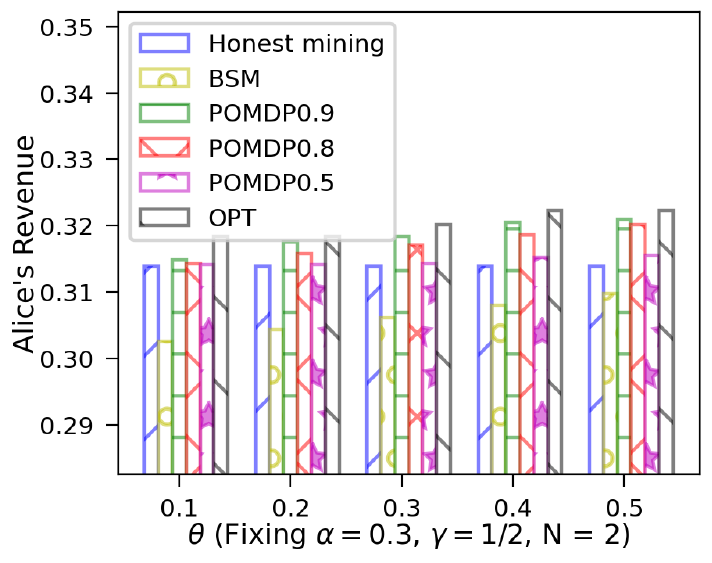}\\
			\caption{The revenue for Alice when $\alpha=0.3$, $N=2$ with different $\theta$.}
			\label{fig:theta21}
		\end{minipage}
				\begin{minipage}[t]{0.33\linewidth}
			\setlength\abovecaptionskip{0.5pt}
			\setlength\belowcaptionskip{-1pt}
			\centering
			\includegraphics[width=0.8\textwidth]{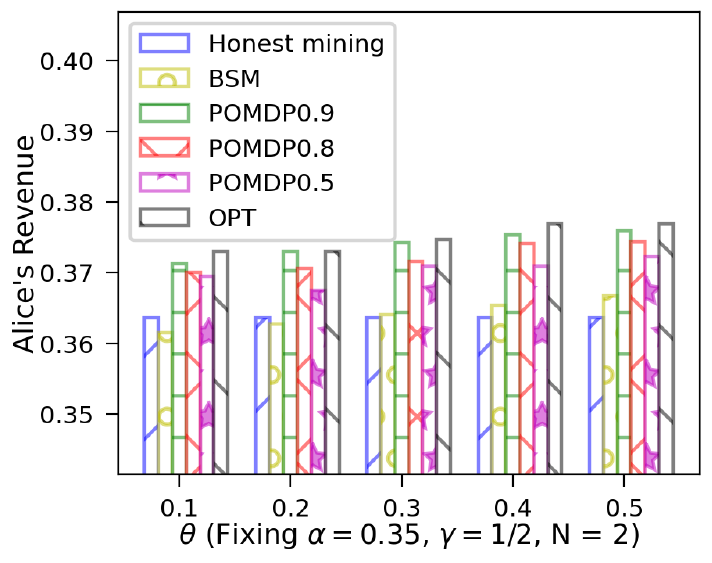}\\
			\caption{The revenue for Alice when $\alpha=0.35$, $N=2$ with different $\theta$.}
			\label{fig:theta22}
		\end{minipage}
				\begin{minipage}[t]{0.33\linewidth}
			\setlength\abovecaptionskip{0.5pt}
			\setlength\belowcaptionskip{-1pt}
			\centering
			\includegraphics[width=0.8\textwidth]{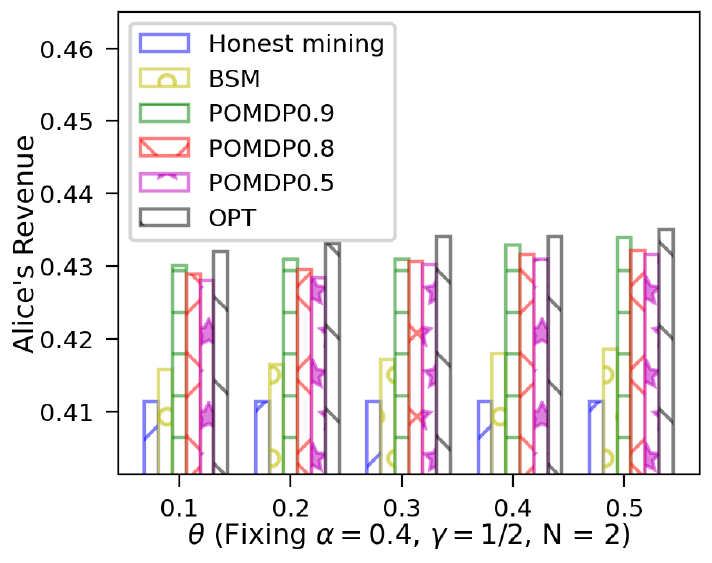}\\
			\caption{The revenue for Alice when $\alpha=0.4$, $N=2$ with different $\theta$.}
			\label{fig:theta23}
		\end{minipage}
		\hspace{0.2cm}
		\begin{minipage}[t]{0.33\linewidth}
			\setlength\abovecaptionskip{0.5pt}
			\setlength\belowcaptionskip{-1pt}
			\centering
			\includegraphics[width=0.8\textwidth]{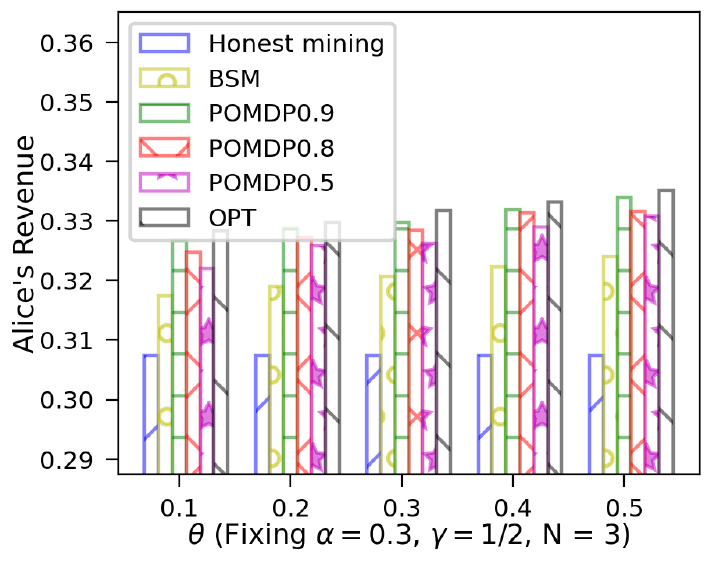}\\
			\caption{The revenue for Alice when $\alpha=0.3$, $N=3$ with different $\theta$.}
			\label{fig:theta31}
		\end{minipage}
				\begin{minipage}[t]{0.33\linewidth}
			\setlength\abovecaptionskip{0.5pt}
			\setlength\belowcaptionskip{-1pt}
			\centering
			\includegraphics[width=0.8\textwidth]{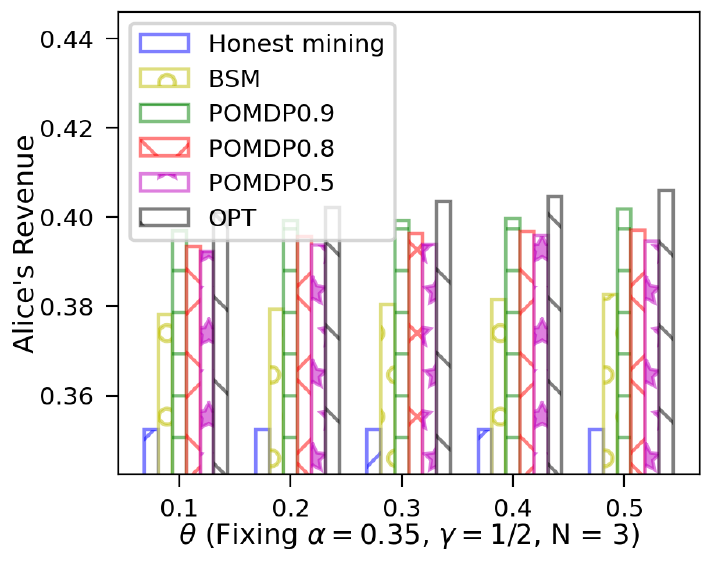}\\
			\caption{The revenue for Alice when $\alpha=0.35$, $N=3$ with different $\theta$.}
			\label{fig:theta32}
		\end{minipage}
				\begin{minipage}[t]{0.33\linewidth}
			\setlength\abovecaptionskip{0.5pt}
			\setlength\belowcaptionskip{-1pt}
			\centering
			\includegraphics[width=0.8\textwidth]{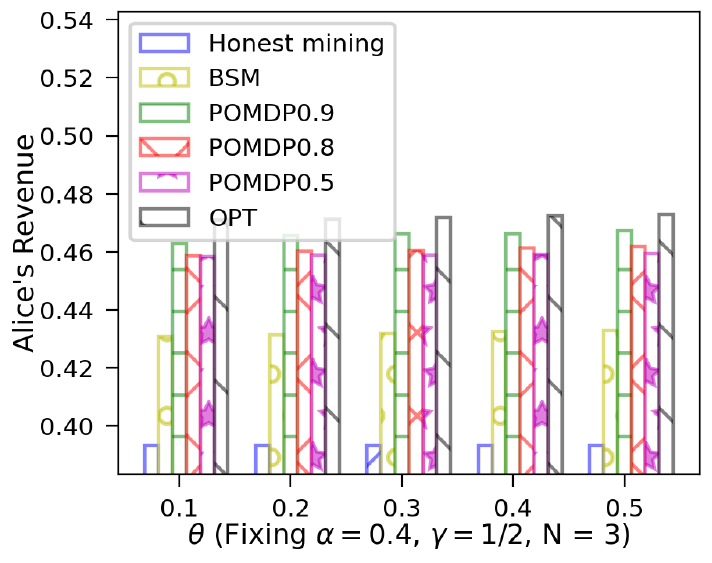}\\
			\caption{The revenue for Alice when $\alpha=0.4$, $N=3$ with different $\theta$.}
			\label{fig:theta33}
		\end{minipage}
		\vspace{-0.7cm}
		\hspace{0.2cm}
	\end{figure*}
\begin{observation} {\textit{When Alice uses the POMDP-based policy, and Bob uses the basic selfish mining, Alice has a much lower profitable threshold compared with the basic selfish mining. Her revenue is no less than the honest mining and the basic selfish mining, and is close to the \emph{OPT} policy with complete state information.}}
\end{observation}
We evaluate Alice's profitable threshold and revenue when she uses the \emph{POMDP}-based mining policy. Fig. \ref{fig:OPTthreshold2} and \ref{fig:OPTthreshold3} compare the profitable thresholds of \emph{BSM}, \emph{MDP}-based (\emph{OPT}) and \emph{POMDP}-based (POMDP) mining strategies when $N =2,3 $ and $\alpha_2$ increases from 0.285 to 0.42. An interesting finding is that both \emph{OPT} and POMDP strategies have much smaller profitable thresholds compared with \emph{BSM}. For instance, the profitable threshold is 0.02 when $\alpha_2 = 0.34$ and $N=2$, and is 0.08 when $\alpha_2 = 0.3$ and $N=3$. The reason is the following. When Bob is playing \emph{BSM} and Bob's Hash power is much larger than Alice's, Alice will choose to mine honestly. If there is a fork between Bob and Henry, Alice insists on mining on the public chain that contains her own blocks. Alice's policy is equivalent to decreasing $\gamma_2$ in the situation of a single attacker. Therefore, Bob will find it difficult to gain more revenues, and Alice as well as Henry benefits from Bob's losses. 
In addition, with the increase of Bob's Hash power, Alice's profitable threshold becomes higher. 
{ A cross-comparison between Fig. \ref{fig:OPTthreshold2} and \ref{fig:OPTthreshold3} shows that a larger $N$ makes Alice hard to compete with Bob if $\alpha_1 \leq \alpha_2$. When Bob's Hash power is 0.36, Alice's profitable threshold is 0.08 when $N=2$ and is about 21.6\% when $N=3$.} 

The \emph{POMDP}-based policy can improve Alice's revenue under different situations. We first direct our attention towards examining the influence of miners' Hash power on the profitability of \emph{POMDP-}based policy. Setting $\gamma_1=\gamma_1=1/2$ and $\theta_1=\theta_2=1/3$, Fig. \ref{fig:POMDP21}$\sim$\ref{fig:POMDP23} plot Alice's revenues using Honest mining, \emph{BSM}, \emph{OPT} and \emph{POMDP}-based mining at $N=2$; Fig. \ref{fig:POMDP31}$\sim$\ref{fig:POMDP33} show those revenues with $N=3$. In each set of experiments, we fix Bob's Hash power ($\alpha_2$) and change Alice's Hash power ($\alpha_1$) from 0.20 to 0.40. As for the \emph{POMDP}-based policy, three cases are considered, in which the probability of generating a block is 0.9, 0.8 or 0.5 at each time slot. 
One can easily observe that the \emph{POMDP}-based policy has comparable revenues with the honest and the \emph{OPT} mining policies when Alice's Hash power is relatively small, such as the case of $\alpha_1=0.25$ in Fig. \ref{fig:POMDP2} and Fig. \ref{fig:POMDP23}. While the basic selfish mining seems very ``stubborn'', causing Alice's revenue much lower than the honest mining in this situation. The revenue of the \emph{POMDP}-based policy is significantly higher than that of \emph{BSM} and honest mining when Alice's hash power is large, such as the case of $\alpha_1=0.4$ in Fig. \ref{fig:POMDP21}.

Subsequently, we explore the potential benefits of various $\gamma$ and $\theta$. Fig. \ref{fig:gamma21} $\sim$ Fig. \ref{fig:gamma33} plots Alice's revenue using honest mining, \emph{BSM}, \emph{OPT} and \emph{POMDP}-based mining at $N=2$ and $N=3$ with different $\gamma$; Fig. \ref{fig:theta21} $\sim$ Fig. \ref{fig:theta33} show those revenues with different $\theta$. In each set of experiments, we fix the two attackers' Hash power ($\alpha_1=\alpha_2$) to 0.3, 0.35 and 0.4. We then vary $\gamma_1=\gamma_2=\gamma$ from 0.2 to 1 with $\theta_1=\theta_2=1/3$. At the same time, we also explore the situations that $\theta_1=\theta_2=\theta$ is from 0.1 to 0.5 with $\gamma_1=\gamma_2=1/2$. Evidently, it can be observed that the \emph{POMDP}-based policy has demonstrated superior performance compared with both \emph{BSM} and honest mining across all scenarios.
Meanwhile, as the parameters $\gamma$ and $\theta$ exhibit an increase, a corresponding increase in Alice's revenue of \emph{POMDP}-based policy is discernible.
In conclusion, the \emph{PODMP}-based policy generates equal or higher revenues than honest mining and basic selfish mining, and approaches the performance of the \emph{OPT} policy.


The designed online algorithm can effectively calculate the optimal revenue of $\mathcal{M}_{OP}$. We compare the number of iterations required to execute the binary search algorithm in \cite{ref:optimal selfish mining} and Algorithm \ref{alg:POMDP1}. The reduction ratios are summarized in Table \ref{table:rate}. Under different Hash power combinations and different action slots, the efficiency of Algorithm \ref{alg:POMDP1} is significantly higher. When $N=2$, $\alpha_1=0.3$, $\alpha_2=0.3$ and $p=0.5$, we can save 66.6\% of the computing time. It takes a long time to simulate half a million times to obtain the revenue for each $\rho$. The improvement of our search algorithm plays a significant role in solving $\mathcal{M}_{PO}$ rapidly.

\subsection{Selfish Mining on Multiple Difficulty Adjustment Periods}

\begin{figure*}[t]

		\hspace{0.2cm}
		\begin{minipage}[t]{0.32\linewidth}
			\setlength\abovecaptionskip{0.5pt}
			\setlength\belowcaptionskip{-1pt}
			\centering
			\includegraphics[width=1\textwidth]{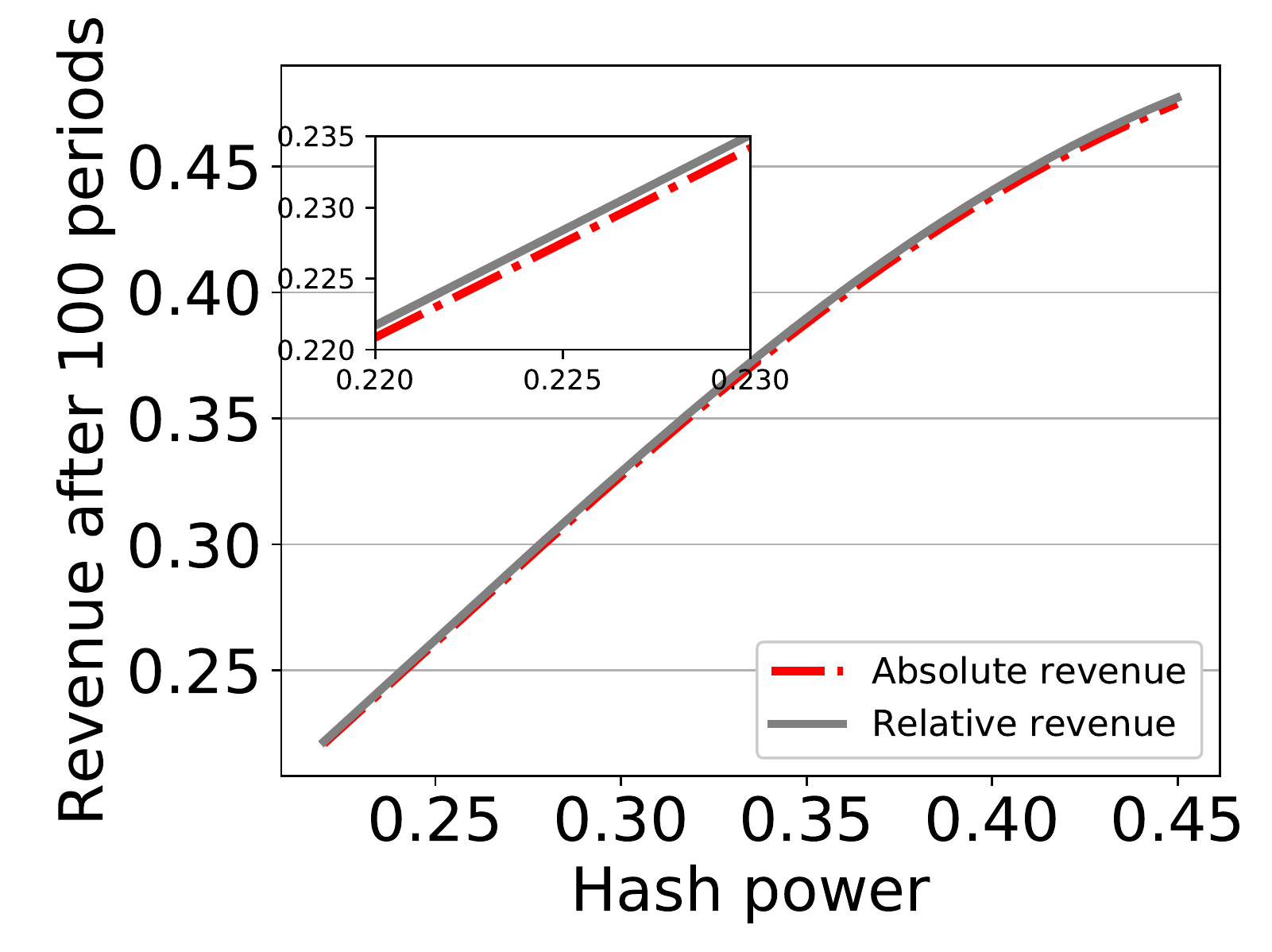}\\
			\caption{The theoretical relative revenue and absolute revenue when $\alpha_1=\alpha_2$ after 100 periods.}
			\label{fig:theory relative absolute100}
		\end{minipage}
				\begin{minipage}[t]{0.32\linewidth}
			\setlength\abovecaptionskip{0.5pt}
			\setlength\belowcaptionskip{-1pt}
			\centering
			\includegraphics[width=1\textwidth]{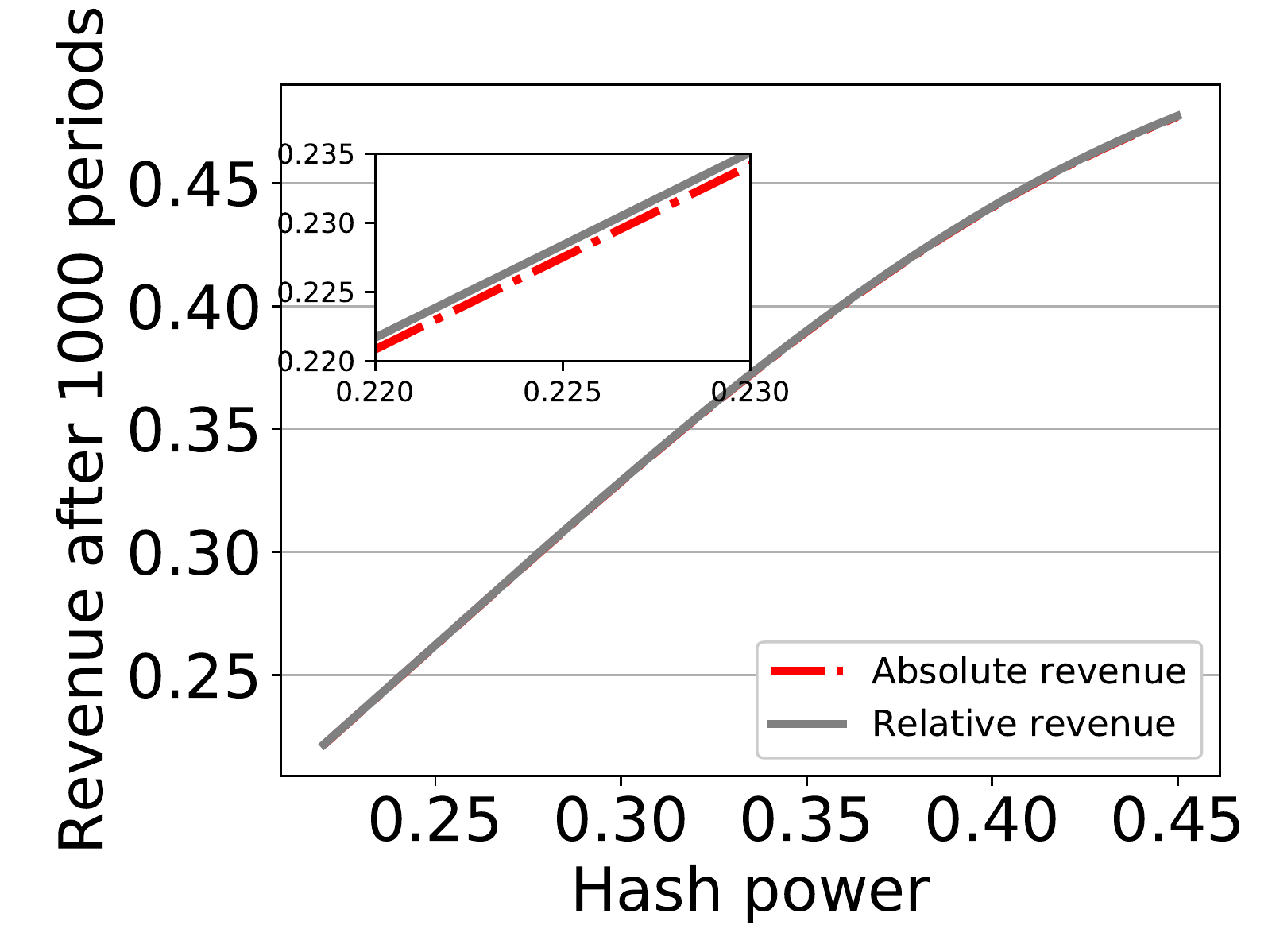}\\
			\caption{The theoretical relative revenue and absolute revenue when $\alpha_1=\alpha_2$ after 1000 periods.}
			\label{fig:theory relative absolute1000}
		\end{minipage}
		\begin{minipage}[t]{0.32\linewidth}
			\setlength\abovecaptionskip{0.5pt}
			\setlength\belowcaptionskip{-1pt}
			\centering
			\includegraphics[width=1\textwidth]{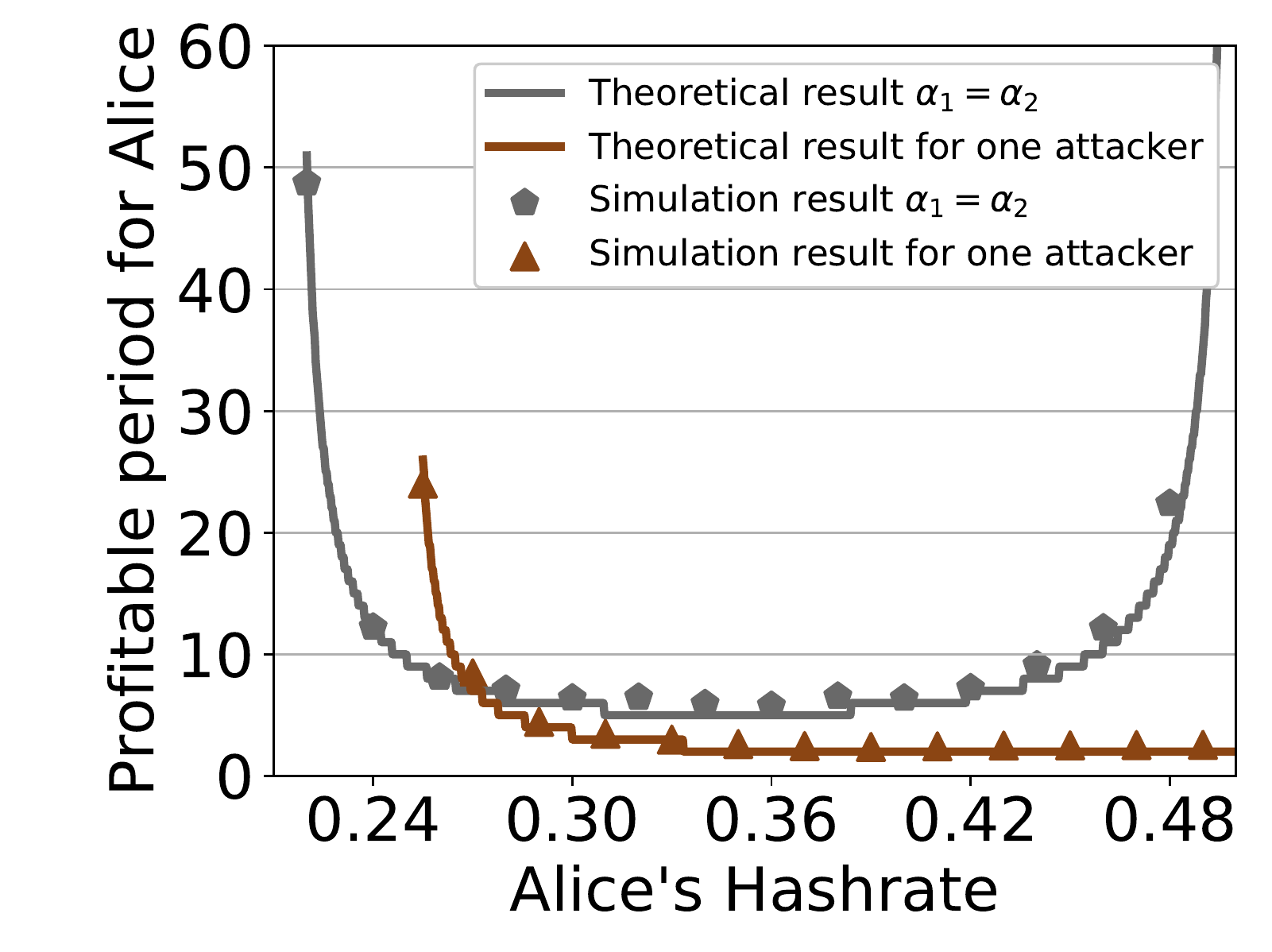}\\
			\caption{Profitable time and Hash power.}
			\label{fig:round}
		\end{minipage}
		\hspace{0.2cm}
		\vspace{-0.7cm}
	\end{figure*}

\begin{observation} {\textit {A selfish miner obtains a smaller absolute revenue than that of honest mining during the first difficulty adjustment period regardless of his Hash power. However, he might gain profit after a number of periods that is related to the selfish miners' Hash power.}}
\end{observation}
Fig. \ref{fig:relative} shows the \emph{relative revenue} and \emph{absolute revenue} of attackers with the same Hashrate 33\% and $N=4$ in each DAA period. The \emph{relative revenue} and \emph{absolute revenue} are equal within the allowable range of error. Therefore, the \emph{relative revenue} can play the same role as \emph{absolute revenue} in representing benefit.
	Fig. \ref{fig:theory relative absolute100} and Fig. \ref{fig:theory relative absolute1000} show the theoretical \emph{relative revenue} and \emph{absolute revenue} after 100 and 1000 DAA periods when $\alpha_1=\alpha_2$ and $N=4$. It can be observed that the \emph{absolute revenue} is always less than the \emph{relative revenue}, but the difference is small. When $\alpha_1=0.22$, the \emph{relative revenue} is 0.2217 and the \emph{absolute revenue} is 0.2209. This difference decreases to 0.0001 after 1000 periods. That means the difference between \emph{relative revenue} and \emph{absolute revenue} decreases with the increase of the attack time.  

	\begin{table}[h]
		\centering  
		\caption{The time reduction ratio of 
		Alg. \ref{alg:POMDP1} compared to traditional algorithm.}
		{\begin{tabular}{|c|c|c|c|}
			\hline
			{Hash power}&{$p$}&{EFF IMP}\\
			\hline
			\multirow{1.5}*{$\alpha_1=0.3,\alpha_2=0.3$}&{$0.9$}&{53.3\%}\\
			\cline{2-3}
			~&{$0.8$}&{46.6\%}\\
			\cline{2-3}
			\multirow{1}*{$N=2$}&{$0.5$}&{66.6\%}\\
			\hline
			\multirow{1.5}*{$\alpha_1=0.35,\alpha_2=0.35$}&{$0.9$}&{80\%}\\
			\cline{2-3}
			~&{$0.8$}&{60\%}\\
			\cline{2-3}
			\multirow{1}*{$N=2$}&{$0.5$}&{46.7\%}\\
			\hline
			\multirow{2}*{$\alpha_1=0.28,\alpha_2=0.28$}&{$0.9$}&{60\%}\\
			\cline{2-3}
			~&{$0.8$}&{46.7\%}\\
			\cline{2-3}
			\multirow{1}*{$N=3$}&{$0.5$}&{46.7\%}\\
			\hline
			
						\multirow{2}*{$\alpha_1=0.3,\alpha_2=0.3$}&{$0.9$}&{73.3\%}\\
			\cline{2-3}
			~&{$0.8$}&{46.7\%}\\
			\cline{2-3}
			\multirow{1}*{$N=3$}&{$0.5$}&{40\%}\\
			\hline
			\end{tabular}}
			\label{table:rate}
			\vspace{-0.7cm}
			\end{table}
	As Eq. \eqref{eq:revenue each round} shows, when Alice has more Hash powers, she can get illegal revenue earlier. However, if both of the two attackers have a large Hashrate, they will benefit late. Fig. \ref{fig:round} shows the simulation results and the theoretical results of profitable time under different Hash powers with symmetric attackers. The horizontal axis represents the attack's Hash power and the ordinate represents the attackers' profitable time, also the curves are theoretical results and the dots are simulation results.
	 The selfish mining is profitable after 51 DAA rounds (i.e. 714 days in Bitcoin) if the Hash power of selfish miners are both 22\% (slightly higher than the profitable threshold). This delay decreases to 
		5 rounds (i.e. 70 days in Bitcoin) as their Hash power accrues to 33\%, which is still very long. 
		As the attackers gained more computing power, Alice's main competitor became Bob rather than Henry. The benefit time begins to increase for Alice and Bob.
		When there is only one attacker and she has 25.5\% Hashrates, the attacker will obtain extra revenue after 26 rounds (about 364 days). 
		
	It shows that when attackers' Hashrate is relatively small, it takes a rather long period to gain profit. When the two attackers all have large Hashrates, it also takes a long period to obtain extra revenue. That means in the real system, it is a little bit hard to perform attack. If the global Hashrate increases, we can also use this formula to calculate when to stop the attack before we can benefit the most.

\section{related work}
\label{sec:related}

Selfish mining attack is one of the core challenges of blockchain consensus that has been extensively studied in the past several years. We hereby describe the recent advances in selfish mining policies and their analytical or experimental performance.

\emph{One attacker.} Since the advent of \cite{ref:majority}, there have been many studies on different forms of selfish mining attacks. 
Nayak et al. \cite{ref:stubborn} proposed the stubborn mining which improves about 13.94\% revenues compared with the basic selfish mining attack. A more intelligent selfish mining strategy has been proposed in \cite{ref:optimal selfish mining} based on the Markov Decision Process and it decreases the profitable threshold to 23.21\%. Tao et al. \cite{ref:hidden} described semi-selfish mining attack on the basis of selfish mining based on hidden Markov decision process, which not only ensures the benefit of the attack, but also reduces the forking rate. 
Negy et al. \cite{ref:selfish mining re-examined} introduced \textit{intermittent selfish mining} and showed that the intermittent selfish miner above 37\% hash power earns more coins per time unit even when $\gamma=0$.
Davidson et al. \cite{ref:DAA1} simulated the profitability of selfish mining under several difficulty adjustment algorithms used in popular cryptocurrencies. 
Selfish mining attack takes on different properties in the Ethereum system because of the uncle block. 
Grunspan and Ritz introduced the selfish mining attack in Ethereum and found that Ethereum is more vulnerable to selfish mining than Bitcoin \cite{ref:ethereum selfish mining1} \cite{ref:ethereum selfish mining2}. 
 Zhang et al. studied the selfish mining, double-spending and feather-forking in different blockchain systems based on MDP. They obtained that no PoW protocol achieves
ideal chain quality or is resistant against all three attacks \cite{ref:lay down}.

At the same time, selfish mining attack can also be combined with other attacks to achieve greater benefits. 
Gervais et al. devised optimal strategies for double-spending and selfish mining while taking into account real world constraints such as network propagation, different block sizes, block generation intervals, information propagation mechanism, and the impact of eclipse attacks \cite{ref:on the security}.
Kwon et al. \cite{ref:FAW} proposed FAW attack which combines the selfish mining attack and withholding attack. The reward for an FAW attacker is no less than that for a BWH attacker. Gao et al. extended the work of Kwon et al. proposing the power adjusting withholding attack (PAW) and bribery selfish mining attack (BSW) \cite{ref:PAW}. They showed PAW could avoid the ``miner's dilemma" in BWH attack and BSW increases revenues by 10\% compared with traditional SM \cite{ref:miner's dilemma}. However, BSW will introduce ``venal miner's dilemma". To avoid the ``venal miner's dilemma", Yang et al. proposed the IPBSM attack which assumed all attackers take the optimal bribery selfish mining \cite{ref:IPBSM}. 

\emph{Multiple attackers.} The participation of multiple attackers in the system will greatly change the benefits of selfish mining attacks. 
Ruan et al. simulated the multiple strategic miners employing strategies other than honest mining and extended the attackers' strategy by proposing a new strategy set \emph{publish-n} \cite{ref:ruanna}. Their results show a lower profitable threshold in the presence of multiple attackers compared to that with a single attacker.
Xia et al. explored the impact of multiple miners and propagation delay on selfish mining\cite{ref:gamma multiple attacker}. They used simulations to show that 
the large selfish miner with more mining power can obtain extra revenue while other smaller attackers cannot when there are multiple selfish miners.
Charlie et al. proposed SquirRL which is a framework for using deep reinforcement learning to analyze attacks on blockchain incentive mechanisms. The revenue of SquirRL is greater than that of the Markov decision attackers when there are multiple selfish mining attackers \cite{ref:squirRL}.
Francisco et al. proposed semi-selfish mining when there are two attackers and modeled this attack \cite{ref:semi selfish mining}. They focused on the simplest selfish mining strategy in which selfish miners never maintain private chains of length greater than 2. Their result shows the Nash equilibrium under different Hash powers and the threshold for each policy based on the game theory. Also, they modeled the situation when the number of attackers is more than two. However, they do not get the closed-form result.  

Further research is undertaken to explore the correlation between the number of attackers and the security of the system. Sebastian et al. proposed the simulation results that the profitable threshold decreases in proportion to the number of selfish miners \cite{ref:on the preliminary investgation}. Meanwhile, they found the existence of Nash equilibria where many miners use selfish mining strategy and gain extra revenue simultaneously. Zhang et al. simulated the situation when there were multiple selfish mining attackers in the system \cite{ref:n attackers1}. It shows there are scenarios where it is enough to have 12\% mining power to benefit from selfish mining but also that having more than seven selfish miners which benefit simultaneously is highly unlikely. Azimy et al. designed a Bitcoin network simulator and used it to simulate different configurations of miners \cite{{ref:competive selfish mining}}. Their finding shows that in almost all of the configurations, with the presence of a more powerful selfish miner, selfish mining actually decreases the revenue of the weaker selfish miners and also helps the stronger selfish miner.

\section{Conclusion}

In this paper, we first study how the existence of multiple misbehaving miners influences the profitability of basic selfish mining. By establishing the Markov chain model to describe the action of attackers and honest miners, we can obtain the minimum profitable threshold is symmetric 21.48\%, which decreases as the number of symmetric attackers increases. If there are two asymmetric selfish mining attackers in the system, the profitable threshold of one attacker decreases first and then increases with the increase of the other attacker's Hash power. We validate this in both models and experiments. We next investigate the revenues of the attackers when one executes the basic selfish mining and the other implements the strategic mining. A new mining strategy is designed for the miners with incomplete information based on \textit{POMDP}. We can obtain revenue by the new strategy no less than honest mining and basic selfish mining. Considering the difficulty adjustment, we model the transient process and acquire the closed-form solution of the profitable time. It can be discovered that the profitable time is large when the attacker's Hash power is low. Moreover, there is a negative correlation between the profitable time and the attackers' mining power.

	\vspace{-1cm}
\begin{IEEEbiography}[{\includegraphics[width=1in,height=1.25in,clip,keepaspectratio]{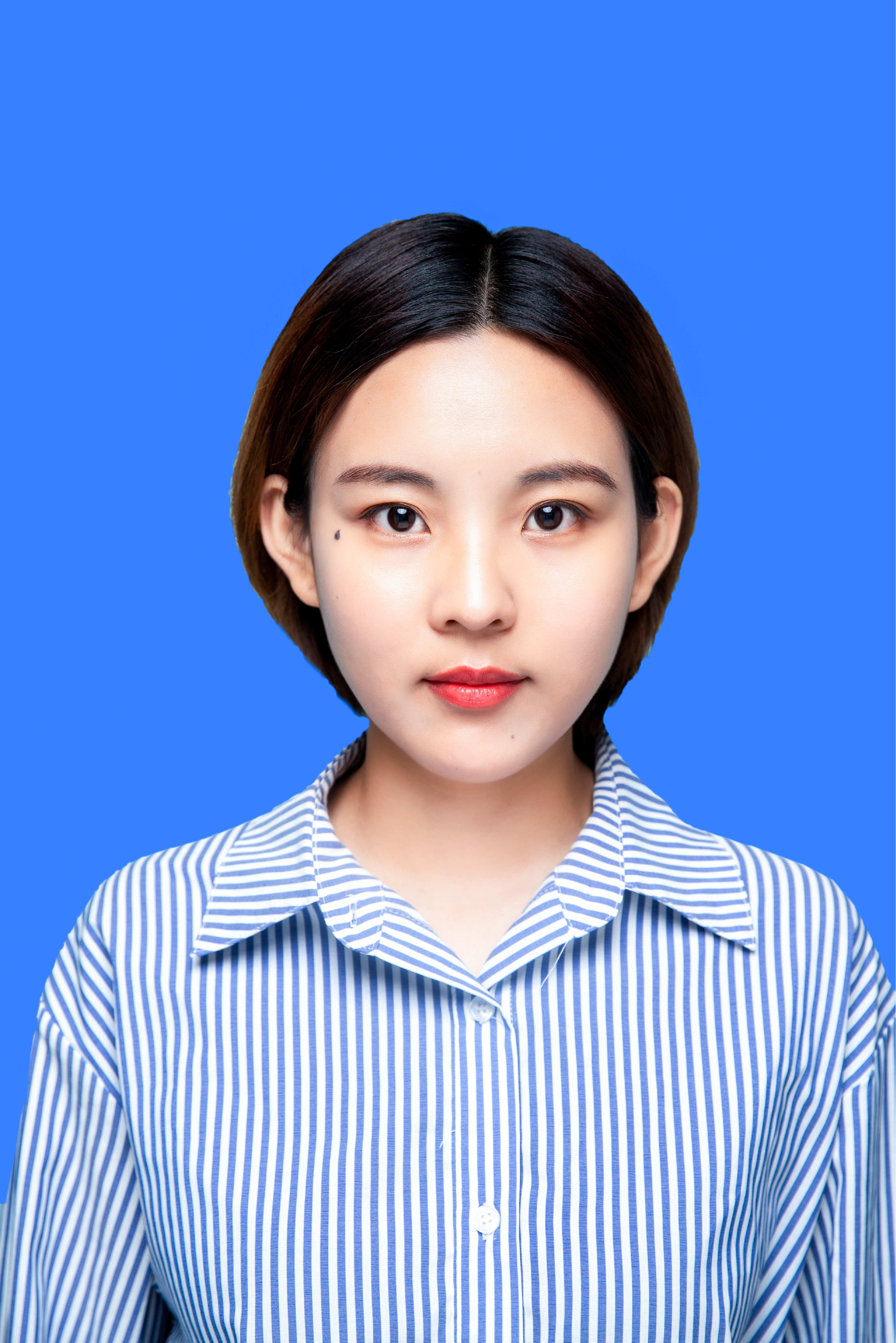}}]{Qianlan Bai}
		is a Ph.D. candidate in
		computer science at Shanghai Key Laboratory of Intelligent Information Processing, School of Computer Science, Fudan University, Shanghai, 200433, China. Her research interests include economic analysis and security analysis about blockchain.
	\end{IEEEbiography}

	\vspace{-1cm}

	\begin{IEEEbiography}[{\includegraphics[width=1in,height=1.25in,clip,keepaspectratio]{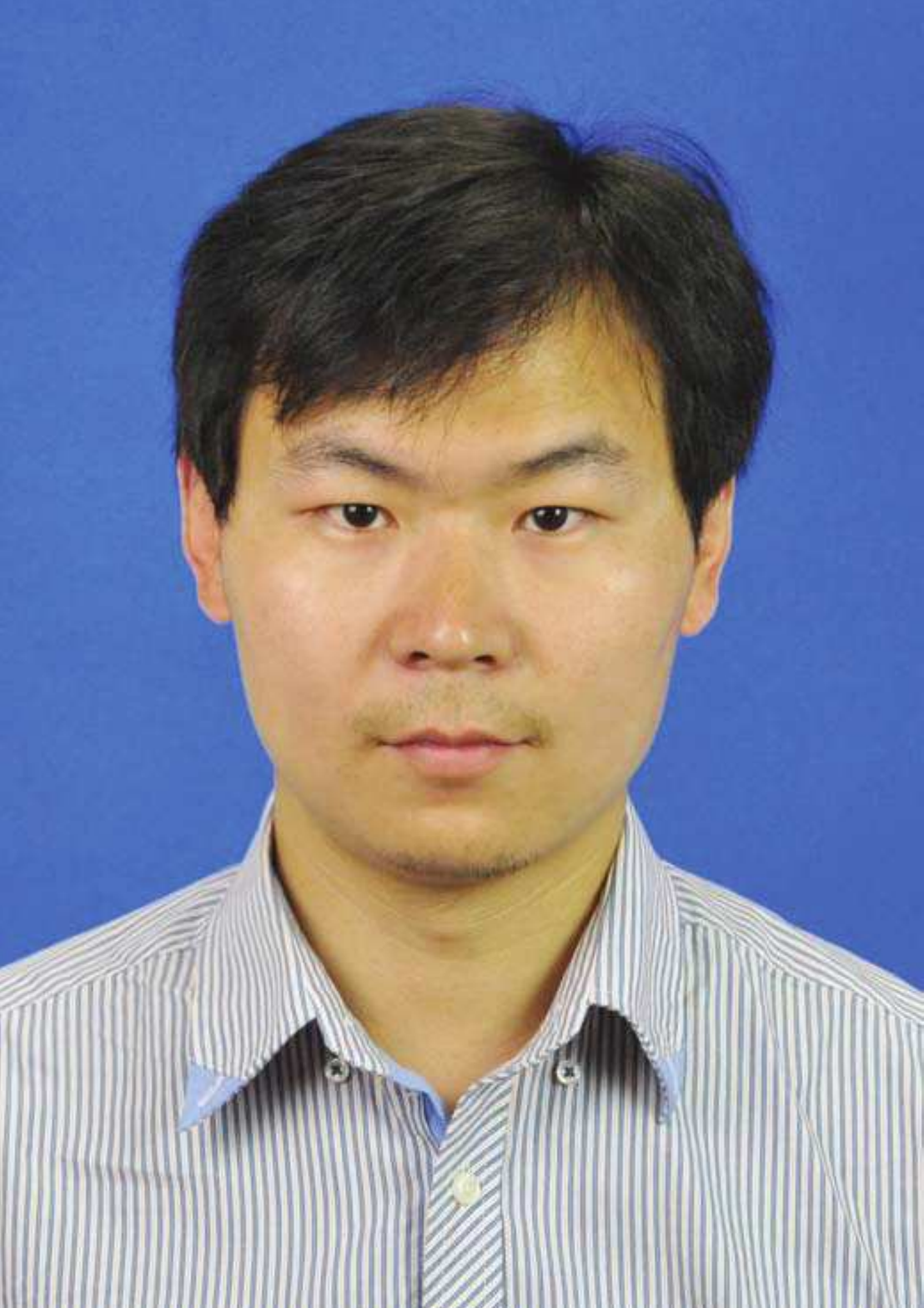}}]{Yuedong Xu}
		is a tenured professor in the School of 
		Information Science and Technology, Fudan University, 
		China. He received PhD from The Chinese University
		of Hong Kong, M.S. degree from
		Huazhong University of Science and Technology and 
		B.S. degree from Anhui University. 
		From late 2009 to 2012, he was a postdoc with INRIA Sophia
		Antipolis and Universite d'Avignon, France.
		His areas of interest include performance evaluation, 
		optimization, security, data analytics and economic analysis of communication networks and mobile computing.
		
	\end{IEEEbiography}
	\vspace{-1cm}
	\begin{IEEEbiography}[{\includegraphics[width=1in,height=1.25in,clip,keepaspectratio]{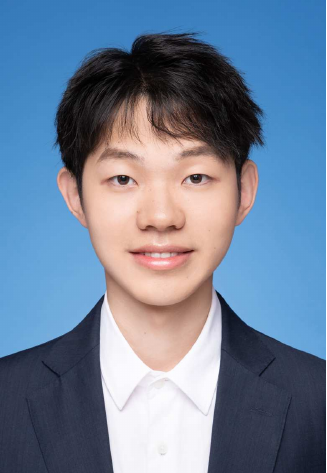}}]{Nianyi Liu}
	is an undergraduate student majored in Intelligence Science and Technology in the School of Information Science and Technology, Fudan University, China.
		
	\end{IEEEbiography}

\vspace{-1cm}
	\begin{IEEEbiography}[{\includegraphics[width=1in,height=1.25in,clip,keepaspectratio]{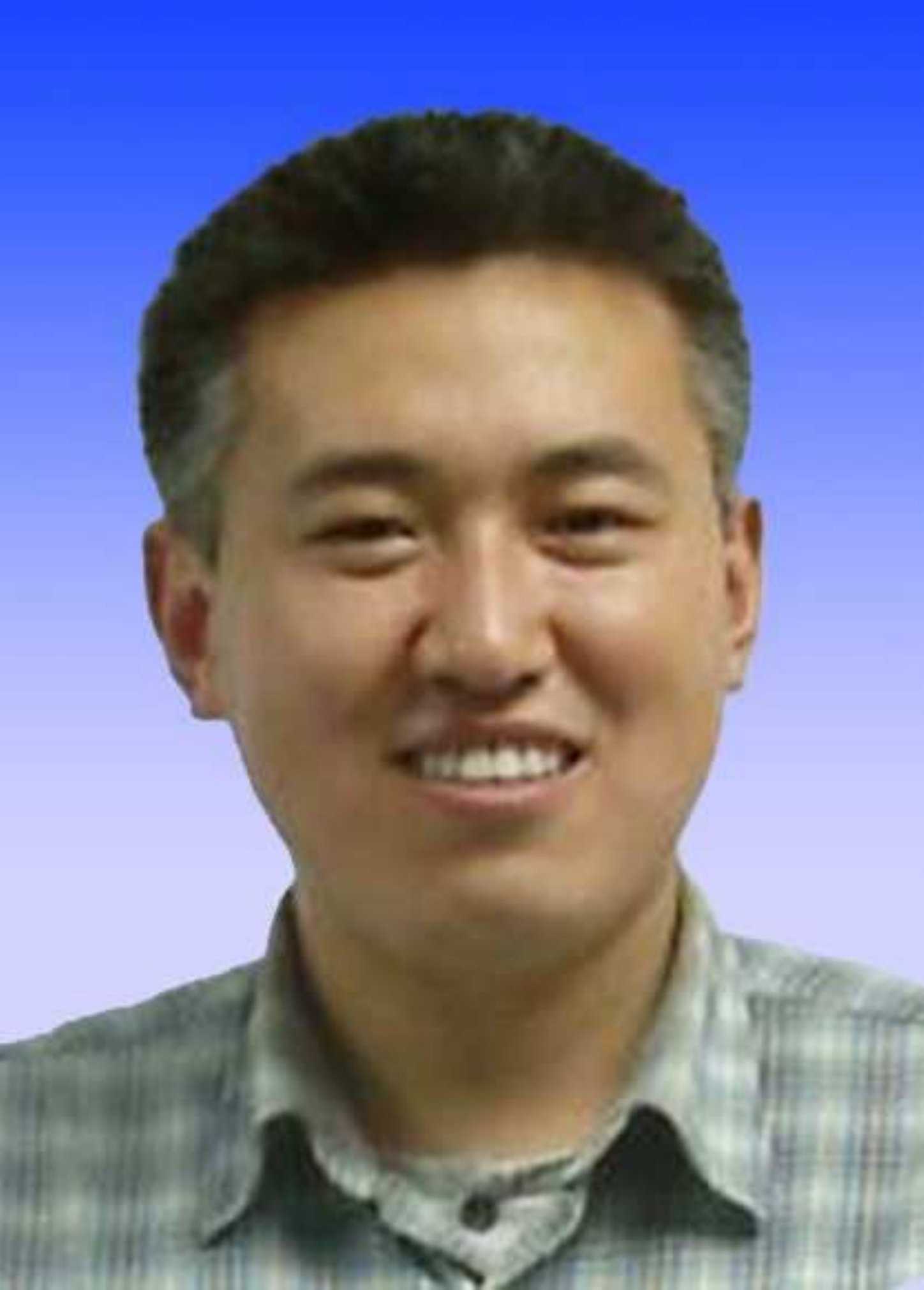}}]{Xin Wang}
		born in 1973, received his B.S. degree in Infor-mation Theory and MS Degree in Communication and Elec-tronic Systems from Xidian University, China in 1994 and 1997, respectively. He received his Ph.D. degree in Computer Science from Shizuoka University, Japan, in 2002. He is cur-rently a professor in Fudan University, Shanghai, China. His research interests include quality of network service, next-generation network architecture, mobile Internet and network coding. He is a Distinguished Member of CCF and a member of IEEE.
		
	\end{IEEEbiography}
\clearpage
\section*{Appendix}

\subsection{Tie-breaking with three public chains}
\label{appendix:three-ties}

\begin{figure}[!ht]
		\vspace{-0.2in}
		\centering
		\includegraphics[width=0.3\textwidth]{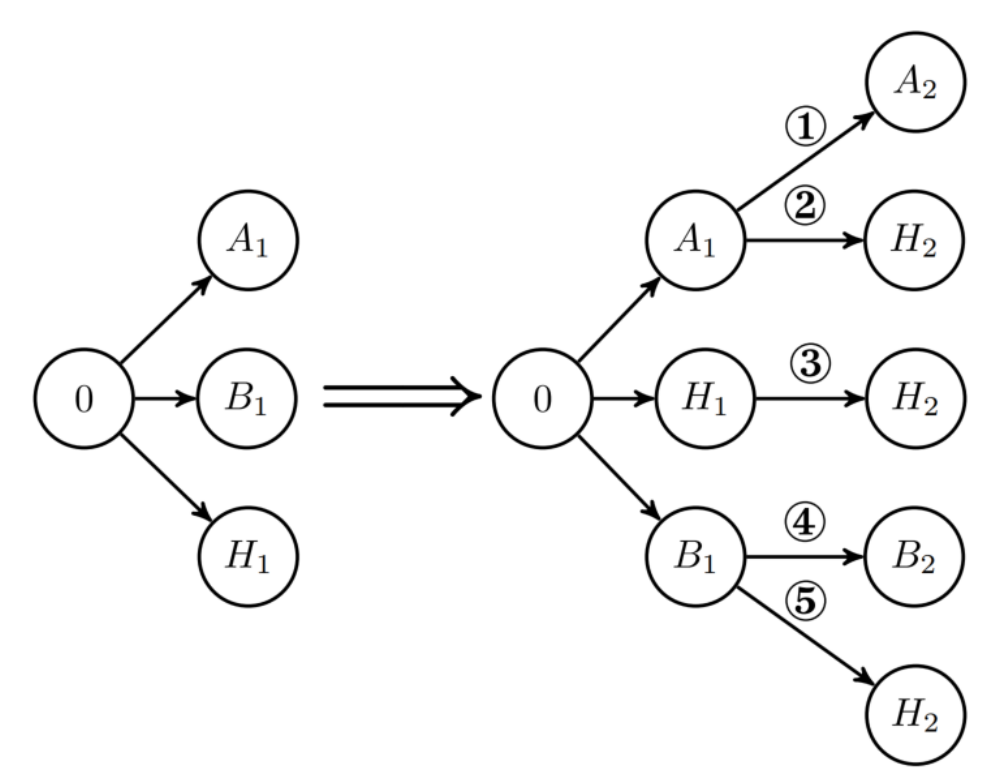}
		\caption{Tie-breaking case with three public chains.}
		\label{fig:tie_three}
	\end{figure}

For the situation that each of Alice and Bob hides one private block, they will publish their private chains instantly after Henry finds a new block. As shown in Fig. \ref{fig:tie_three}, there exist three competing public chains. Alice will mine after $A_1$ and Bob will mine after $B_1$ for sure; Henry is not aware of which chain is maliciously forked so that he may mine on each public chain. There are also five possible situations. The risk-avoiding release, together with two tie-breaking solutions, constitutes all the dynamics of private and public chains. 

\subsection{Scaling to $N > 2$}
\label{appendix:N4}

\begin{figure*}[htb]
		\begin{minipage}[t]{0.6\linewidth}
\setlength\abovecaptionskip{3pt}
\setlength\belowcaptionskip{-1pt}
\centering
		\includegraphics[width=1\textwidth]{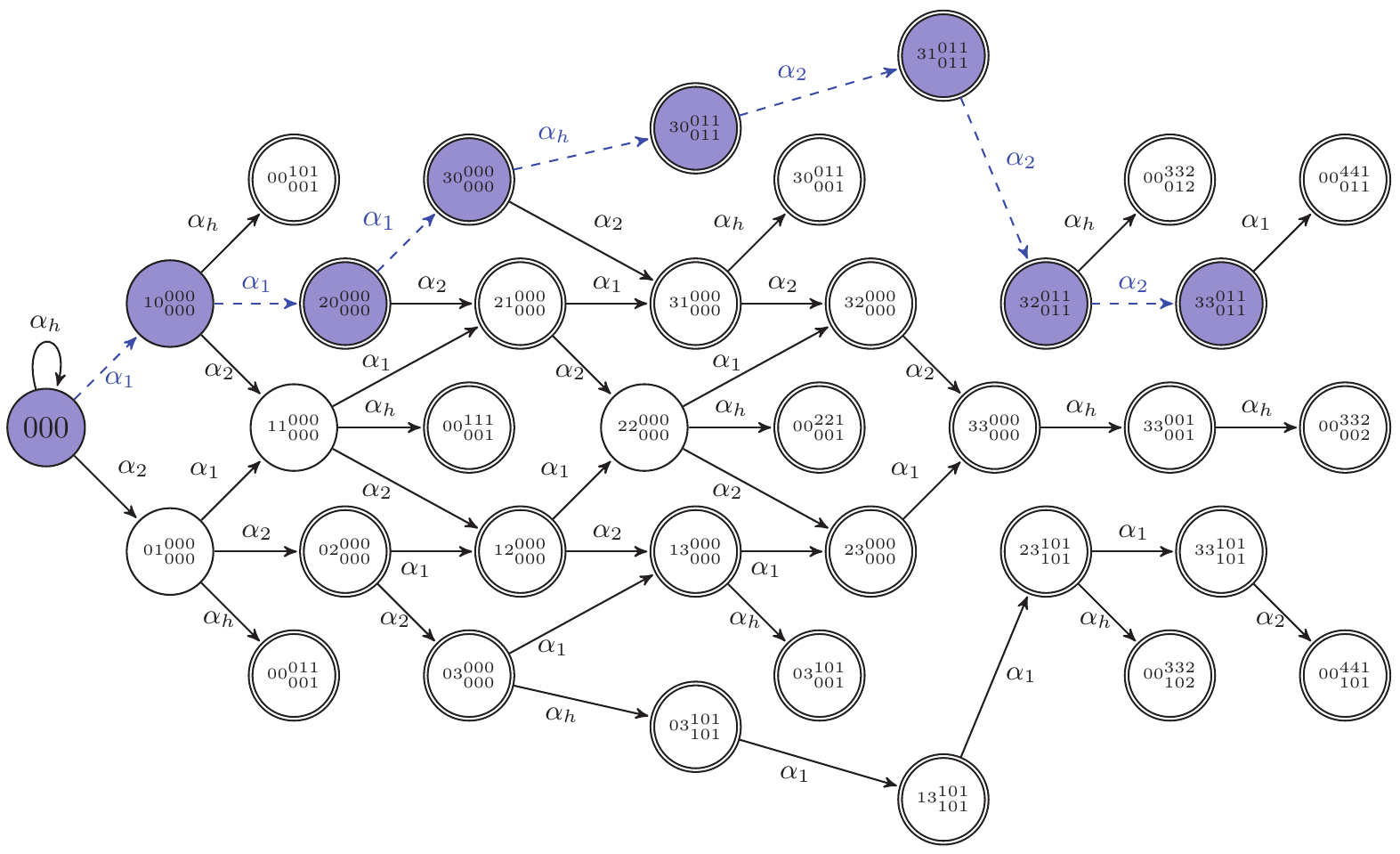}
		\caption{State machine with $N=4$.}
		\label{fig:state machine 4}
\end{minipage}
		\begin{minipage}[t]{0.34\linewidth}
\setlength\abovecaptionskip{3pt}
\setlength\belowcaptionskip{-1pt}
\centering
		\includegraphics[width=1\textwidth]{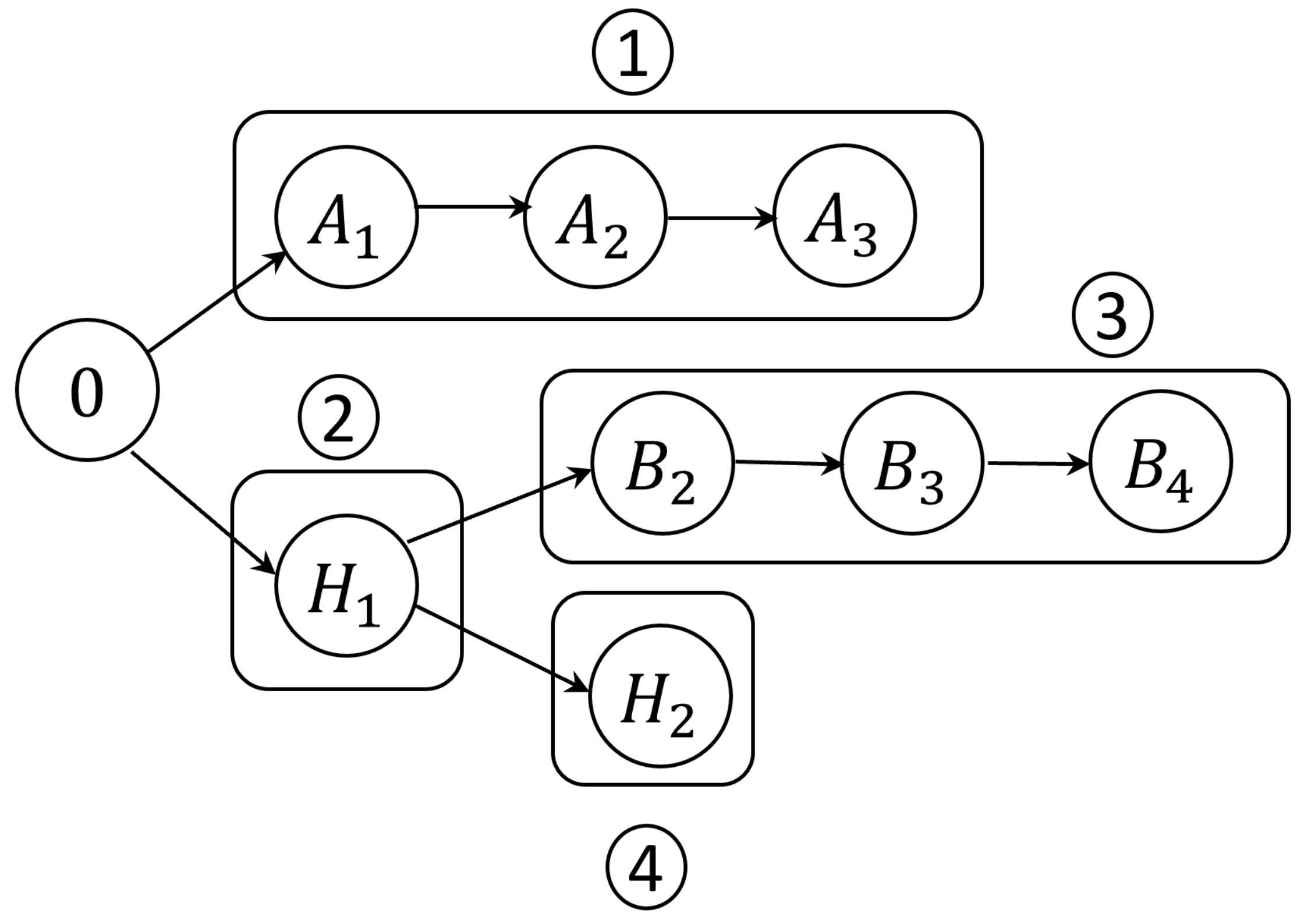}
		\caption{A path sample with $N=4$.}
		\label{fig:state machine 4 example}
\end{minipage}
	\end{figure*}

We next model the revenues of all miners when each private chain can hide more than one block. Especially, as the maximum number of private blocks for one attacker is no less than four (i.e. $N$=4), the ``chain reaction" occurs and the resulting finite state machine becomes very much complicated. A state should include not only the lengths of all chains, but also the interleaving of blocks on them. We confine our study to three miners: Alice, Bob and Henry, and investigate the case $N=4$ without loss of generality. The main difficulty hindering the mathematical analysis is that Alice and Bob have different beliefs in the starting position of the current mining round. Besides, the blocks in the winning chain may belong to Henry and Alice/Bob so that we need to memorize them in order to compute their revenues. In contrast, both Alice and Bob always have the common starting position in the racing without chain reaction (i.e. state $(0,0,0)$ in Fig.\ref{fig:state machine 4}).

The state transitions with $N=4$ are expressed in Fig. \ref{fig:state machine 4} where each state consists of eight parameters. The current mining round starts at the leftmost node where no miner has discovered a block. The notation $h_1$ indicates the distance between the starting position believed by Alice and the real starting position. Similarly, $h_2$ and $h_3$ indicate these distances of Bob and Henry. We record $h_1$, $h_2$ and $h_3$ because the blocks between the real and the believed starting positions influence which chain will prevail finally and how the revenues are calculated. The notation $l_1$ (resp. $l_2$) represents the number of unreleased blocks at Alice's (resp. Bob's) private chain. $\mu_1$ denotes the number of Henry's blocks between the real starting position and the Alice-believed starting position.
$\mu_2$ and $\mu_3$ are defined for Bob and Henry in the same way. Combined them together, we define a Markov state as ${l_1l_2}^{h_1h_2h_3}_{\mu_1\mu_2\mu_3}$ that is also applicable to the situation with $N>4$. 

We hereby present a concrete example of state transition. The related states in Fig. \ref{fig:state machine 4} are marked in blue, and their transitions are illustrated in Fig. \ref{fig:state machine 4 example} separately. At stage \circled{1}, Alice has mined three blocks stealthily so that the system state jumps from $0$ to $30_{000}^{000}$ through $10^{000}_{000}$ and $20_{000}^{000}$. At stage \circled{2}, Henry mines a valid block $H_1$ and publishes it to the public chain immediately. The system state then jumps from $30_{000}^{000}$ to $30_{011}^{011}$. Bob has mined three blocks after $H_1$ at stage \circled{3} and the system state moves to $33_{011}^{011}$. So far, neither Alice nor Bob will release their private blocks. At stage \circled{4}, Henry discovers a new block $H_2$ that triggers the release action of Alice. After Alice publishes all her blocks to obsolete Henry's chain, Bob finds that the public chain is catching up. As a consequence, Bob publishes all his blocks and wins the competition finally, i.e. the system state returning to the starting position. In this round, Bob receives three block rewards and Henry receives one block reward. 

According to the transitive probability, the revenue for each miner can be represented as:
		\begin{align}
	&{\pi_{000}}=1/(1+\alpha_1+\alpha_2+\alpha_1\alpha_3+\alpha_1^2+2\alpha_2\alpha_1+\alpha_2^2+\alpha_1^3\nonumber\\&+\alpha_2\alpha_3
	+3\alpha_2\alpha_1^2+2\alpha_1\alpha_2\alpha_3+3\alpha_1\alpha_2^2+\alpha_2^3+\alpha_1^3\alpha_3\nonumber\\
	&+4\alpha_1^3\alpha_2+6\alpha_1^2\alpha_2^2+4\alpha_1\alpha_2^3+\alpha_2^3\alpha_3+5\alpha_1^3\alpha_2\alpha_3+10\alpha_1^3\alpha_2^2\nonumber\\
	&+6\alpha_1^2\alpha_2^2\alpha_3+10\alpha_1^2\alpha_2^3+5\alpha_1\alpha_2^3\alpha_3+\alpha_1^3\alpha_2^2\alpha_3\nonumber+20\alpha_1^3\alpha_2^3\\
	&+\alpha_1^3\alpha_2^2\alpha_3^2+22\alpha_1^3\alpha_2^3\alpha_3\nonumber+\alpha_1^4\alpha_2^3\alpha_3+20\alpha_1^3\alpha_2^3\alpha_3^2\\
	&
	+\alpha_1^2\alpha_2^3\alpha_3+\alpha_1^2\alpha_2^3\alpha_3^2+\alpha_1^3\alpha_2^4\alpha_3);
	\end{align}
	\begin{align}
	{R_a}=&{\pi_{000}}\cdot[4\alpha_1^4\left(1+\alpha_h\right)+3\alpha_1^3\alpha_h^2+16\alpha_1^4\alpha_2+4\alpha_1^2\alpha_h\nonumber\\
	&+40\alpha_1^4\alpha_2^2\left(1+2\alpha_2\right)+\alpha_1\alpha_2\alpha_h\left(1+\gamma_1+2\theta_{1}\alpha_h\right)\nonumber\\
	&+10\alpha_1^2\alpha_2\alpha_h+20\alpha_1^3\alpha_2\alpha_h\left(3\alpha_2+\alpha_1\right)+15\alpha_1^3\alpha_2\alpha_h^2\nonumber\\
	&+4\alpha_1^4\alpha_2^2\alpha_h\left(1+\alpha_h\right)+4\alpha_1^4\alpha_2^3\alpha_h^2\left(\beta_1+20\right)\nonumber\\
	&+5\alpha_1^5\alpha_2^3\alpha_h+4\alpha_1^4\alpha_2^3\alpha_h\left(\alpha_2+21\right)+3\alpha_1^3\alpha_2^4\alpha_h^2\beta_1\nonumber\\
	&+\alpha_1\alpha_h^2\gamma_1+12\alpha_1^2\alpha_2^2\alpha_h^2\beta_1+\alpha_1^2\alpha_2^2\alpha_h^3\beta_1\left(3\alpha_1+2\alpha_2\right)\nonumber\\
	&+6\alpha_1^3\alpha_2^3\alpha_h^2\left(10\alpha_h\beta_1+1\right)];
	\end{align}
	\begin{align}
	{R_b}=&{\pi_{000}}[4\alpha_2^4\left(1+\alpha_h\right)+3\alpha_2^3\alpha_h^2+16\alpha_1\alpha_2^4+4\alpha_2^2\alpha_h\nonumber\\
	&+40\alpha_1^2\alpha_2^4\left(1+2\alpha_1\right)+\alpha_1\alpha_2\alpha_h\left(1+\gamma_2+2\theta_{2}\alpha_h\right)\nonumber\\
	&+10\alpha_1\alpha_2^2\alpha_h+20\alpha_1\alpha_2^3\alpha_h\left(3\alpha_1+\alpha_2\right)+15\alpha_1\alpha_2^3\alpha_h^2\nonumber\\
	&+4\alpha_1^2\alpha_2^4\alpha_h\left(1+\alpha_h\right)+4\alpha_1^3\alpha_2^4\alpha_h^2\left(\beta_2+20\right)\nonumber\\
	&+5\alpha_1^3\alpha_2^5\alpha_h+4\alpha_1^3\alpha_2^4\alpha_h\left(\alpha_1+21\right)+3\alpha_1^4\alpha_2^3\alpha_h^2\beta_2\nonumber\\
	&+\alpha_2\alpha_h^2\gamma_2+12\alpha_1^2\alpha_2^2\alpha_h^2\beta_2+\alpha_1^2\alpha_2^2\alpha_h^3\beta_2\left(2\alpha_1+3\alpha_2\right)\nonumber\\
	&+6\alpha_1^3\alpha_2^3\alpha_h^2\left(10\alpha_h\beta_2+1\right)];
	\end{align}
	\begin{align}
	{R_h}=&{\pi_{000}}[\alpha_1\alpha_h^2\left(2-\gamma_1 \right)+\alpha_2\alpha_h^2\left(2-\gamma_2 \right)+\alpha_1^2\alpha_2^3\alpha_h^3\left(2\beta_1\!\!+\!\!\beta_2 \right)\nonumber\\
	&+2\alpha_1\alpha_2\alpha_h^2\left(2-\theta_{1}-\theta_{2} \right)+\alpha_1^2\alpha_2^2\alpha_h^2\left(6+4\alpha_1\alpha_2\right)\nonumber\\
	&+\alpha_1^3\alpha_2^2\alpha_h^3\left(\beta_1+2\beta_2 \right)+\alpha_1\alpha_2\alpha_h\left(2-\gamma_1-\gamma_2 \right)\nonumber\\
	&+\alpha_1^3\alpha_2^3\alpha_h\left(\alpha_1+\alpha_2\right)+\alpha_1^3\alpha_2^4\alpha_h^2\left(2\beta_1+\beta_2 \right)+\alpha_h\nonumber\\
	&+\alpha_1^4\alpha_2^3\alpha_h^2\left(\beta_1\!\!+\!\!2\beta_2 \right)\!\!+\!\!20\alpha_1^3\alpha_2^3\alpha_h^3\!\!+\!\!2\alpha_1^4\alpha_2^4\alpha_h];\\
	\vspace{0.2cm}
	&\beta_1=\gamma_1/(\gamma_1+\gamma_2)\quad\beta_2=\gamma_2/(\gamma_1+\gamma_2).
	\end{align}
\begin{align}
    \hat{R}_i=\frac{R_i}{R_a+R_b+R_h}, \forall i\in\{a,b,h\}.
\end{align}
The cases with $N>4$ can be analyzed in the same way. We validate in our experiments that the relative revenue tends to converge when $N\geq 4$. If $N$ is too large, the repeated chain-reactions will occur, which aggravates the system instability and increases the possibility of detecting selfish mining attacks. 

	\begin{figure*}
\begin{minipage}[t]{0.33\linewidth}
			\setlength\abovecaptionskip{3pt}
			\setlength\belowcaptionskip{-1pt}
			\centering
			\includegraphics[width=0.9\textwidth]{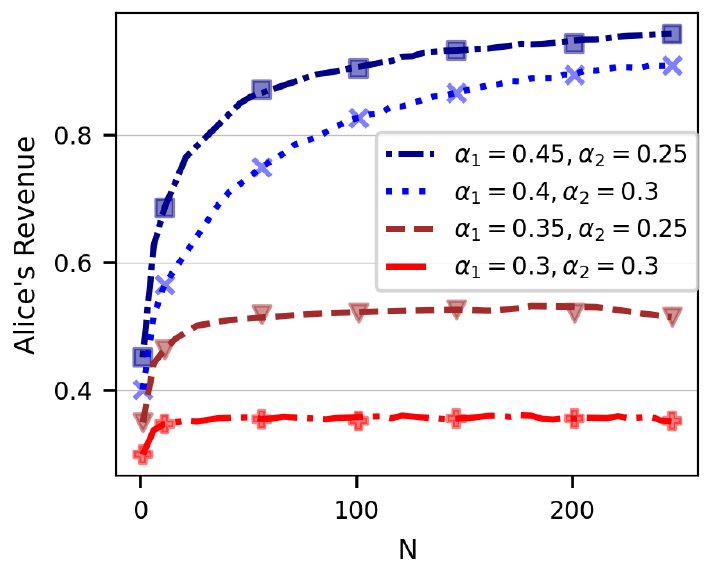}\\
			\caption{Alice's Revenue with $\alpha_1>\max(\alpha_2,\alpha_h)$.}
			\label{fig2}
		\end{minipage}
\begin{minipage}[t]{0.33\linewidth}
			\setlength\abovecaptionskip{3pt}
			\setlength\belowcaptionskip{-1pt}
			\centering
			\includegraphics[width=0.9\textwidth]{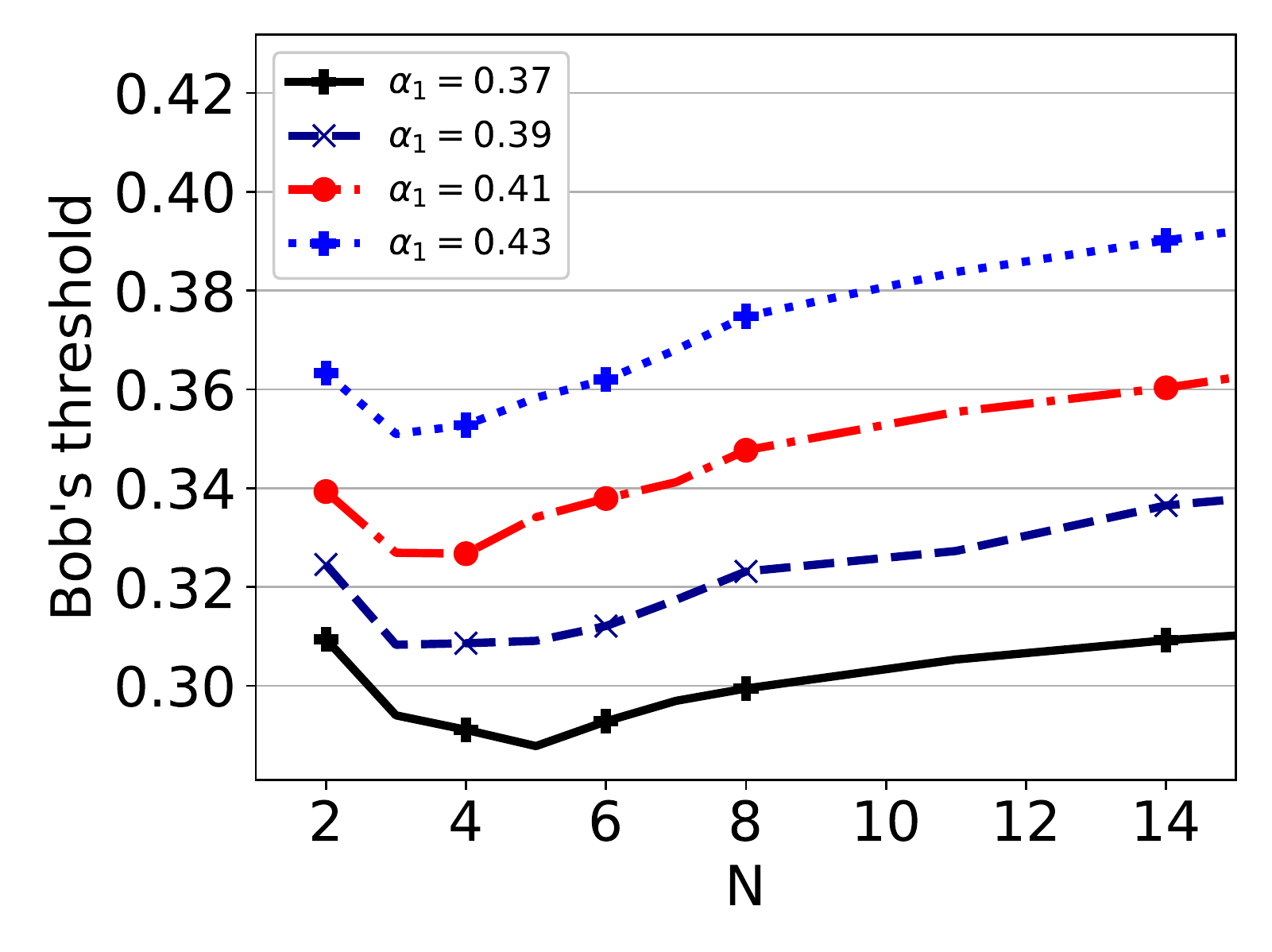}\\
			\caption{Simulated threshold for Bob when $\alpha_1>\max(\alpha_2,\alpha_h)$.}
			\label{fig:threshold decrease increase N}
\end{minipage}
\begin{minipage}[t]{0.33\linewidth}
			\setlength\abovecaptionskip{0.5pt}
			\setlength\belowcaptionskip{-1pt}
			\centering
			\includegraphics[width=0.9\textwidth]{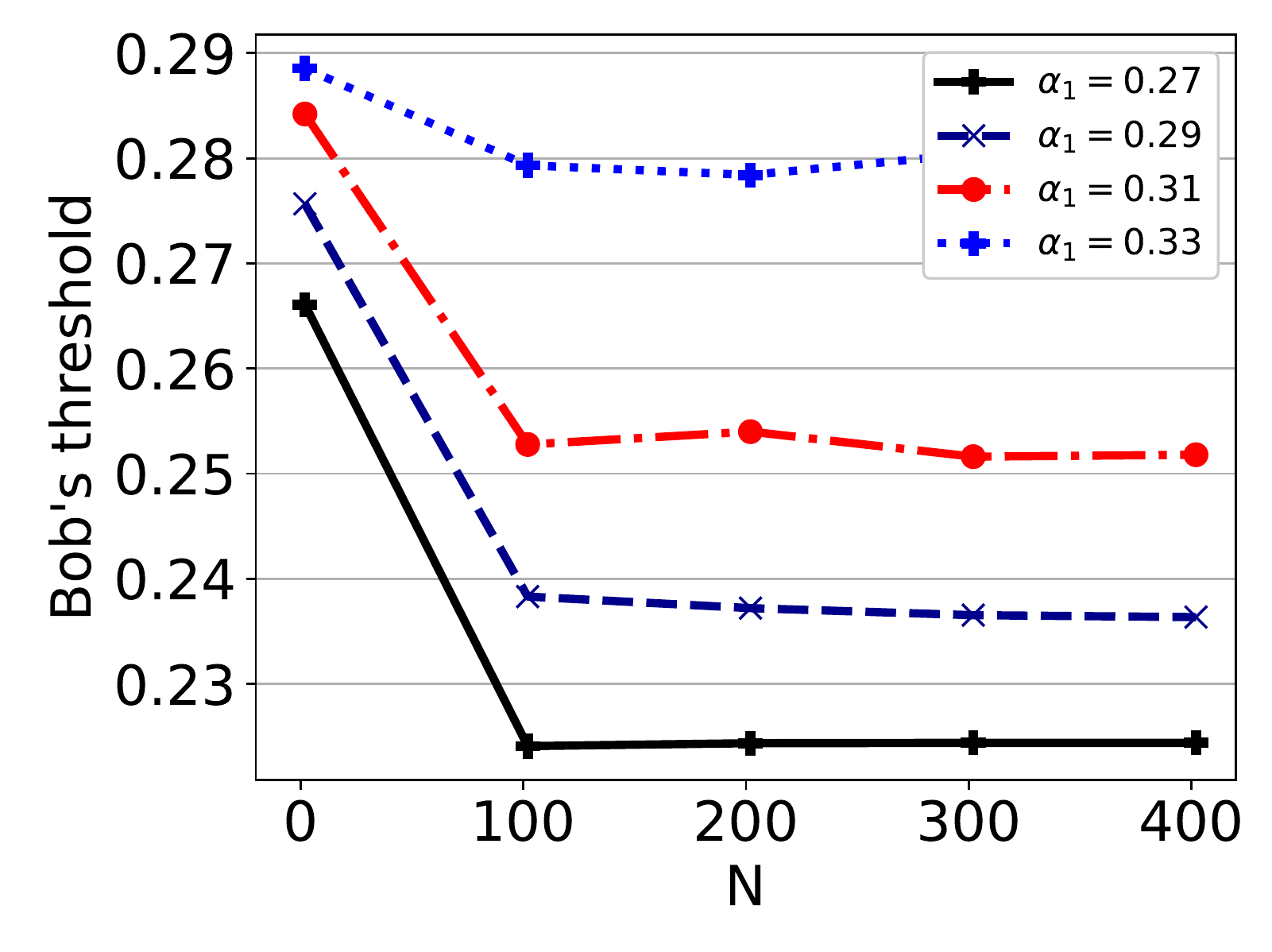}\\
			\caption{Simulated threshold for Bob when $\alpha_h>max(\alpha_1,\alpha_2)$.}
			\label{fig:threshold decrease converge N}
\end{minipage}		
\end{figure*}

\subsection{Profitable threshold under asymmetric attackers}
\label{appendix:observation3}

\begin{observation}{\textit{
If $\alpha_h<\max(\alpha_1,\alpha_2)$, the revenue of the attacker with more Hash power among three miners will increase as $N$ increases. If $\alpha_h>\max(\alpha_1,\alpha_2)$, the revenue of the attacker with more Hash power will increase first and then remain stable as $N$ increases. The common benefit Hash power region of two attackers increases at first and then decreases with the increase of $N$.
}}
\end{observation}

We explore the revenue of the attackers with more Hash power than the honest miner under different $N$ when $\alpha_h<\max(\alpha_1,\alpha_2)$.	
 In Fig. \ref{fig2},  we show Alice's revenue of Alice at four situations: the Hash powers of Alice, Bob and Henry are (0.45, 0.25, 0.3), (0.4, 0.3, 0.3), (0.35, 0.25, 0.4) and (0.3, 0.3, 0.4) that are labeled as situations 1, 2, 3 and 4. Situations 1 and 2 manifest that the revenue of the attacker with more Hash power than others increases as $N$ increases. The attacker can obtain more than 90\% of the revenue, achieving the similar effect as the 51\% attack, even though she does not have 51\% of Hash power. Situation 3 and 4 show that Alice's revenue converges as $N$ increases if Henry has the largest Hash power. The revenue of Alice converges to 0.524 in situation 3 and 0.357 in situation 4 with $N=200$. This implies that limiting the attacker's Hash power will prevent the attacker from obtaining too many revenues when $N$ is large.
	
 We then investigate Bob's profitable threshold when Alice has a different Hash power and $N$ is large. Fig. \ref{fig:threshold decrease increase N} shows Bob's thresholds under different $N$ when Alice has the largest Hash power, i.e. $\alpha_1> \max\{\alpha_2, \alpha_h\}$. Bob's profitable threshold decreases first and then increases as $N$ increases. That means even if Bob can obtain extra revenue when $N$ is small, he can not obtain extra revenue when $N$ is large enough. Under this circumstance, Alice's revenue can obtain far more than her Hash power proportion, and her attack becomes meaningless because this system can not attract other miners even other selfish mining attackers. Fig. \ref{fig:threshold decrease converge N} shows Bob's profitable threshold will decrease first and then converge as $N$ increases when Henry has more Hash power than Alice and Bob. This suggests that limiting the attacker's Hash power also encourages other miners including other selfish mining attackers to continue mining even if $N$ is large. According to these analyses, the attacker will not hide too many blocks even if he has more Hash power than others.

 	\begin{figure*}[htb]
		\begin{minipage}[t]{1\linewidth}
			\setlength\abovecaptionskip{0.5pt}
			\setlength\belowcaptionskip{-1pt}
			\centering
			\includegraphics[width=1\textwidth]{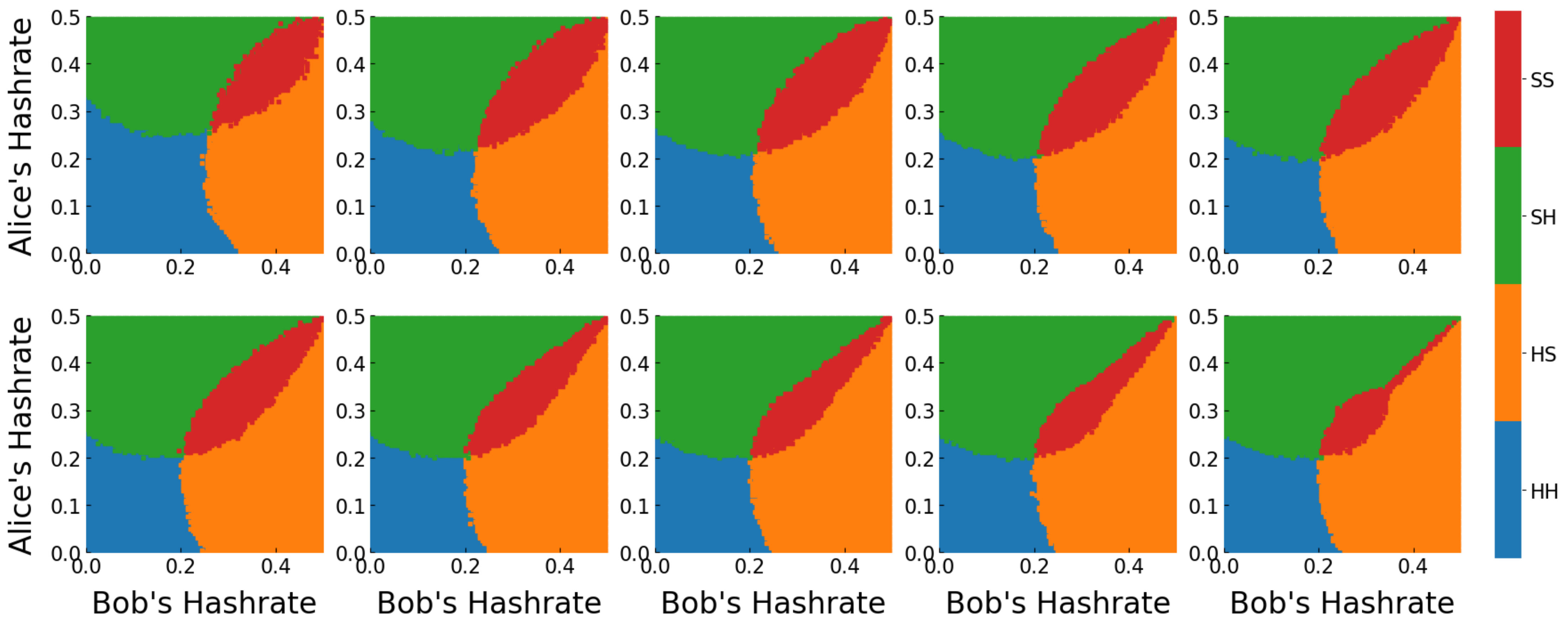}\\
			\caption{Profitable regions of both selish mining attackers with $N=2,3,4,5,7,8,15,25,35,\infty$.}
					\label{fig:common}
		\end{minipage}
		\vspace{-0.5cm}
\end{figure*}
  \begin{figure*}

		\begin{minipage}[h]{0.5\linewidth}
			\setlength\abovecaptionskip{0.5pt}
			\setlength\belowcaptionskip{-1pt}
			\centering
			\includegraphics[width=0.6\textwidth]{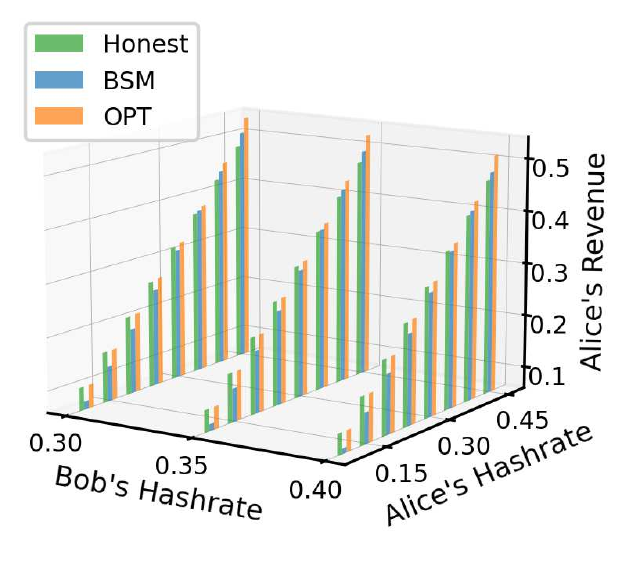}\\
			\caption{Optimal revenue for Alice when $N=2$.}
			\label{fig:OPTN2}
		\end{minipage}
		\hspace{0.2cm}
		\begin{minipage}[h]{0.5\linewidth}
			\setlength\abovecaptionskip{0.5pt}
			\setlength\belowcaptionskip{-1pt}
			\centering
			\includegraphics[width=0.6\textwidth]{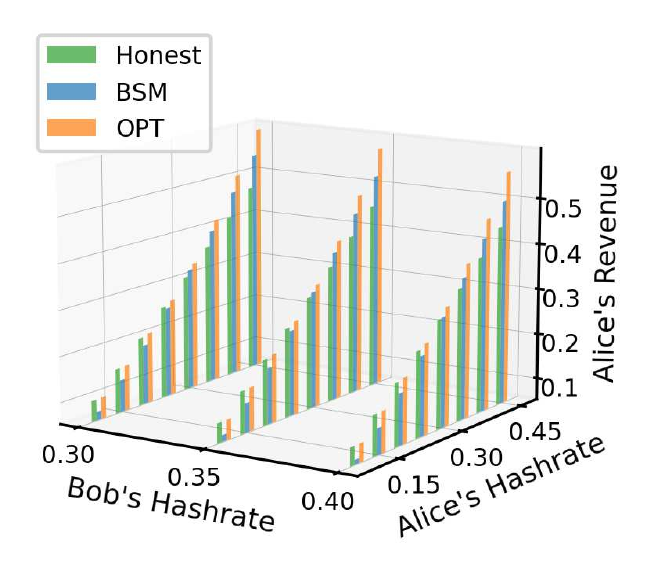}\\
			\caption{Optimal revenue for Alice when $N=3$.}
			\label{fig:OPTN3}
		\end{minipage}
		\end{figure*}
We now analyze the interplay between Alice's and Bob's  profitable thresholds. Fig. \ref{fig:common} shows the profitable regions for each attacker under different $N$. The blue part indicates that neither attacker can gain additional revenue if they perform the \emph{BSM} attack, and the red part indicates that both attackers can gain additional revenue through selfish mining. The green (resp. orange) part represents the situation that only Alice (resp. Bob) can obtain extra revenue. The intersection of four regions is actually the profitable threshold with symmetric Hash powers of Alice and Bob that decreases over $N$ and converges gradually. The common profitable region (in red) first expands and then shrinks as $N$ increases. The reason is the following. A large red region basically says that both Alice and Bob are profitable with \emph{BSM} even if their Hash powers are asymmetric to some extent. When $N$ is very small, Alice and Bob can hide only a couple of blocks so that their ability of wasting Henry's Hash power is restrained. With the increases of $N$, their selfish mining ability becomes more powerful, and thus could have more chances of obsoleting the public chain even if each of them has a smaller Hash power than Henry. Meanwhile, given the restriction of $N$, both Alice and Bob may receive a certain amount of extra revenues, resulting in a larger common profitable region. When $N$ is large, the stronger attacker is inclined to dominating the selfish mining. If Alice's Hash power is larger than Bob's, Bob will find it difficult to compete with Alice so that Bob's profitable threshold is getting higher. If Alice's Hash power is the largest, it is similar to 51\% attack. Bob's selfish mining is profitable only when Alice's and Bob's Hash powers are sufficiently close, causing the common profitable region to shrink to a line segment.

\subsection{MDP-based mining experiment}
\label{appendix:MDP}



We will compute Alice's optimal policy and the corresponding revenue of \emph{MDP}-based (\emph{OPT}) mining. Let the error parameter $\epsilon=0.00001$ and the execution number $\xi_0=500000$. Fig. \ref{fig:OPTN2} illustrates the optimal revenue obtained by Alice when Alice's Hash power is $\alpha_1$ $\in$ $\{0.10$, $0.15$, $0.20$, $0.25$, $0.30$, $0.35$, $0.40$, $0.45\}$, Bob's Hash power is $\alpha_2$ $\in$ $\{0.30$, $0.35$, $0.40\}$ and $N=2$. The maximum length of the longest public chain is set as $(N+1)$. One can observe that Alice's optimal mining and honest mining yield the same revenue when her mining power is relatively small, e.g. $\alpha_1 = 0.1$, $\alpha_2 = 0.40$ and $\alpha_h = 0.50$. Under these situations, the optimal mining policy is exactly the honest mining, while \emph{BSM} underperforms the honest mining significantly. On the contrary, when Alice's Hash power is large, e.g., $\alpha_1 = 0.40$, $\alpha_2 = 0.40$ and $\alpha_h = 0.20$, the optimal mining policy results in a higher revenue than the basic selfish mining, and the basic selfish mining is better than the honest mining. Fig. \ref{fig:OPTN3} shows a set of similar experiments except for $N=3$. When $\alpha_1 = 0.30$, $\alpha_2 = 0.30$ and $\alpha_h = 0.4$, Alice's optimal mining policy obviously outperforms both \emph{BSM} and the honest mining.

We describe a concrete example of optimal mining policy at a few representative states for simplicity. 
Table \ref{table:policy1} shows Alice's optimal strategy with $\alpha_1$$=$$0.2$, $\alpha_2$$=$$0.4$ and $\alpha_h$$=$$0.4$. Alice has less Hash power than Bob and Henry. She chooses to adopt the public chain if the length of her chain is shorter than that of Bob or Henry. If the length of the chain (i.e. $l_1+h_1$) she is mining on is equal to that of the public chain (i.e. $h_3$), and is able to ``fork'', she will release $l_1$ private blocks to seize the chance of winning. If both Alice and Bob hold a private block, she will choose to hide the private block and continue to mine after her own block. By performing the optimal policy, Alice can earn an additional 0.08\% revenue, even though Bob has 40\% Hash power.

\begin{table}[ht]
		\vspace{-0.4cm}
		\centering  
		\caption{Optimal policy for ($\alpha_1=0.2,\alpha_2=0.4,N=2$).}
		\begin{tabular}{|c|c|c|c|}
			\hline  
			    state & action \\
    \hline
    $ (l_2+h_2)>(l_1+h_1)$& $adopt$ \\
    \hline
        $ h_3>(l_1+h_1)$& $adopt$ \\
    \hline
    $(l_1+h_1)=(l_2+h_2)=h3, fork=r$ & $l_1$\\
			\hline 
	 $l_1=l_2=1,h_3=0$ & $0$\\
	 \hline
	 $l_1=1,l_2=0$&$l_1$\\
	 \hline
	  $(l_1+h_1)>(l_2+h_2)$ & $l_1$\\
\hline		
		\end{tabular}
		\label{table:policy1}
		\vspace{-0.3cm}
\end{table}

\newpage
\onecolumn
      	\begin{table*}[h]
		\centering  
		\caption{A description of the transition and reward matrices $\mathcal{P}$ and $\mathcal{R}$ in the decision problem $M$.}
		\resizebox{1\textwidth}{!}{
		\begin{tabular}{|c|c|c|c|c|c|c|}
			\hline  
\multirow{24}*{$l_2+h_2-h_3>2$}&\multirow{6}*{$adopt$}&\multirow{3}*{$loc!=2$}&{$\alpha_1$}&{$(1,ir,1,l_2,h_3,h_2,h_3,\mu_3,\mu_2,\mu_3)$}&\multirow{3}*{$(0,0,0)$}\\
~&~&~&{$\alpha_2$}&{$(1,ir,0,l_2+1,h_3,h_2,h_3,\mu_3,\mu_2,\mu_3)$}&~\\
~&~&~&{$\alpha_h$}&{$(1,r,0,l_2,h_3,h_2,h_3+1,\mu_3,\mu_2,\mu_3+1)$}&~\\
			\cline{3-6}	
			~&~&\multirow{3}*{$loc=2$}&{$\alpha_1$}&{$(1,ir,1,l_2,h_3-h_2,0,h_3-h_2,h_3-h_2,0,h_3-h_2)$}&\multirow{3}*{$(0,h_2-\mu_2,\mu_2)$}\\
~&~&~&{$\alpha_2$}&{$(1,ir,0,l_2+1,h_3-h_2,0,h_3-h_2,h_3-h_2,0,h_3-h_2)$}&~\\
~&~&~&{$\alpha_h$}&{$(1,r,0,l_2,h_3-h_2,0,h_3-h_2+1,h_3-h_2,0,h_3-h_2+1)$}&~\\
\cline{2-6}

			~&\multirow{3}*{$h_1+action\leq h_3$}&\multirow{1}*{$ r,action \neq (h_3-h_1)$}&{$\alpha_1$}&{$(loc,ir,l_1-action+1,l_2,h_1+action,h_2,h_3,\mu_1,\mu_2,\mu_3)$}&\multirow{3}*{$(0,0,0)$}\\
~&~&\multirow{2}*{$ir$}&{$\alpha_2$}&{$(loc,ir,l_1-action,l_2+1,h_1+action,h_2,h_3,\mu_1,\mu_2,\mu_3)$}&~\\
~&~&~&{$\alpha_h$}&{$(loc,r,l_1-action,l_2,h_1+action,h_2,h_3+1,\mu_1,\mu_2,\mu_3+1)$}&~\\
			\cline{2-6}

			~&\multirow{4}*{$h_1+action=h_3$}&\multirow{2}*{$r, action>0$}&{$\alpha_1$}&{$(loc,f_{13},l_1-action+1,l_2,h_3,h_2,h_3,\mu_1,\mu_2,\mu_3)$}&\multirow{4}*{$(0,0,0)$}\\
~&~&~&{$\alpha_2$}&{$(loc,f_{13},l_1-action,l_2+1,h_3,h_2,h_3,\mu_1,\mu_2,\mu_3)$}&~\\
~&~&\multirow{2}*{$f_{13}, action=0$}&{$\alpha_h\gamma_1$}&{$(1,r,l_1-action,l_2,h_3,h_2,h_3+1,\mu_1,\mu_2,\mu_1+1)$}&~\\
~&~&~&{$\alpha_h(1-\gamma_1)$}&{$(loc,r,l_1-action,l_2,h_3,h_2,h_3+1,\mu_1,\mu_2,\mu_3+1)$}&~\\
			\cline{2-6}

			~&\multirow{3}*{$h_3<h_1+action<l_2+h_2-1$}&\multirow{3}*{$ $}&{$\alpha_1$}&{$(1,ir,l_1-action+1,l_2,h_1+action,h_2,h_1+action,\mu_1,\mu_2,\mu_1)$}&\multirow{3}*{$(0,0,0)$}\\
~&~&~&{$\alpha_2$}&{$(1,ir,l_1-action,l_2+1,h_1+action,h_2,h_1+action,\mu_1,\mu_2,\mu_1)$}&~\\
~&~&~&{$\alpha_h$}&{$(1,r,l_1-action,l_2,h_1+action,h_2,h_1+action+1,\mu_1,\mu_2,\mu_1+1)$}&~\\
			\cline{2-6}	
			
			~&\multirow{3}*{$h_1+action=l_2+h_2-1$}&\multirow{3}*{$ $}&{$\alpha_1$}&{$(2,r,l_1+1-action,0,h_1+action,h_2+l_2,h_2+l_2,\mu_1,\mu_2,\mu_2)$}&\multirow{3}*{$(0,0,0)$}\\
~&~&~&{$\alpha_2$}&{$(2,r,l_1-action,1,h_1+action,h_2+l_2,h_2+l_2,\mu_1,\mu_2,\mu_2)$}&~\\
~&~&~&{$\alpha_h$}&{$(2,r,l_1-action,0,h_1+action,h_2+l_2+1,h_2+l_2+1,\mu_1,\mu_2+1,\mu_2+1)$}&~\\
			\cline{2-6}	
			
			~&\multirow{3}*{$h_1+action=l_2+h_2$}&\multirow{3}*{$ $}&{$\alpha_1$}&{$(loc,f_{12},l_1-action+1,0,h_1+action,h_2+l_2,h_2+l_2,\mu_1,\mu_2,\mu_3)$}&\multirow{3}*{$(0,0,0)$}\\
~&~&~&{$\alpha_2$}&{$(2,r,l_1-action,0,h_1+action,h_2+l_2+1,h_2+l_2+1,\mu_1,\mu_2,\mu_2)$}&~\\
~&~&~&{$\alpha_h\beta_2$}&{$(2,r,l_1-action,0,h_1+action,h_2+l_2+1,h_2+l_2+1,\mu_1,\mu_2+1,\mu_2+1)$}&~\\
\cline{4-6}
~&~&~&{$\alpha_h\beta_1$}&{$(2,r,l_1-action,0,0,1,1,0,1,1)$}&$(h_1+action-\mu_1,0,\mu_1)$\\
			\cline{2-6}	
			
			~&\multirow{3}*{$h_1+action>l_2+h_2$}&\multirow{3}*{$ $}&{$\alpha_1$}&{$(3,ir,l_1-action+1,0,0,0,0,0,0,0)$}&\multirow{3}*{$(h_1+action-\mu_1,0,\mu_1)$}\\
~&~&~&{$\alpha_2$}&{$(3,ir,l_1-action,1,0,0,0,0,0,0)$}&~\\
~&~&~&{$\alpha_h$}&{$(3,r,l_1-action,0,0,1,1,0,1,1)$}&~\\
			\hline

			\multirow{24}*{$l_2+h_2-h_3=2$}&\multirow{6}*{$adopt$}&\multirow{3}*{$loc!=2$}&{$\alpha_1$}&{$(1,ir,1,l_2,h_3,h_2,h_3,\mu_3,\mu_2,\mu_3)$}&\multirow{3}*{$(0,0,0)$}\\
~&~&~&{$\alpha_2$}&{$(1,ir,0,l_2+1,h_3,h_2,h_3,\mu_3,\mu_2,\mu_3)$}&~\\
~&~&~&{$\alpha_h$}&{$(2,r,0,0,h_3,h_2+l_2,h_2+l_2,\mu_3,\mu_2,\mu_2)$}&~\\
			\cline{3-6}	
			~&~&\multirow{3}*{$loc=2$}&{$\alpha_1$}&{$(1,ir,1,l_2,h_3-h_2,0,h_3-h_2,h_3-h_2,0,h_3-h_2)$}&\multirow{3}*{$(0,h_2-\mu_2,\mu_2)$}\\
~&~&~&{$\alpha_2$}&{$(1,ir,0,l_2+1,h_3-h_2,0,h_3-h_2,h_3-h_2,0,h_3-h_2)$}&~\\
~&~&~&{$\alpha_h$}&{$(2,r,0,0,h_3-h_2,l_2,l_2,h_3-h_2,0,0)$}&~\\
\cline{2-6}

			~&\multirow{3}*{$h_1+action\leq h_3$}&\multirow{2}*{$ r,action \neq (h_3-h_1)$}&{$\alpha_1$}&{$(loc,ir,l_1-action+1,l_2,h_1+action,h_2,h_3,\mu_1,\mu_2,\mu_3)$}&\multirow{3}*{$(0,0,0)$}\\
~&~&~&{$\alpha_2$}&{$(loc,ir,l_1-action,l_2+1,h_1+action,h_2,h_3,\mu_1,\mu_2,\mu_3)$}&~\\
~&~&$ir$&{$\alpha_h$}&{$(2,r,l_1-action,0,h_1+action,h_2+l_2,h_2+l_2,\mu_1,\mu_2,\mu_2)$}&~\\
			\cline{2-6}

			~&\multirow{4}*{$h_1+action=h_3$}&\multirow{1.5}*{$r, action>0$}&{$\alpha_1$}&{$(loc,f_{13},l_1-action+1,l_2,h_3,h_2,h_3,\mu_1,\mu_2,\mu_3)$}&\multirow{4}*{$(0,0,0)$}\\
~&~&\multirow{1.5}*{$f_{13}, action=0$}&{$\alpha_2$}&{$(loc,f_{13},l_1-action,l_2+1,h_3,h_2,h_3,\mu_1,\mu_2,\mu_3)$}&~\\
~&~&~&{$\alpha_h$}&{$(2,r,l_1-action,0,h_1+action,h_2+l_2,h_2+l_2,\mu_1,\mu_2,\mu_2)$}&~\\
			\cline{2-6}

			~&\multirow{3}*{$h_1+action=l_2+h_2-1$}&\multirow{3}*{$ $}&{$\alpha_1$}&{$(2,r,l_1+1-action,0,h_1+action,h_2+l_2,h_2+l_2,\mu_1,\mu_2,\mu_2)$}&\multirow{3}*{$(0,0,0)$}\\
~&~&~&{$\alpha_2$}&{$(2,r,l_1-action,1,h_1+action,h_2+l_2,h_2+l_2,\mu_1,\mu_2,\mu_2)$}&~\\
~&~&~&{$\alpha_h$}&{$(2,r,l_1-action,0,h_1+action,h_2+l_2+1,h_2+l_2+1,\mu_1,\mu_2+1,\mu_2+1)$}&~\\
			\cline{2-6}	
		
			~&\multirow{3}*{$h_1+action=l_2+h_2$}&\multirow{3}*{$ $}	&{$\alpha_1$}&{$(loc,f_{12},l_1-action+1,0,h_1+action,h_2+l_2,h_2+l_2,\mu_1,\mu_2,\mu_3)$}
		&\multirow{3}*{$(0,0,0)$}\\
~&~&~&{$\alpha_2$}&{$(2,r,l_1-action,0,h_1+action,h_2+l_2+1,h_2+l_2+1,\mu_1,\mu_2,\mu_2)$}&~\\
~&~&~&{$\alpha_h\beta_2$}&{$(2,r,l_1-action,0,h_1+action,h_2+l_2+1,h_2+l_2+1,\mu_1,\mu_2+1,\mu_2+1)$}&~\\
\cline{4-6}
~&~&~&{$\alpha_h\beta_1$}&{$(2,r,l_1-action,0,0,1,1,0,1,1)$}&$(h_1+action-\mu_1,0,\mu_1)$\\
			\cline{2-6}	
		
			~&\multirow{3}*{$h_1+action>l_2+h_2$}&\multirow{3}*{$ $}&{$\alpha_1$}&{$(3,ir,l_1-action+1,0,0,0,0,0,0,0)$}&\multirow{3}*{$(h_1+action-\mu_1,0,\mu_1)$}\\
~&~&~&{$\alpha_2$}&{$(3,ir,l_1-action,1,0,0,0,0,0,0)$}&~\\
~&~&~&{$\alpha_h$}&{$(3,r,l_1-action,0,0,1,1,0,1,1)$}&~\\
			\cline{2-5}	
			\hline
			\multirow{12}*{$l_2+h_2-h_3=1$}&\multirow{3}*{$adopt$}&\multirow{3}*{$loc=2,loc=3$}&{$\alpha_1$}&{$(1,ir,1,l_2,h_3-h_2,0,h_3-h_2,h_3-h_2,0,h_3-h_2)$}&\multirow{3}*{$(0,h_2-\mu_2,\mu_2)$}\\
~&~&~&{$\alpha_2$}&{$(1,ir,0,l_2+1,h_3-h_2,0,h_3-h_2,h_3-h_2,0,h_3-h_2)$}&~\\
~&~&~&{$\alpha_h$}&{$(loc,f_{23},0,0,h_3-h_2,h_3-h_2+1,h_3-h_2+1,0,0,1)$}&~\\
\cline{2-6}

			~&\multirow{3}*{$h_1+action\leq h_3$}&\multirow{3}*{$ $}&{$\alpha_1$}&{$(loc,ir,l_1-action+1,l_2,h_1+action,h_2,h_3,\mu_1,\mu_2,\mu_3)$}&\multirow{3}*{$(0,0,0)$}\\
~&~&~&{$\alpha_2$}&{$(loc,ir,l_1-action,l_2+1,h_1+action,h_2,h_3,\mu_1,\mu_2,\mu_3)$}&~\\
~&~&~&{$\alpha_h$}&{$(loc,f_{23},l_1-action,0,h_1+action,h_3+1,h_3+1,\mu_1,\mu_2,\mu_3+1)$}&~\\
			\cline{2-6}

			~&\multirow{3}*{$h_1+action=l_2+h_2$}&\multirow{3}*{$ $}&{$\alpha_1$}&{$(loc,f_{12},l_1-action+1,0,h_1+action,h_2+l_2,h_2+l_2,\mu_1,\mu_2,\mu_3)$}&\multirow{3}*{$(0,0,0)$}\\
~&~&~&{$\alpha_2$}&{$(2,r,l_1-action,0,h_1+action,h_2+l_2+1,h_2+l_2+1,\mu_1,\mu_2,\mu_2)$}&~\\
~&~&~&{$\alpha_h\beta_2$}&{$(2,r,l_1-action,0,h_1+action,h_2+l_2+1,h_2+l_2+1,\mu_1,\mu_2+1,\mu_2+1)$}&~\\
\cline{4-6}
~&~&~&{$\alpha_h\beta_1$}&{$(2,r,l_1-action,0,0,1,1,0,1,1)$}&$(h_1+action-\mu_1,0,\mu_1)$\\
			\cline{2-6}	
			
			~&\multirow{3}*{$h_1+action>l_2+h_2$}&\multirow{3}*{$ $}&{$\alpha_1$}&{$(3,ir,l_1-action+1,0,0,0,0,0,0,0)$}&\multirow{3}*{$(h_1+action-\mu_1,0,\mu_1)$}\\
~&~&~&{$\alpha_2$}&{$(3,ir,l_1-action,1,0,0,0,0,0,0)$}&~\\
~&~&~&{$\alpha_h$}&{$(3,r,l_1-action,0,0,1,1,0,1,1)$}&~\\
			\hline
					\end{tabular}}
		\label{table:MDP}
	\end{table*}

	      	\begin{table*}[h]
		\centering  
		\resizebox{1\textwidth}{!}{
		\begin{tabular}{|c|c|c|c|c|c|c|}
		\hline
\multirow{54}*{$l_2+h_2=h_3$}&\multirow{16}*{$adopt$}&\multirow{5}*{$f_{23},f_{123},loc=2$}&{$\alpha_1\gamma_2$}&{$(1,ir,0,0,0,0,0,0,0,0)$}&{$(1,h_2-\mu_2,\mu_2)$}\\
~&~&~&{$\alpha_1(1-\gamma_2)$}&{$(3,ir,0,0,0,0,0,0,0,0)$}&$(1,h_3-\mu_3,\mu_3)$\\
~&~&~&{$\alpha_2$}&{$(2,ir,0,0,0,0,0,0,0,0)$}&$(0,h_2+1-\mu_2,\mu_2)$\\
~&~&~&{$\alpha_h\gamma_2$}&{$(2,r,0,0,0,0,0,0,0,0)$}&$(0,h_2-\mu_2,\mu_2+1)$\\
~&~&~&{$\alpha_h(1-\gamma_2)$}&{$(3,r,0,0,0,0,0,0,0,0)$}&$(0,h_3-\mu_3,\mu_3+1)$\\
\cline{3-6}
~&~&\multirow{3}*{$f_{23},f_{123},loc!=2$}&{$\alpha_1\gamma_2$}&{$(1,ir,0,0,0,0,0,0,0,0)$}&{$(1,h_2-\mu_2,\mu_2)$}\\
~&~&~&{$\alpha_1(1-\gamma_2)$}&{$(3,ir,0,0,0,0,0,0,0,0)$}&$(1+h_3-\mu_3,0,\mu_3)$\\
~&~&~&{$\alpha_2$}&{$(2,ir,0,0,0,0,0,0,0,0)$}&$(0,h_2+1-\mu_2,\mu_2)$\\
~&~&~&{$\alpha_h\gamma_2$}&{$(2,r,0,0,0,0,0,0,0,0)$}&$(0,h_2-\mu_2,\mu_2+1)$\\
~&~&~&{$\alpha_h(1-\gamma_2)$}&{$(3,r,0,0,0,0,0,0,0,0)$}&$(h_3-\mu_3,0,\mu_3+1)$\\
\cline{3-6}

~&~&\multirow{3}*{$(r,ir,loc=2),f_{12}$}&{$\alpha_1$}&{$(3,ir,1,0,0,0,0,0,0,0)$}&\multirow{3}*{$(0,h_2-\mu_2,\mu_2)$}\\
~&~&~&{$\alpha_2$}&{$(3,ir,0,1,0,0,0,0,0,0)$}&~\\
~&~&~&{$\alpha_h$}&{$(3,r,0,0,0,1,1,0,1,1)$}&~\\
\cline{3-6}

~&~&\multirow{3}*{$(r,ir,loc!=2),f_{13}$}&{$\alpha_1$}&{$(3,ir,1,0,0,0,0,0,0,0)$}&\multirow{3}*{$(h_3-\mu_3,0,\mu_3)$}\\
~&~&~&{$\alpha_2$}&{$(3,ir,0,1,0,0,0,0,0,0)$}&~\\
~&~&~&{$\alpha_h$}&{$(3,r,0,0,0,1,1,0,1,1)$}&~\\
\cline{3-6}

\cline{2-6}

~&\multirow{21}*{$action=0$}&\multirow{4}*{$f_{12}$}&{$\alpha_1$}&{$(loc,fork,l_1+1,l_2,h_1,h_2,h_3,\mu_1,\mu_2,\mu_3)$}&$(0,0,0)$\\
~&~&~&{$\alpha_2$}&{$(2,r,l_1,l_2,h_1,h_2+1,h_2+1,\mu_1,\mu_2,\mu_2)$}&$(0,0,0)$\\
~&~&~&{$\alpha_h\beta_1$}&{$(2,r,l_1,0,0,1,1,0,1,1)$}&$(h_1-\mu_1,0,\mu_1)$\\
~&~&~&{$\alpha_h\beta_2$}&{$(2,r,l_1,l_2,h_1,h_2+1,h_2+1,\mu_1,\mu_2+1,\mu_2+1)$}&$(0,0,0)$\\
\cline{3-6}

~&~&\multirow{3}*{$f_{13}$}&{$\alpha_1$}&{$(1oc,fork,l_1+1,l_2,h_1,h_2,h_3,\mu_1,\mu_2,\mu_3)$}&{$(0,0,0)$}\\
~&~&~&{$\alpha_2\gamma_1$}&{$(2,r,l_1,l_2,0,1,1,0,0,0)$}&$(h_1-\mu_1,0,\mu_1)$\\
~&~&~&{$\alpha_2(1-\gamma_1)$}&{$(2,r,l_1,l_2,h_1,h_3+1,h_3+1,\mu_1,\mu_3,\mu_3)$}&$(0,0,0)$\\
~&~&~&{$\alpha_h\gamma_1$}&{$(2,r,l_1,0,0,1,1,0,1,1)$}&$(h_1-\mu_1,0,\mu_1)$\\
~&~&~&{$\alpha_h(1-\gamma_1)$}&{$(2,r,l_1,l_2,h_1,h_3+1,h_3+1,\mu_1,\mu_3+1,\mu_3+1)$}&$(0,0,0)$\\
\cline{3-6}

~&~&\multirow{3}*{$f_{23}$}&{$\alpha_1$}&{$(loc,fork,l_1+1,l_2,h_1,h_2,h_3,\mu_1,\mu_2,\mu_3)$}&\multirow{4}*{$(0,0,0)$}\\
~&~&~&{$\alpha_2$}&{$(2,r,l_1,0,h_1,h_2+1,h_2+1,\mu_1,\mu_2,\mu_2)$}&~\\
~&~&~&{$\alpha_h\gamma_2$}&{$(2,r,l_1,0,h_1,h_2+1,h_2+1,\mu_1,\mu_2+1,\mu_2+1)$}&~\\
~&~&~&{$\alpha_h(1-\gamma_2)$}&{$(2,r,l_1,0,h_1,h_2+1,h_2+1,\mu_1,\mu_3+1,\mu_3+1)$}&~\\
\cline{3-6}

~&~&\multirow{3}*{$f_{123}$}&{$\alpha_1$}&{$(loc,fork,l_1+1,l_2,h_1,h_2,h_3,\mu_1,\mu_2,\mu_3)$}&{$(0,0,0)$}\\
~&~&~&{$\alpha_2$}&{$(2,r,l_1,0,h_1,h_2+1,h_2+1,\mu_1,\mu_2,\mu_2)$}&$(0,0,0)$\\
~&~&~&{$\alpha_h\theta_1$}&{$(2,r,l_1,0,0,1,1,0,1,1)$}&$(h_1-\mu_1,0,\mu_1)$\\
~&~&~&{$\alpha_h\theta_2$}&{$(2,r,l_1,0,h_1,h_3+1,h_3+1,\mu_1,\mu_2+1,\mu_2+1)$}&$(0,0,0)$\\
~&~&~&{$\alpha_h(1-\theta_1-\theta_2)$}&{$(2,r,l_1,0,h_1,h_3+1,h_3+1,\mu_1,\mu_3+1,\mu_3+1)$}&$(0,0,0)$\\
\cline{3-6}

~&~&\multirow{3}*{$r,ir$}&{$\alpha_1$}&{$(loc,ir,l_1+1,l_2,h_1,h_2,h_3,\mu_1,\mu_2,\mu_3)$}&\multirow{3}*{$(0,0,0)$}\\
~&~&~&{$\alpha_2$}&{$(loc,ir,l_1,l_2+1,h_1,h_2,h_3,\mu_1,\mu_2,\mu_3)$}&~\\
~&~&~&{$\alpha_h$}&{$(loc,ir,l_1,l_2,h_1,h_2+1,h_3+1,\mu_1,\mu_2+1,\mu_3+1)$}&~\\

\cline{2-6}

~&\multirow{12}*{$action=h_3-h_1>0$}&\multirow{5}*{$f_{23}$}&{$\alpha_1$}&{$(loc,f_{123},l_1-action+1,l_2,h_1+action,h_2,h_3,\mu_1,\mu_2,\mu_3)$}&{$(0,0,0)$}\\
~&~&~&{$\alpha_2$}&{$(2,r,l_1-action,0,h_1+action,h_2+1,h_3+1,\mu_1,\mu_2,\mu_2)$}&$(0,0,0)$\\
~&~&~&{$\alpha_h\theta_1$}&{$(2,r,l_1-action,l_2,0,1,1,0,1,1)$}&$(h_1-\mu_1,0,\mu_1)$\\
~&~&~&{$\alpha_h\theta_2$}&{$(2,r,l_1-action,l_2,h_1+action,h_2+1,h_3+1,\mu_1,\mu_2+1,\mu_2+1)$}&$(0,0,0)$\\
~&~&~&{$\alpha_h(1-\theta_1-\theta_2)$}&{$(2,r,l_1-action,l_2,h_1+action,h_2+1,h_3+1,\mu_1,\mu_3+1,\mu_3+1)$}&$(0,0,0)$\\
\cline{3-6}

~&~&\multirow{3}*{$r,h_2=\mu_2$}&{$\alpha_1$}&{$(loc,f_{13},l_1-action+1,l_2,h_1+action,h_2,h_3,\mu_1,\mu_2,\mu_3)$}&{$(0,0,0)$}\\
~&~&~&{$\alpha_2\gamma_1$}&{$(2,r,l_1-action,l_2,0,1,1,0,0,0)$}&$(h_1+action-\mu_1,0,\mu_1)$\\
~&~&~&{$\alpha_2(1-\gamma_1)$}&{$(2,r,l_1-action,l_2,h_1+action,h_2+1,h_3+1,\mu_1,\mu_3,\mu_3)$}&$(0,0,0)$\\
~&~&~&{$\alpha_h\gamma_1$}&{$(2,r,l_1-action,l_2,0,1,1,0,1,1)$}&$(h_1+action-\mu_1,0,\mu_1)$\\
~&~&~&{$\alpha_h(1-\gamma_1)$}&{$(2,r,l_1-action,l_2,h_1+action,h_2+1,h_3+1,\mu_1,\mu_3+1,\mu_3+1)$}&$(0,0,0)$\\
\cline{3-6}

~&~&\multirow{3}*{$r,h_2!=\mu_2$}&{$\alpha_1$}&{$(loc,f_{12},l_1-action+1,l_2,h_1+action,h_2,h_3,\mu_1,\mu_2,\mu_3)$}&{$(0,0,0)$}\\
~&~&~&{$\alpha_2$}&{$(2,r,l_1-action,l_2,h_1+action,h_2+1,h_3+1,\mu_1,\mu_2,\mu_2)$}&$(0,0,0)$\\
~&~&~&{$\alpha_h\beta_1$}&{$(2,r,l_1-action,l_2,0,1,1,0,1,1)$}&$(h_1+action-\mu_1,0,\mu_1)$\\
~&~&~&{$\alpha_h\beta_2$}&{$(2,r,l_1-action,l_2,h_1+action,h_2+1,h_3+1,\mu_1,\mu_3+1,\mu_3+1)$}&$(0,0,0)$\\
\cline{2-6}

			~&\multirow{3}*{$h_1+action>l_2+h_2$}&\multirow{3}*{$ $}&{$\alpha_1$}&{$(3,ir,l_1-action+1,0,0,0,0,0,0,0)$}&\multirow{3}*{$(h_1+action-\mu_1,0,\mu_1)$}\\
~&~&~&{$\alpha_2$}&{$(3,ir,l_1-action,1,0,0,0,0,0,0)$}&~\\
~&~&~&{$\alpha_h$}&{$(3,r,l_1-action,0,0,1,1,0,1,1)$}&~\\
			
			\hline

		\end{tabular}}
	\end{table*}

	      	\begin{table*}[h]
		\centering  
		\caption{A description of the transition and reward matrices $\mathcal{P}$ and $\mathcal{R}$ in the decision problem $\mathcal{M}_{PO}$.}
		\resizebox{1\textwidth}{!}{
		\begin{tabular}{|c|c|c|c|c|c|c|}
			\hline  
\multirow{31}*{$l_2+h_2-h_3>2$}&\multirow{6}*{$adopt$}&\multirow{4}*{$loc!=2$}&{$(1-p)$}&{$(loc,ir,0,l_2,h_3,h_2,h_3,\mu_3,\mu_2,\mu_3)$}&\multirow{4}*{$(0,0,0)$}\\
~&~&~&{$\alpha_1p$}&{$(1,ir,1,l_2,h_3,h_2,h_3,\mu_3,\mu_2,\mu_3)$}&~\\
~&~&~&{$\alpha_2p$}&{$(1,ir,0,l_2+1,h_3,h_2,h_3,\mu_3,\mu_2,\mu_3)$}&~\\
~&~&~&{$\alpha_hp$}&{$(1,r,0,l_2,h_3,h_2,h_3+1,\mu_3,\mu_2,\mu_3+1)$}&~\\
			\cline{3-6}	
						~&~&\multirow{4}*{$loc=2$}&{$(1-p)$}&{$(3,ir,0,l_2,h_3-h_2,0,h_3-h_2,h_3-h_2,0,h_3-h_2)$}&\multirow{3}*{$(0,h_2-\mu_2,\mu_2)$}\\
			~&~&~&{$\alpha_1p$}&{$(1,ir,1,l_2,h_3-h_2,0,h_3-h_2,h_3-h_2,0,h_3-h_2)$}&~\\
~&~&~&{$\alpha_2$p}&{$(1,ir,0,l_2+1,h_3-h_2,0,h_3-h_2,h_3-h_2,0,h_3-h_2)$}&~\\
~&~&~&{$\alpha_hp$}&{$(1,r,0,l_2,h_3-h_2,0,h_3-h_2+1,h_3-h_2,0,h_3-h_2+1)$}&~\\
\cline{2-6}

			~&\multirow{3}*{$h_1+action\leq h_3$}&\multirow{2}*{$r,action \neq (h_3-h_1) $}&{$1-p$}&{$(loc,ir,l_1-action,l_2,h_1+action,h_2,h_3,\mu_1,\mu_2,\mu_3)$}&\multirow{4}*{$(0,0,0)$}\\
			~&~&~&{$\alpha_1p$}&{$(loc,ir,l_1-action+1,l_2,h_1+action,h_2,h_3,\mu_1,\mu_2,\mu_3)$}&~\\
~&~&\multirow{2}*{$ir$}&{$\alpha_2p$}&{$(loc,ir,l_1-action,l_2+1,h_1+action,h_2,h_3,\mu_1,\mu_2,\mu_3)$}&~\\
~&~&~&{$\alpha_hp$}&{$(loc,r,l_1-action,l_2,h_1+action,h_2,h_3+1,\mu_1,\mu_2,\mu_3+1)$}&~\\
			\cline{2-6}

			~&\multirow{4}*{$h_1+action=h_3$}&~&{$(1-p)$}&{$(loc,f_{13},l_1-action,l_2,h_3,h_2,h_3,\mu_1,\mu_2,\mu_3)$}&\multirow{5}*{$(0,0,0)$}\\
			~&~&{$r,action>0$}&{$\alpha_1p$}&{$(loc,f_{13},l_1-action+1,l_2,h_3,h_2,h_3,\mu_1,\mu_2,\mu_3)$}&~\\
~&~&~&{$\alpha_2p$}&{$(loc,f_{13},l_1-action,l_2+1,h_3,h_2,h_3,\mu_1,\mu_2,\mu_3)$}&~\\
~&~&\multirow{2}*{$f_{13},action=0$}&{$\alpha_hp\gamma_1$}&{$(1,r,l_1-action,l_2,h_3,h_2,h_3+1,\mu_1,\mu_2,\mu_1+1)$}&~\\
~&~&~&{$\alpha_hp(1-\gamma_1)$}&{$(loc,r,l_1-action,l_2,h_3,h_2,h_3+1,\mu_1,\mu_2,\mu_3+1)$}&~\\
			\cline{2-6}

			~&\multirow{4}*{$h_3<h_1+action<l_2+h_2-1$}&\multirow{4}*{$ $}&{$(1-p)$}&{$(1,ir,l_1-action,l_2,h_1+action,h_2,h_1+action,\mu_1,\mu_2,\mu_1)$}&\multirow{3}*{$(0,0,0)$}\\
			~&~&&{$\alpha_1p$}&{$(1,ir,l_1-action+1,l_2,h_1+action,h_2,h_1+action,\mu_1,\mu_2,\mu_1)$}&~\\
~&~&~&{$\alpha_2p$}&{$(1,ir,l_1-action,l_2+1,h_1+action,h_2,h_1+action,\mu_1,\mu_2,\mu_1)$}&~\\
~&~&~&{$\alpha_hp$}&{$(1,r,l_1-action,l_2,h_1+action,h_2,h_1+action+1,\mu_1,\mu_2,\mu_1+1)$}&~\\
			\cline{2-6}	
			
			~&\multirow{4}*{$h_1+action=l_2+h_2-1$}&\multirow{4}*{$ $}&{$(1-p)$}&{$(2,ir,l_1-action,0,h_1+action,h_2+l_2,h_2+l_2,\mu_1,\mu_2,\mu_2)$}&\multirow{4}*{$(0,0,0)$}\\
			~&~&~&{$\alpha_1p$}&{$(2,r,l_1+1-action,0,h_1+action,h_2+l_2,h_2+l_2,\mu_1,\mu_2,\mu_2)$}&~\\
~&~&~&{$\alpha_2p$}&{$(2,r,l_1-action,1,h_1+action,h_2+l_2,h_2+l_2,\mu_1,\mu_2,\mu_2)$}&~\\
~&~&~&{$\alpha_hp$}&{$(2,r,l_1-action,0,h_1+action,h_2+l_2+1,h_2+l_2+1,\mu_1,\mu_2+1,\mu_2+1)$}&~\\
			\cline{2-6}	
		
			~&\multirow{6}*{$h_1+action=l_2+h_2$}&\multirow{3}*{$ $}&{$1-p$}&{$(loc,f_{12},l_1-action,0,h_1+action,h_2+l_2,h_2+l_2,\mu_1,\mu_2,\mu_3)$}&\multirow{3}*{$(0,0,0)$}\\
			~&~&~&{$\alpha_1p$}&{$(loc,f_{12},l_1-action+1,0,h_1+action,h_2+l_2,h_2+l_2,\mu_1,\mu_2,\mu_3)$}&~\\
~&~&~&{$\alpha_2p$}&{$(2,r,l_1-action,0,h_1+action,h_2+l_2+1,h_2+l_2+1,\mu_1,\mu_2,\mu_2)$}&~\\
~&~&~&{$\alpha_hp\beta_2$}&{$(2,r,l_1-action,0,h_1+action,h_2+l_2+1,h_2+l_2+1,\mu_1,\mu_2+1,\mu_2+1)$}&~\\
\cline{4-6}
~&~&~&{$\alpha_hp\beta_1$}&{$(2,r,l_1-action,0,0,1,1,0,1,1)$}&$(h_1+action-\mu_1,0,\mu_1)$\\
			\cline{2-6}	
			
			~&\multirow{4}*{$h_1+action>l_2+h_2$}&\multirow{4}*{$ $}&{$(1-p)$}&{$(3,ir,l_1-action,0,0,0,0,0,0,0)$}&\multirow{4}*{$(h_1+action-\mu_1,0,\mu_1)$}\\
			~&~&~&{$\alpha_1p$}&{$(3,ir,l_1-action+1,0,0,0,0,0,0,0)$}&~\\
~&~&~&{$\alpha_2p$}&{$(3,ir,l_1-action,1,0,0,0,0,0,0)$}&~\\
~&~&~&{$\alpha_hp$}&{$(3,r,l_1-action,0,0,1,1,0,1,1)$}&~\\
			\hline

			\multirow{31}*{$l_2+h_2-h_3=2$}&\multirow{8}*{$adopt$}&\multirow{3}*{$loc!=2$}&{$(1-p)$}&{$(loc,ir,0,l_2,h_3,h_2,h_3,\mu_3,\mu_2,\mu_3)$}&\multirow{3}*{$(0,0,0)$}\\
			~&~&~&{$\alpha_1p$}&{$(1,ir,1,l_2,h_3,h_2,h_3,\mu_3,\mu_2,\mu_3)$}&~\\
~&~&~&{$\alpha_2p$}&{$(1,ir,0,l_2+1,h_3,h_2,h_3,\mu_3,\mu_2,\mu_3)$}&~\\
~&~&~&{$\alpha_hp$}&{$(2,r,0,0,h_3,h_2+l_2,h_2+l_2,\mu_3,\mu_2,\mu_2)$}&~\\
			\cline{3-6}	
			~&~&\multirow{4}*{$loc=2$}&{$(1-p)$}&{$(3,ir,0,l_2,h_3-h_2,0,h_3-h_2,h_3-h_2,0,h_3-h_2)$}&\multirow{4}*{$(0,h_2-\mu_2,\mu_2)$}\\
			~&~&~&{$\alpha_1p$}&{$(1,ir,1,l_2,h_3-h_2,0,h_3-h_2,h_3-h_2,0,h_3-h_2)$}&~\\
~&~&~&{$\alpha_2p$}&{$(1,ir,0,l_2+1,h_3-h_2,0,h_3-h_2,h_3-h_2,0,h_3-h_2)$}&~\\
~&~&~&{$\alpha_hp$}&{$(2,r,0,0,h_3-h_2,l_2,l_2,h_3-h_2,0,0)$}&~\\
\cline{2-6}

			~&\multirow{4}*{$h_1+action \leq h_3$}&\multirow{2}*{$r,action \neq (h_3-h_1) $}&{$(1-p)$}&{$(loc,ir,l_1-action,l_2,h_1+action,h_2,h_3,\mu_1,\mu_2,\mu_3)$}&\multirow{3}*{$(0,0,0)$}\\
			~&~&~&{$\alpha_1p$}&{$(loc,ir,l_1-action+1,l_2,h_1+action,h_2,h_3,\mu_1,\mu_2,\mu_3)$}&~\\
~&~&\multirow{2}*{$ir $}&{$\alpha_2p$}&{$(loc,ir,l_1-action,l_2+1,h_1+action,h_2,h_3,\mu_1,\mu_2,\mu_3)$}&~\\
~&~&~&{$\alpha_hp$}&{$(2,r,l_1-action,0,h_1+action,h_2+l_2,h_2+l_2,\mu_1,\mu_2,\mu_2)$}&~\\
			\cline{2-6}

			~&\multirow{4}*{$h_1+action=h_3$}&\multirow{1.5}*{$r,action>0$}&{$(1-p)$}&{$(loc,f_{13},l_1-action,l_2,h_3,h_2,h_3,\mu_1,\mu_2,\mu_3)$}&\multirow{4}*{$(0,0,0)$}\\
			~&~&~&{$\alpha_1p$}&{$(loc,f_{13},l_1-action+1,l_2,h_3,h_2,h_3,\mu_1,\mu_2,\mu_3)$}&~\\
~&~&\multirow{1.5}*{$f_{13},action=0$}&{$\alpha_2p$}&{$(loc,f_{13},l_1-action,l_2+1,h_3,h_2,h_3,\mu_1,\mu_2,\mu_3)$}&~\\
~&~&~&{$\alpha_hp$}&{$(2,r,l_1-action,0,h_1+action,h_2+l_2,h_2+l_2,\mu_1,\mu_2,\mu_2)$}&~\\
			\cline{2-6}

			~&\multirow{4}*{$h_1+action=l_2+h_2-1$}&\multirow{3}*{$ $}&{$(1-p)$}&{$(2,ir,l_1-action,0,h_1+action,h_2+l_2,h_2+l_2,\mu_1,\mu_2,\mu_2)$}&\multirow{3}*{$(0,0,0)$}\\
			~&~&~&{$\alpha_1p$}&{$(2,r,l_1+1-action,0,h_1+action,h_2+l_2,h_2+l_2,\mu_1,\mu_2,\mu_2)$}&~\\
~&~&~&{$\alpha_2p$}&{$(2,r,l_1-action,1,h_1+action,h_2+l_2,h_2+l_2,\mu_1,\mu_2,\mu_2)$}&~\\
~&~&~&{$\alpha_hp$}&{$(2,r,l_1-action,0,h_1+action,h_2+l_2+1,h_2+l_2+1,\mu_1,\mu_2+1,\mu_2+1)$}&~\\
			\cline{2-6}	
		
			~&\multirow{5}*{$h_1+action=l_2+h_2$}&\multirow{5}*{$ $}&{$(1-p)$}&{$(loc,f_{12},l_1-action,0,h_1+action,h_2+l_2,h_2+l_2,\mu_1,\mu_2,\mu_3)$}&\multirow{3}*{$(0,0,0)$}\\
			~&~&~&{$\alpha_1p$}&{$(loc,f_{12},l_1-action+1,0,h_1+action,h_2+l_2,h_2+l_2,\mu_1,\mu_2,\mu_3)$}&~\\
~&~&~&{$\alpha_2p$}&{$(2,r,l_1-action,0,h_1+action,h_2+l_2+1,h_2+l_2+1,\mu_1,\mu_2,\mu_2)$}&~\\
~&~&~&{$\alpha_hp\beta_2$}&{$(2,r,l_1-action,0,h_1+action,h_2+l_2+1,h_2+l_2+1,\mu_1,\mu_2+1,\mu_2+1)$}&~\\
\cline{4-6}
~&~&~&{$\alpha_hp\beta_1$}&{$(2,r,l_1-action,0,0,1,1,0,1,1)$}&$(h_1+action-\mu_1,0,\mu_1)$\\
			\cline{2-6}	
		
			~&\multirow{4}*{$h_1+action>l_2+h_2$}&\multirow{4}*{$ $}&{$(1-p)$}&{$(3,ir,l_1-action,0,0,0,0,0,0,0)$}&\multirow{4}*{$(h_1+action-\mu_1,0,\mu_1)$}\\
			~&~&~&{$\alpha_1p$}&{$(3,ir,l_1-action+1,0,0,0,0,0,0,0)$}&~\\
~&~&~&{$\alpha_2p$}&{$(3,ir,l_1-action,1,0,0,0,0,0,0)$}&~\\
~&~&~&{$\alpha_hp$}&{$(3,r,l_1-action,0,0,1,1,0,1,1)$}&~\\
			\cline{2-5}	
			\hline

			\multirow{17}*{$l_2+h_2-h_3=1$}&\multirow{4}*{$adopt$}&\multirow{3}*{$loc=2,loc=3$}&{$(1-p)$}&{$(1,ir,0,l_2,h_3-h_2,0,h_3-h_2,h_3-h_2,0,h_3-h_2)$}&\multirow{4}*{$(0,h_2-\mu_2,\mu_2)$}\\
	~&~&~&{$\alpha_1p$}&{$(1,ir,1,l_2,h_3-h_2,0,h_3-h_2,h_3-h_2,0,h_3-h_2)$}&~\\
~&~&~&{$\alpha_2p$}&{$(1,ir,0,l_2+1,h_3-h_2,0,h_3-h_2,h_3-h_2,0,h_3-h_2)$}&~\\
~&~&~&{$\alpha_hp$}&{$(loc,f_{23},0,0,h_3-h_2,h_3-h_2+1,h_3-h_2+1,0,0,1)$}&~\\
\cline{2-6}

			~&\multirow{4}*{$h_1+action\leq h_3$}&\multirow{4}*{$ $}&{$(1-p)$}&{$(loc,ir,l_1-action,l_2,h_1+action,h_2,h_3,\mu_1,\mu_2,\mu_3)$}&\multirow{4}*{$(0,0,0)$}\\
			~&~&~&{$\alpha_1p$}&{$(loc,ir,l_1-action+1,l_2,h_1+action,h_2,h_3,\mu_1,\mu_2,\mu_3)$}&~\\
~&~&~&{$\alpha_2p$}&{$(loc,ir,l_1-action,l_2+1,h_1+action,h_2,h_3,\mu_1,\mu_2,\mu_3)$}&~\\
~&~&~&{$\alpha_hp$}&{$(loc,f_{23},l_1-action,0,h_1+action,h_3+1,h_3+1,\mu_1,\mu_2,\mu_3+1)$}&~\\
			\cline{2-6}

			~&\multirow{4}*{$h_1+action=l_2+h_2$}&\multirow{4}*{$ $}&{$(1-p)$}&{$(loc,f_{12},l_1-action,0,h_1+action,h_2+l_2,h_2+l_2,\mu_1,\mu_2,\mu_3)$}&\multirow{4}*{$(0,0,0)$}\\
			~&~&~&{$\alpha_1p$}&{$(loc,f_{12},l_1-action+1,0,h_1+action,h_2+l_2,h_2+l_2,\mu_1,\mu_2,\mu_3)$}&~\\
~&~&~&{$\alpha_2p$}&{$(2,r,l_1-action,0,h_1+action,h_2+l_2+1,h_2+l_2+1,\mu_1,\mu_2,\mu_2)$}&~\\
~&~&~&{$\alpha_hp\beta_2$}&{$(2,r,l_1-action,0,h_1+action,h_2+l_2+1,h_2+l_2+1,\mu_1,\mu_2+1,\mu_2+1)$}&~\\
\cline{4-6}
~&~&~&{$\alpha_hp\beta_1$}&{$(2,r,l_1-action,0,0,1,1,0,1,1)$}&$(h_1+action-\mu_1,0,\mu_1)$\\
			\cline{2-6}	
			
			~&\multirow{4}*{$h_1+action>l_2+h_2$}&\multirow{4}*{$ $}&{$(1-p)$}&{$(3,ir,l_1-action,0,0,0,0,0,0,0)$}&\multirow{4}*{$(h_1+action-\mu_1,0,\mu_1)$}\\
			~&~&~&{$\alpha_1p$}&{$(3,ir,l_1-action+1,0,0,0,0,0,0,0)$}&~\\
~&~&~&{$\alpha_2p$}&{$(3,ir,l_1-action,1,0,0,0,0,0,0)$}&~\\
~&~&~&{$\alpha_hp$}&{$(3,r,l_1-action,0,0,1,1,0,1,1)$}&~\\
			\hline
					\end{tabular}}
		\label{table:POMDP}
	\end{table*}

	      	\begin{table*}[h]
		\centering  
		\resizebox{1\textwidth}{!}{
		\begin{tabular}{|c|c|c|c|c|c|c|}
		\hline
\multirow{67}*{$l_2+h_2=h_3$}&\multirow{16}*{$adopt$}&\multirow{6}*{$f_{23},f_{123},loc=2$}&{$(1-p)$}&{$(loc,f_{23},0,l_2,0,\mu_3-\mu_2,\mu_3-\mu_2,0,0,\mu_3-\mu_2)$}&{$(0,h_3-\mu_3,\mu_2)$}\\
~&~&~&{$\alpha_1p\gamma_2$}&{$(1,ir,0,0,0,0,0,0,0,0)$}&{$(1,h_2-\mu_2,\mu_2)$}\\
~&~&~&{$\alpha_1p(1-\gamma_2)$}&{$(3,ir,0,0,0,0,0,0,0,0)$}&$(1,h_3-\mu_3,\mu_3)$\\
~&~&~&{$\alpha_2p$}&{$(2,ir,0,0,0,0,0,0,0,0)$}&$(0,h_2+1-\mu_2,\mu_2)$\\
~&~&~&{$\alpha_hp\gamma_2$}&{$(2,r,0,0,0,0,0,0,0,0)$}&$(0,h_2-\mu_2,\mu_2+1)$\\
~&~&~&{$\alpha_hp(1-\gamma_2)$}&{$(3,r,0,0,0,0,0,0,0,0)$}&$(0,h_3-\mu_3,\mu_3+1)$\\
\cline{3-6}
~&~&\multirow{6}*{$f_{23},f_{123},loc!=2$}&{$(1-p)$}&{$(loc,f_{23},0,l_2,h_3,h_2,h_3,\mu_3,\mu_2,\mu_3)$}&{$(0,0,0)$}\\
~&~&~&{$\alpha_1p\gamma_2$}&{$(1,ir,0,0,0,0,0,0,0,0)$}&{$(1,h_2-\mu_2,\mu_2)$}\\
~&~&~&{$\alpha_1p(1-\gamma_2)$}&{$(3,ir,0,0,0,0,0,0,0,0)$}&$(1+h_3-\mu_3,0,\mu_3)$\\
~&~&~&{$\alpha_2p$}&{$(2,ir,0,0,0,0,0,0,0,0)$}&$(0,h_2+1-\mu_2,\mu_2)$\\
~&~&~&{$\alpha_hp\gamma_2$}&{$(2,r,0,0,0,0,0,0,0,0)$}&$(0,h_2-\mu_2,1+\mu_2)$\\
~&~&~&{$\alpha_hp(1-\gamma_2)$}&{$(3,r,0,0,0,0,0,0,0,0)$}&$(h_3-\mu_3,0,\mu_3+1)$\\
\cline{3-6}

~&~&\multirow{4}*{$(r,ir,loc=2),f_{12}$}&{$(1-p)$}&{$(3,ir,0,0,0,0,0,0,0,0)$}&\multirow{4}*{$(0,h_2-\mu_2,\mu_2)$}\\
~&~&~&{$\alpha_1p$}&{$(3,ir,1,0,0,0,0,0,0,0)$}&~\\
~&~&~&{$\alpha_2p$}&{$(3,ir,0,1,0,0,0,0,0,0)$}&~\\
~&~&~&{$\alpha_hp$}&{$(3,r,0,0,0,1,1,0,1,1)$}&~\\
\cline{3-6}

~&~&\multirow{4}*{$(r,ir,loc!=2),f_{13}$}&{$(1-p)$}&{$(3,ir,0,0,0,0,0,0,0,0)$}&\multirow{4}*{$(h_3-\mu_3,0,\mu_3)$}\\

~&~&~&{$\alpha_1p$}&{$(3,ir,1,0,0,0,0,0,0,0)$}&~\\
~&~&~&{$\alpha_2p$}&{$(3,ir,0,1,0,0,0,0,0,0)$}&~\\
~&~&~&{$\alpha_hp$}&{$(3,r,0,0,0,1,1,0,1,1)$}&~\\
\cline{3-6}

\cline{2-6}

~&\multirow{26}*{$action=0$}&\multirow{5}*{$f_{12}$}&{$(1-p)$}&{$(loc,fork,l_1,l_2,h_1,h_2,h_3,\mu_1,\mu_2,\mu_3)$}&{$(0,0,0)$}\\
~&~&~&{$\alpha_1p$}&{$(loc,fork,l_1+1,l_2,h_1,h_2,h_3,\mu_1,\mu_2,\mu_3)$}&$(0,0,0)$\\
~&~&~&{$\alpha_2p$}&{$(2,r,l_1,l_2,h_1,h_2+1,h_2+1,\mu_1,\mu_2,\mu_2)$}&$(0,0,0)$\\
~&~&~&{$\alpha_hp\beta_1$}&{$(2,r,l_1,l_2,0,1,1,0,1,1)$}&$(h_1-\mu_1,0,\mu_1)$\\
~&~&~&{$\alpha_hp\beta_2$}&{$(2,r,l_1,l_2,h_1,h_2+1,h_3+1,\mu_1,\mu_2+1,\mu_2+1)$}&$(0,0,0)$\\
\cline{3-6}

~&~&\multirow{6}*{$f_{13}$}&{$(1-p)$}&{$(1oc,fork,l_1,l_2,h_1,h_2,h_3,\mu_1,\mu_2,\mu_3)$}&{$(0,0,0)$}\\

~&~&~&{$\alpha_1p$}&{$(1oc,fork,l_1+1,l_2,h_1,h_2,h_3,\mu_1,\mu_2,\mu_3)$}&{$(0,0,0)$}\\
~&~&~&{$\alpha_2p\gamma_1$}&{$(2,r,l_1,l_2,0,1,1,0,0,0)$}&$(h_1-\mu_1,0,\mu_1)$\\
~&~&~&{$\alpha_2p(1-\gamma_1)$}&{$(2,r,l_1,l_2,h_1,h_3+1,h_3+1,\mu_1,\mu_3,\mu_3)$}&$(0,0,0)$\\
~&~&~&{$\alpha_hp\gamma_1$}&{$(2,r,l_1,0,0,1,1,0,1,1)$}&$(h_1-\mu_1,0,\mu_1)$\\
~&~&~&{$\alpha_hp(1-\gamma_1)$}&{$(2,r,l_1,l_2,h_1,h_3+1,h_3+1,\mu_1,\mu_3+1,\mu_3+1)$}&$(0,0,0)$\\
\cline{3-6}

~&~&\multirow{5}*{$f_{23}$}&{$(1-p)$}&{$(loc,fork,l_1,l_2,h_1,h_2,h_3,\mu_1,\mu_2,\mu_3)$}&\multirow{5}*{$(0,0,0)$}\\
~&~&~&{$\alpha_1p$}&{$(loc,fork,l_1+1,l_2,h_1,h_2,h_3,\mu_1,\mu_2,\mu_3)$}&~\\
~&~&~&{$\alpha_2p$}&{$(2,r,l_1,0,h_1,h_2+1,h_2+1,\mu_1,\mu_2,\mu_2)$}&~\\
~&~&~&{$\alpha_hp\gamma_2$}&{$(2,r,l_1,0,h_1,h_2+1,h_2+1,\mu_1,\mu_2+1,\mu_2+1)$}&~\\
~&~&~&{$\alpha_hp(1-\gamma_2)$}&{$(2,r,l_1,0,h_1,h_2+1,h_2+1,\mu_1,\mu_3+1,\mu_3+1)$}&~\\
\cline{3-6}

~&~&\multirow{6}*{$f_{123}$}&{$(1-p)$}&{$(loc,fork,l_1,l_2,h_1,h_2,h_3,\mu_1,\mu_2,\mu_3)$}&{$(0,0,0)$}\\

~&~&~&{$\alpha_1p$}&{$(loc,fork,l_1+1,l_2,h_1,h_2,h_3,\mu_1,\mu_2,\mu_3)$}&{$(0,0,0)$}\\
~&~&~&{$\alpha_2p$}&{$(2,r,l_1,0,h_1,h_2+1,h_2+1,\mu_1,\mu_2,\mu_2)$}&$(0,0,0)$\\
~&~&~&{$\alpha_hp\theta_1$}&{$(2,r,l_1,0,0,1,1,0,1,1)$}&$(h_1-\mu_1,0,\mu_1)$\\
~&~&~&{$\alpha_hp\theta_2$}&{$(2,r,l_1,0,h_1,h_3+1,h_3+1,\mu_1,\mu_2+1,\mu_2+1)$}&$(0,0,0)$\\
~&~&~&{$\alpha_hp(1-\theta_1-\theta_2)$}&{$(2,r,l_1,0,h_1,h_3+1,h_3+1,\mu_1,\mu_3+1,\mu_3+1)$}&$(0,0,0)$\\
\cline{3-6}

~&~&\multirow{3}*{$r,ir$}&{$(1-p)$}&{$(loc,ir,l_1,l_2,h_1,h_2,h_3,\mu_1,\mu_2,\mu_3)$}&\multirow{4}*{$(0,0,0)$}\\
~&~&~&{$\alpha_1p$}&{$(loc,ir,l_1+1,l_2,h_1,h_2,h_3,\mu_1,\mu_2,\mu_3)$}&~\\
~&~&~&{$\alpha_2p$}&{$(loc,ir,l_1,l_2+1,h_1,h_2,h_3,\mu_1,\mu_2,\mu_3)$}&~\\
~&~&~&{$\alpha_hp$}&{$(loc,ir,l_1,l_2,h_1,h_2+1,h_3+1,\mu_1,\mu_2+1,\mu_3+1)$}&~\\

\cline{2-6}

~&\multirow{12}*{$action=h_3-h_1>0$}&\multirow{7}*{$f_{23}$}&{$(1-p)$}&{$(loc,f_{123},l_1-action,l_2,h_1+action,h_2,h_3,\mu_1,\mu_2,\mu_3)$}&{$(0,0,0)$}\\
~&~&~&{$\alpha_1p$}&{$(loc,f_{123},l_1-action+1,l_2,h_1+action,h_2,h_3,\mu_1,\mu_2,\mu_3)$}&{$(0,0,0)$}\\
~&~&~&{$\alpha_2p$}&{$(2,r,l_1-action,0,h_1+action,h_2+1,h_3+1,\mu_1,\mu_2,\mu_2)$}&$(0,0,0)$\\
~&~&~&{$\alpha_hp\theta_1$}&{$(2,r,l_1-action,l_2,0,1,1,0,1,1)$}&$(h_1-\mu_1,0,\mu_1)$\\
~&~&~&{$\alpha_hp\theta_2$}&{$(2,r,l_1-action,l_2,h_1+action,h_2+1,h_3+1,\mu_1,\mu_2+1,\mu_2+1)$}&$(0,0,0)$\\
~&~&~&{$\alpha_hp(1-\theta_1-\theta_2)$}&{$(2,r,l_1-action,l_2,h_1+action,h_2+1,h_3+1,\mu_1,\mu_3+1,\mu_3+1)$}&$(0,0,0)$\\
\cline{3-6}

~&~&\multirow{6}*{$r,h_2=\mu_2$}&{$(1-p)$}&{$(loc,f_{13},l_1-action,l_2,h_1+action,h_2,h_3,\mu_1,\mu_2,\mu_3)$}&{$(0,0,0)$}\\
~&~&~&{$\alpha_1p$}&{$(loc,f_{13},l_1-action+1,l_2,h_1+action,h_2,h_3,\mu_1,\mu_2,\mu_3)$}&{$(0,0,0)$}\\
~&~&~&{$\alpha_2p\gamma_1$}&{$(2,r,l_1-action,l_2,0,1,1,0,0,0)$}&$(h_1+action-\mu_1,0,\mu_1)$\\
~&~&~&{$\alpha_2p(1-\gamma_1)$}&{$(2,r,l_1-action,l_2,h_1+action,h_2+1,h_3+1,\mu_1,\mu_3,\mu_3)$}&$(0,0,0)$\\
~&~&~&{$\alpha_hp\gamma_1$}&{$(2,r,l_1-action,l_2,0,1,1,0,1,1)$}&$(h_1+action-\mu_1,0,\mu_1)$\\
~&~&~&{$\alpha_hp(1-\gamma_1)$}&{$(2,r,l_1-action,l_2,h_1+action,h_2+1,h_3+1,\mu_1,\mu_3+1,\mu_3+1)$}&$(0,0,0)$\\
\cline{3-6}

~&~&\multirow{3}*{$r,h_2!=\mu_2$}&{$(1-p)$}&{$(loc,f_{12},l_1-action,l_2,h_1+action,h_2,h_3,\mu_1,\mu_2,\mu_3)$}&{$(0,0,0)$}\\
~&~&~&{$\alpha_1p$}&{$(loc,f_{12},l_1-action+1,l_2,h_1+action,h_2,h_3,\mu_1,\mu_2,\mu_3)$}&{$(0,0,0)$}\\
~&~&~&{$\alpha_2p$}&{$(2,r,l_1-action,l_2,h_1+action,h_2+1,h_3+1,\mu_1,\mu_2,\mu_2)$}&$(0,0,0)$\\
~&~&~&{$\alpha_hp\beta_1$}&{$(2,r,l_1-action,l_2,0,1,1,0,1,1)$}&$(h_1+action-\mu_1,0,\mu_1)$\\
~&~&~&{$\alpha_hp\beta_2$}&{$(2,r,l_1-action,l_2,h_1+action,h_2+1,h_3+1,\mu_1,\mu_3+1,\mu_3+1)$}&$(0,0,0)$\\
\cline{2-6}

~&\multirow{4}*{$h_1+action>l_2+h_2$}&\multirow{4}*{$ $}&{$(1-p)$}&{$(3,ir,l_1-action,0,0,0,0,0,0,0)$}&\multirow{4}*{$(h_1+action-\mu_1,0,\mu_1)$}\\
~&~&~&{$\alpha_1p$}&{$(3,ir,l_1-action+1,0,0,0,0,0,0,0)$}&~\\
~&~&~&{$\alpha_2p$}&{$(3,ir,l_1-action,1,0,0,0,0,0,0)$}&~\\
~&~&~&{$\alpha_hp$}&{$(3,r,l_1-action,0,0,1,1,0,1,1)$}&~\\
\hline

		\end{tabular}}
	\end{table*}


\end{document}